\documentclass[twocolumn]{aastex63}
\usepackage{xspace}
\usepackage{pifont}
\usepackage{multirow}

\newcommand{\cmark}{\ding{51}}%
\newcommand{\xmark}{\ding{55}}%

\received{July 10, 2020}
\revised{July 10, 2020}
\accepted{\today}

\newcommand{\Ha}{H$\alpha$\xspace}

\newcommand{\kms}{$\rm km \, s^{-1}$\xspace}
\newcommand{\kube}{{\sc kubeviz}\xspace}
\newcommand{\sinopsis}{{\sc sinopsis}\xspace}
\newcommand{\ma}{$\rm M_\ast$\xspace}
\newcommand{\ms}{$\rm M_\odot$\xspace}

\submitjournal{ApJ}

\shorttitle{GASP galaxies in low-dense environments}
\shortauthors{B. Vulcani et al.}
\graphicspath{{./}{}}

\begin{document}

\title{GASP XXXIII. The ability of spatially resolved data to distinguish among the different physical mechanisms affecting galaxies in low-density environments}

\correspondingauthor{Benedetta Vulcani}
\email{benedetta.vulcani@inaf.it}

\author[0000-0003-0980-1499]{Benedetta Vulcani}
\affiliation{INAF- Osservatorio astronomico di Padova, Vicolo Osservatorio 5, 35122 Padova, Italy}

\author{Bianca M. Poggianti}
\affiliation{INAF- Osservatorio astronomico di Padova, Vicolo Osservatorio 5, 35122 Padova, Italy}

\author{Alessia Moretti}
\affiliation{INAF- Osservatorio astronomico di Padova, Vicolo Osservatorio 5, 35122 Padova, Italy}

\author{Andrea Franchetto}
\affiliation{Dipartimento di Fisica \& Astronomia ``Galileo Galilei'', Universit\`a di Padova, vicolo Osservatorio 3, 35122, Padova, Italy}
\affiliation{INAF- Osservatorio astronomico di Padova, Vicolo Osservatorio 5, 35122 Padova, Italy}

\author{Cecilia Bacchini}
\affiliation{INAF- Osservatorio astronomico di Padova, Vicolo Osservatorio 5, 35122 Padova, Italy}

\author{Sean McGee}
\affiliation{University of Birmingham School of Physics and Astronomy, Edgbaston, Birmingham B15 2TT, England}

\author{Yara~L. Jaff\'e }
\affiliation{Instituto de F\'isica y Astronom\'ia, Universidad de Valpara\'iso, Avda. Gran Breta\~na 1111 Valpara\'iso, Chile}

\author{Matilde Mingozzi}
\affiliation{INAF- Osservatorio astronomico di Padova, Vicolo Osservatorio 5, 35122 Padova, Italy}
\affiliation{Space Telescope Science Institute, 3700 San Martin Drive, Baltimore, MD 21218}

\author{Ariel Werle}
\affiliation{INAF- Osservatorio astronomico di Padova, Vicolo Osservatorio 5, 35122 Padova, Italy}

\author{Neven Tomi\v{c}i\'{c}}
\affiliation{INAF- Osservatorio astronomico di Padova, Vicolo Osservatorio 5, 35122 Padova, Italy}

\author{Jacopo Fritz}
\affiliation{Instituto de Radioastronom\'ia y Astrof\'isica, UNAM, Campus Morelia, A.P. 3-72, C.P. 58089, Mexico}

\author{Daniela Bettoni}
\affiliation{INAF- Osservatorio astronomico di Padova, Vicolo Osservatorio 5, 35122 Padova, Italy}

\author{Anna Wolter}
\affiliation{INAF- Osservatorio Astronomico di Brera, via Brera 28, I-20121 Milano, Italy}

\author{Marco Gullieuszik }
\affiliation{INAF- Osservatorio astronomico di Padova, Vicolo Osservatorio 5, 35122 Padova, Italy}

\begin{abstract}
Galaxies inhabit a wide range of environments and therefore are affected by different physical mechanisms. Spatially resolved maps combined with the knowledge of the hosting environment are very powerful to classify galaxies by physical process. In the context of the GAs Stripping Phenomena in galaxies (GASP), we present a study of 27 non-cluster galaxies: 24 of them were selected for showing asymmetries and disturbances in the optical morphology, suggestive of gas stripping, three of them are passive galaxies and were included to characterize the final stages of galaxy evolution. We therefore provide a panorama of the different processes taking place in low-density environments. The analysis of VLT/MUSE data allows us to separate galaxies into the following categories: Galaxy-galaxy interactions (2 galaxies), mergers (6),  ram pressure stripping (4), cosmic web stripping (2), cosmic web enhancement (5), gas accretion (3), starvation (3). In one galaxy we identify the combination of merger and ram pressure stripping. Only 6/27 of these galaxies have just a tentative classification.  We then investigate where these galaxies are located on  scaling relations determined for a sample of undisturbed galaxies.
Our analysis shows the successes and limitations of a visual optical selection in identifying the processes that deplete galaxies of their gas content and probes the power of IFU data in pinning down the acting mechanism. 
\end{abstract}



\section{Introduction} \label{sec:intro}
Galaxies can be found in environments of very different nature, ranging from very low (isolated galaxies) to high density (massive cluster cores), with filaments and groups laying in between these extreme conditions. Both the properties of galaxies and the relative fractions of different galaxy types depend on the environment \citep[e.g.,][]{Oemler1974, Dressler1980, Goto2003, Blanton2005}. In low density environments, galaxies tend to be blue, star-forming and late-type, while dense environments are dominated by red, early-type galaxies. 
The dynamics of the system, e.g. the relative speed with which galaxies move, or the characteristics of the intergalactic medium, e.g. the presence of hot gas, may influence the evolution of galaxies and thus modify their properties. Within this scenario, different physical processes have been proposed to  affect the evolutionary history of galaxies, in most cases leading to a suppression of the star formation. Among those turning galaxies into passive, some are expected to affect both the stellar and the gas components, and they can substantially change the structure of galaxies and even cause significant loss of mass. Examples include galaxy-galaxy interactions, mergers and  galaxy harassment \citep[e.g.,][]{toomre72, Moore1998}. Some others are instead expected to leave the stellar component mostly unaltered and are due to the presence of gas in the intergalactic medium that can substantially affect the gas in galaxies through mechanisms such as ram pressure \citep[e.g.,][]{Gunn1972, abadi99} or starvation \citep[e.g.,][]{Larson1980, Balogh2000, Kawata2008}. 

There are also
processes  that feed galaxies with gas, replenishing their fuel for star formation and therefore causing star formation enhancement. Gas accretion from the external regions has been observed in nearby galaxies \citep{Sancisi2008} and also predicted by cosmological models \citep{Semelin2005}. Note though that some of the aforementioned processes that eventually lead to star formation quenching can also induce short bursts of star formation: ram pressure stripping via the  compression of the available gas \citep{BekkiCouch2003, Gavazzi2003, Vulcani2018_L, Tomicic2018}; gas rich mergers through accumulation of cold gas \citep[e.g.,][]{Fujita2003, Brinchmann2004, Ostriker2011}.

The relative influence of these processes depends on several physical parameters that vary from one environment to another. For example, galaxy mergers are rare in clusters because of the large velocity dispersions of the systems and are instead favoured in less dense environments \citep[e.g.][]{mihos04}; in contrast,  ram pressure stripping is  expected to be more effective in the cluster cores because of the large velocities and higher densities of the intracluster medium, while its role in less dense environments has not been characterized in detail yet. 

While much of the literature on the impact of the hosting environments on galaxy star formation has been confined to studies of galaxy clusters, a systematic  characterization and census of all the different mechanisms in the less dense environments is  still lacking. 

A step towards  achieving an understanding of the role of the environment on galaxy evolution is the study of the spatially resolved properties of galaxies: each of the aforementioned processes are indeed expected to leave a different imprint on the gas and star distributions. Data delivering maps of galaxy properties are hence extremely useful and the advent of integral field spectroscopy (IFS) has  increased considerably our understanding of the processes that govern galaxy transformations.

Few studies have so far specifically focused on the analysis of spatially resolved properties of galaxies in the field, and never compared the signatures of the different mechanisms in homogeneous samples. 

\cite{Privon2017} investigated the properties of an interacting dwarf pair found in isolation in the local universe  with Very Large Telescope/Multi Unit Spectroscopic Explorer (VLT/MUSE) optical IFU observations, finding that starbursts in low-mass galaxy mergers may be triggered by large-scale ISM compression, and thus may be more distributed than in high mass systems.
\cite{Fossati2019} used VLT/MUSE observations to map the extended ionized gas in between the members of a group infalling into a cluster at z = 0.021. Their analysis highlights the coexistence  of different mechanisms: the group is shaped by pre-processing produced by gravitational interactions in the local group environment combined with ram pressure stripping by the global cluster halo.
\cite{DuartePuertas2019} observed the Stephan's Quintet, the prototypical compact group of galaxies in the local Universe, with the imaging Fourier transform spectrometer SITELLE, at the Canada-France-Hawaii-Telescope, to perform a deep search for intergalactic star formation, shedding light on the complicated history of this system. 

\cite{Schaefer2019TheGroups} used a statistical approach to explore the radial distribution of star formation in galaxies in the Sydney Australian Astronomical Observatory Multi-object Integral Field Spectrograph \citep[SAMI,][]{Croom2012} Galaxy  Survey as a function of their local group environment, finding that the dynamical mass of the parent halo of a galaxy is a good predictor of environmental quenching. 

In this context, the Gas Stripping Phenomena in galaxies (GASP\footnote{\url{http://web.oapd.inaf.it/gasp/index.html}}) survey is helping to shed light on gas removal processes as a function of environment and to understand in what environmental conditions such processes are efficient. The survey is based on VLT/MUSE observations and therefore delivers maps of many galaxy properties, allowing us to characterize both the ionized gas and the stellar components.  GASP explores a wide range of environments, from galaxy clusters to groups and poor groups, filaments and galaxies in isolation. Its targets are located in dark matter halos with masses spanning four orders of magnitude ($10^{11}-10^{15} M_\odot$).

Most of the GASP papers have focused on cluster galaxies \citep{Poggianti2017, Poggianti2019b, Bellhouse2017, Bellhouse2019, Fritz2017, Gullieuszik2017, Moretti2017, George2018, George2019} characterizing mainly ram pressure stripping events and connecting them to cluster properties \citep{Jaffe2018, Gullieuszik2020}. A few papers have also investigated peculiar galaxies in the field \citep{Vulcani2017c, Vulcani2018_b} and characterized galaxies in filaments \citep{Vulcani2019_fil} and galaxies belonging to the same group \citep{Vulcani2018_g}, highlighting the variety of mechanisms that take place in these environments. 

The goal of this paper is to present a comprehensive characterization of all the field galaxies  with peculiar properties in GASP, trying to determine the main mechanisms altering their properties and providing a panorama of the different processes taking place in low-density environments. As the selection of the GASP targets was biased towards galaxies with signs of possible stripping  (i.e. unilateral debris, see Sec.\ref{sec:data}), the sample is not suitable for performing general statistics of the most probable processes occurring in the field. We will however state the `success' of an optical selection in identifying gas removal processes for field galaxies and probe the power of IFU data in pinning down the acting mechanism. We will also study whether galaxies that are affected by environmental processes follow the fundamental scaling relations of undisturbed galaxies. 

{  This study is based on a data set of excellent quality for the large spatial extent of the Field-Of-View (FoV) and the broad wavelength coverage. This latter in particular allows us to study the properties of both the ionised gas and the stellar components out to large galactocentric distance in a  homogeneous way. These are the main advantages with respect to existing  surveys obtained with other optical IFU instruments, such as  CALIFA \citep{Sanchez2012}, SAMI \citep{Bryant2015}, and MaNGA \citep{Bundy2015},\footnote{{  We refer to Tab. 3 in \cite{Bundy2015} for a detailed comparison of the different IFU Surveys.}} or optical Fabry–Perot interferometry such as GHASP \citep{Garrido2002}. Unfortunately, these differences between GASP and other IFU surveys hamper a straightforward and homogeneous comparison with other samples,  preventing us from increasing the number of galaxies in our study. }

The paper is structured as follows: Sec. \ref{sec:analysis} presents the data analysis and Sec. \ref{sec:data}  the data sample; Sec. \ref{sec:results1} describes in detail {  the typical signatures of the different mechanisms on galaxies, Sec. \ref{sec:results1_1}} presents some clear examples of the different mechanisms identified in the sample. Sec. \ref{sec:results2} summarizes the main results and investigates some scaling relations. Finally, Sec. \ref{sec:summary} gives a summary of the main results. In Appendix \ref{sec:all} a panorama of all galaxies in the sample is given.

Throughout the paper, we adopt a \cite{Chabrier2003} initial mass function (IMF) in the mass range 0.1-100 M$_{\odot}$. The cosmological constants assumed are $\Omega_m=0.3$, $\Omega_{\Lambda}=0.7$ and H$_0=70$ km s$^{-1}$ Mpc$^{-1}$. 

\section{Data analysis} \label{sec:analysis}

The survey strategy, observations, data reduction and analysis procedure is presented in detail in  \citet{Poggianti2017}. 

Observations were carried out between April 2015 and April 2017  using the MUSE spectrograph located at VLT in Paranal. Each galaxy was observed with clear conditions (seeing $<0\farcs$9). 

Data were reduced with the most recent available version of the MUSE pipeline\footnote{\url{http://www.eso.org/sci/software/pipelines/muse}} and datacubes were  averaged filtered  in the spatial direction with a 5$\times$5 pixel kernel,  corresponding to our worst seeing conditions of 1$^{\prime\prime}$  = 0.7-1.3 kpc at the redshifts of the targets. 

Reduced datacubes were corrected for extinction due to our Galaxy, using the extinction value estimated at the galaxy position \citep{Schlafly2011} and assuming the extinction law from \cite{Cardelli1989}. 

To obtain an emission-only datacube, we subtracted the stellar-only component of each spectrum
derived with our spectrophotometric code \sinopsis \citep{Fritz2017}.  In addition, \sinopsis provided us with spatially resolved estimates of the following stellar population properties: stellar masses; average star formation rate and total mass formed in four age bins (= star formation histories, SFH): young (ongoing star formation) =  $t < 2 \times 10^7$ yr, recent $= 2 \times 10^7 < t< 5.7 \times 10^8$ yr, intermediate-age = $5.7 \times 10^8 <t< 5.7 \times 10^9$ yr, and old =$> 5.7 \times 10^9$ yr; luminosity- weighted stellar ages; reconstructed absolute magnitudes in a wide range of filters.

Emission line fluxes and errors, along with the underlying continuum, gaseous velocities (with respect to given redshift), and velocity dispersions  were derived  using the IDL software \kube \citep{Fossati2016}.  We consider as reliable only spaxels with S/N(\Ha)$>$4.
\Ha luminosities\footnote{{  Throughout the paper, we assume that \Ha emission originates in HII regions and not in diffuse ionized gas. A detailed analysis of the diffuse ionized gas can be found in \citet{Tomicic2021}.}} corrected both for stellar absorption and for dust
extinction were used to compute SFRs, adopting the \cite{Kennicutt1998a}'s relation: $\rm SFR (M_{\odot}
\, yr^{-1}) = 4.6 \times 10^{-42} L_{\rm H\alpha} (erg \, s^{-1})$. 
The extinction was estimated from the Balmer decrement
assuming an intrinsic value $\rm H\alpha/H\beta = 2.86$ and the \cite{Cardelli1989} extinction law. 

Stellar kinematics were extracted from the spectra using the Penalized Pixel-Fitting (pPXF) code \citep{Cappellari2004}, which fit the observed spectra with the stellar population templates by \cite{Vazdekis2010}. We performed the fit of spatially binned spectra based on signal-to-noise ratio (S/N = 10 for most galaxies), as described in \cite{Cappellari2003}, with the weighted Voronoi tessellation modification proposed by \cite{Diehl2006}.  Maps were then smoothed using the two-dimensional local regression techniques (LOESS) as implemented in the Python code developed by M. Cappellari.\footnote{\url{http://www-astro.physics.ox.ac.uk/~mxc/software}}

We employed the standard diagnostic diagrams to separate the regions powered by star formation from regions powered by Active galactic Nuclei (AGN) or Low-Ionization Nuclear Emission Region (LINER) emission \citep{Baldwin1981}.   Only spaxels with a S/N$>$3 in all emission lines involved (including \Ha) were considered. To compute total SFRs, we adopted the [O\textsc{III}]5007/$\rm H\beta$  vs. [N\textsc{II}]6583/$\rm H\alpha$ diagram and the 
division lines by \citet{Kauffmann2003} to select the spaxels whose ionized flux is powered by star formation. 
To characterize the physical processes acting on galaxies, we also inspected the [OIII]5007/$\rm H\beta$ vs [OI]6300/$\rm H\alpha$ diagnostic diagram. 
Among the various line-ratio diagrams, the one based on the [O\textsc{I}] is the most sensitive to physical processes different from Star Formation (e.g. thermal conduction from the surrounding hot ICM, turbulence and shocks) and can therefore be considered as a conservative lower limit of the real star formation budget \citep{Poggianti2019}.

Metallicity of the ionized gas was computed for
each star-forming spaxel using the pyqz Python
module7 \citep{Dopita2013} v0.8.2; we obtained the 
$12 + \log(O/H)$ values by interpolating
from a finite set of diagnostic line ratio grids computed
with the MAPPINGS code \citep[see][for details]{Franchetto2020}. Only spaxels with a S/N$>$3 in all emission lines involved are considered.

Structural parameters (effective radius $R_e$, inclination $i$, ellipticity $\varepsilon$, position angle $PA$) were computed on I-band images by measuring the radius of an ellipse including half of the total light of the galaxy \citep[see][for details]{Franchetto2020}.

To identify the spaxels belonging to the galaxy main body we sliced the 2D image of the near-\Ha continuum obtained with 
\kube. We selected the isophote enclosing essentially all of the galaxy body, down to $\sim$1$\sigma$ above the background level \citep[see][for details]{Poggianti2017}. 
Integrated values are measured as the sum of all the spaxels within this contour. Only for metallicity, the integrated value we provide is the  mean value computed at R$_e$ \citep{Franchetto2020}.

\subsection{Ancillary catalogs - Galaxy environments}
As defining the environment is quite a critical task, especially for the  sparsest galaxies, in what follows we will report the group classification from three different works: our own PM2GC \citep{Calvi2011TheGalaxies}, the \cite{Tempel2014_g} and \cite{Saulder2016} group catalogs. The latter two are both based on  Sloan Digital Sky Survey  data. Specifically, \cite{Tempel2014_g} is based on  SDSS DR10 \citep{york00, Ahn2014}, while \cite{Saulder2016} used the SDSS DR12 \citep[][]{Alam2015}, combined to  the Two Micron All Sky Survey and the 2MASS Redshift Survey \citep[2MASS and 2MRS,][]{Skrutskie2006, Huchra2012}. 
However, (1) a few galaxies do not fall into the SDSS/2MASS footprints (2) besides the group environment we also need to detect companions that might exert tidal forces. We therefore combined MCG  \citep{Driver2005}, SDSS, WINGS/OmegaWINGS \citep{Moretti2014WINGSClusters, Moretti2017} and Hyperleda\footnote{\url{http://leda.univ-lyon1.fr}} redshifts to get a catalog as complete as possible. We also downloaded data from the Hyperleda catalog for galaxies with no redshift available, to further detect possible companions. 

We note that as group finding methods are automatic and meant to study groups from a statistical point of view, sometimes the values obtained to characterize the physical parameters of the single groups are not very reliable and should be taken with caution. 

\section{Galaxy Sample} \label{sec:data}
The GASP program observed 114 galaxies, extracted from three surveys that, together, cover the whole range of environmental conditions at low redshift: WINGS \citep{Fasano2006}, OMEGAWINGS \citep{Gullieuszik2015}, and PM2GC \citep{Calvi2011TheGalaxies}. 94 of these galaxies were selected from the \cite{Poggianti2016JELLYFISHREDSHIFT} catalog of gas stripping candidates as they showed in their B-band images  debris trails, tails, or surrounding debris located on one side of the galaxy.   
\begin{longrotatetable}
\begin{deluxetable*}{llllllllllll}
\tablecaption{Main properties of the galaxies in the sample. \label{tab:sample}}
\tabletypesize{\scriptsize}
\tablehead{
\colhead{ID} & \colhead{z} & 
\colhead{RA} & \colhead{DEC} & 
\colhead{$M_\ast$} & \colhead{SFR} & 
\colhead{$R_e$} & \colhead{$\varepsilon$} & 
\colhead{incl} & \colhead{12+$\log(O/H)@R_e$} & \colhead{mechanism}  & \colhead{flag$_{c}$\tablenotemark{\footnotesize{a}}}\\ 
\colhead{} & \colhead{} & \colhead{[J2000]} & \colhead{[J2000]} & 
\colhead{[$\rm 10^9 M_\odot$]} & \colhead{[$\rm M_\odot/yr$]} & \colhead{[arcsec]} &
\colhead{[deg]} & \colhead{[deg]} & \colhead{} & \colhead{}  & \colhead{}
} 
\startdata
  JO134\tablenotemark{\footnotesize{b}} & 0.0166 & 12:54:38.33 & -30:09:26.5 & 1.1$\pm$0.2 & 0.188$\pm$0.001 & -- & --  & --  & 8.1$\pm$0.1& merger+RPS &1 \\
  JO190\tablenotemark{\footnotesize{b}} & 0.0132 & 22:26:53.62 & -30:53:11.1 & 2.2$\pm$0.6 & 0.370$\pm$0.002 &  -- &  -- & --  & 8.3$\pm$0.1 & merger & 1\\
  JO20\tablenotemark{\footnotesize{b}} & 0.1471 & 01:08:55.06 & +02:14:20.8 & 71$\pm$12 & 12.1$\pm$0.1& -- & -- & -- &8.9$\pm$0.1& merger & 1\\
  P11695 & 0.0464 & 10:46:14.89 & +00:03:00.8 & 12$\pm$2 & 3.27$\pm$0.04 & 4.9$\pm$0.4 & 0.09$\pm$0.02 & 25$\pm$2 &8.71$\pm$0.04& gas accretion (GASP VII) & 1\\
  P12823 & 0.0504 & 10:52:24.04 & -00:06:09.9 & 22$\pm$6 & 1.33$\pm$0.01 & 3.3$\pm$0.2 & 0.56$\pm$0.04 & 65$\pm$3 &9.12$\pm$0.02& interaction & 1\\
  P14672 & 0.0498 & 11:01:55.10 & +00:11:41.1 & 8$\pm$1 & 0.154$\pm$0.05 & 2.9$\pm$0.2 & 0.11$\pm$0.04 & 28$\pm$6 &8.98$\pm$0.05& CWE & 0\\
  P16762 & 0.0487 & 11:10:19.63 & -00:08:34.4 & 35$\pm$5& 0.011$\pm$0.002 & 3.4$\pm$0.2 &  0.493$\pm$0.009  &  60.4$\pm$0.6  &--& starvation & 1\\
  P17048 & 0.049 & 11:12:38.27 & +00:08:01.2 & 4$\pm$1 & 0.544$\pm$0.006 & 3.7$\pm$0.2 & 0.20$\pm$0.03 & 37$\pm$3 &8.44$\pm$0.05& interaction & 1\\
  P18060 & 0.043 & 11:14:59.28 & -00:00:43.2 & 1.2$\pm$0.3 & 0.049$\pm$0.001 & 3.0$\pm$0.3 & 0.36$\pm$0.02 & 51$\pm$1 &8.4$\pm$0.1& CWS & 1\\
  P19482 & 0.0406 & 11:22:31.25 & -00:01:01.6 & 21$\pm$4 & 1.25$\pm$0.02 & 4.7$\pm$0.4 & 0.419$\pm$0.005 & 55$\pm$4 &8.97$\pm$0.05& CWE (GASP XVI) & 1 \\
  P20159 & 0.0489 & 11:24:22.26 & -00:16:34.3 & 3.7$\pm$0.9 & 0.233$\pm$0.003 & 5.2$\pm$0.4 & 0.76$\pm$0.01 & 78.5$\pm$0.8 &8.2$\pm$0.1& RPS &0\\
  P3984 & 0.0464 & 10:14:05.86 & -00:07:37.6 & 2.5$\pm$0.6 & 0.441$\pm$0.007 & 3.6$\pm$0.5 & 0.54$\pm$0.05 & 64$\pm$4 &8.36$\pm$0.09& merger & 1\\
  P40457 & 0.0678 & 13:01:33.06 & -00:04:51.1 & 5$\pm$1 & 0.187$\pm$0.005 & 4.0$\pm$0.6 & 0.61$\pm$0.04 & 68$\pm$3 &8.4$\pm$0.1& gas accretion & 0\\
  P443 & 0.0464 & 09:59:31.53 & -00:15:22.8 & 26$\pm$2 & 0.015$\pm$0.002 & 2.9$\pm$0.3   &  0.30$\pm$0.05  & 46$\pm$4   &--& starvation & 1\\
  P4946 & 0.0621 & 10:18:30.81 & +00:05:05.0 & 29$\pm$6 & 0.288$\pm$0.07 & 2.4$\pm$0.3 & 0.63$\pm$0.03 & 69$\pm$2 &?\tablenotemark{\footnotesize{c}}& gas accretion (GASP XII) & 0\\
  P5055 & 0.061 & 10:18:08.54 & -00:05:03.1 & 51$\pm$9 & 0.870$\pm$0.007 & 6.0$\pm$0.7 & 0.822$\pm$0.005 & 83.0$\pm$0.4 &8.93$\pm$0.03& RPS (GASP XII) & 1\\
  P5169 & 0.0634 & 10:18:13.79 & +00:03:56.6 & 2.6$\pm$0.4& 0.0018$\pm$0.0003 & 2.3$\pm$0.2  & 0.52$\pm$0.04  &  63$\pm$3  &--& starvation (GASP XII) & 1\\
  P5215 & 0.0629 & 10:16:58.24 & -00:14:52.9 & 33$\pm$6 & 0.99$\pm$0.01 & 5.2$\pm$0.3 & 0.266$\pm$0.04 & 43$\pm$4 &9.05$\pm$0.03& RPS/CWE (GASP XII) & 0\\
  P59597 & 0.0495 & 14:17:41.13 & -00:08:39.4 & 5$\pm$1 & 0.283$\pm$0.005 & 5.7$\pm$0.4 & 0.651$\pm$0.009 & 70.9$\pm$0.6 &8.42$\pm$0.06& RPS & 1 \\
  P63661 & 0.055 & 14:32:21.78 & +00:10:41.4 & 21$\pm$5& 1.31$\pm$0.02 & 6.0$\pm$0.6 & 0.26$\pm$0.08 & 43$\pm$7 &8.74$\pm$0.05& CWE (GASP XVI) & 1\\
  P63692 & 0.0561 & 14:31:59.98 & +00:05:03.3 & 0.8$\pm$0.2 & 0.063$\pm$0.001 & 2.7$\pm$0.2 & 0.62$\pm$0.02 & 69$\pm$2 &8.0$\pm$0.1& CWS & 1\\
  P63947 & 0.0562 & 14:31:01.82 & -00:10:56.9 & 2.2$\pm$0.5 & 0.111$\pm$0.002 & 2.9$\pm$0.2 & 0.3$\pm$0.1 & 50$\pm$9 &8.38$\pm$0.09& merger & 1\\
  P8721 & 0.0648 & 10:34:08.65 & +00:00:03.2 & 52$\pm$8 & 2.76$\pm$0.02 & 6.0$\pm$0.2 & 0.68$\pm$0.03 & 73$\pm$2 &8.95$\pm$0.05& CWE (GASP XVI) & 1\\
  P877 & 0.0427 & 10:00:49.96 & -00:08:57.0 & 24$\pm$4 & 1.57$\pm$0.01 & 4.4$\pm$0.3 & 0.17$\pm$0.04 & 34$\pm$4 &9.05$\pm$0.04& merger & 0 \\
  P95080 & 0.0403 & 13:12:08.75 & -00:14:20.3 & 13$\pm$3 & 1.17$\pm$0.02 & 7.5$\pm$0.9 & 0.27$\pm$0.07 & 43$\pm$6 &8.63$\pm$0.07& CWE (GASP XVI) &1\\
  P96244 & 0.0531 & 14:18:35.47 & +00:09:27.8 & 60$\pm$20 & 5.32$\pm$0.03 & 5.9$\pm$0.8 & 0.44$\pm$0.02 & 57$\pm$1 &9.03$\pm$0.04& RPS & 1\\
  P96949 & 0.0511 & 11:54:09.95 & +00:08:18.1 & 25.11$\pm$5  & 1.8$\pm$0.02 & 5.5$\pm$0.5  & 0.51$\pm$0.06  & 62$\pm$4  &8.7$\pm$0.2& merger (GASP VIII)& 1\\
\enddata
\tablenotetext{a}{flag$_c$=1 for certain classification, flag$_c$=0 for uncertain classification.}
\tablenotetext{b}{structural parameters not measured due to the irregularities of the galaxy.}
\tablenotetext{c}{P4946 has no a reliable estimate of the ionized gas metallicity because of the presence of the central AGN that does not allow to have a significant number of star-forming spaxels to use to estimate the metallicity at the effective radius.}
\end{deluxetable*}
\end{longrotatetable}

Galaxies whose optical morphological disturbance was clearly induced by mergers or tidal interactions were deliberately excluded from the target sample.

In this paper we study the field galaxies included in the stripping sample and the few passive ones. Effectively, we excluded from the entire GASP sample  (114 galaxies) cluster stripped galaxies \citep{Vulcani2018_L, Vulcani2020, Vulcani2020b}, field galaxies not showing any disturbance in their \Ha images \citep[control sample galaxies,][]{Vulcani2018_L,Vulcani2019b}, and a galaxy belonging to a background cluster. The  sample analysed in this work therefore includes 27 galaxies. We note that an in depth analysis of 10/27 galaxies has been  already published in previous papers \citep{Vulcani2017c, Vulcani2018_b, Vulcani2018_g, Vulcani2019_fil}, hence in what follows we will just refer to those papers without recapping their main findings.
Those galaxies, however, will be included in Sec.\ref{sec:results2}, where we will characterize the field population as a whole. Table \ref{tab:sample} presents the list of all galaxies, along with some main properties. 

Most of the galaxy names start with ``P'', as they were drawn from the field sample of the PM2GC. There are though few galaxies whose name starts with ``J''. These galaxies were drawn from the imaging of the cluster sample (WINGS/OMEGAWINGS), but had no redshift prior to MUSE observations. 
It was therefore assumed that they were cluster members, but, as will be discussed later on, they are not.

As it will be discussed in the following, according to both the $\rm [OIII]5007/H\beta$  vs. [N II]6583/$\rm H\alpha$ and $\rm [OIII]5007/H\beta$  vs. [OI]6300/$\rm H\alpha$ diagrams only two galaxies host an AGN in their center: JO20 and P4946.

\section{The body of evidence of the different physical processes} \label{sec:results1}
As mentioned in the introduction, galaxies in the different environments  feel different physical processes, which are expected to leave a different imprint on the galaxy spatially resolved properties. 
\begin{figure*}
\centering
\includegraphics[scale=0.5, angle=90]{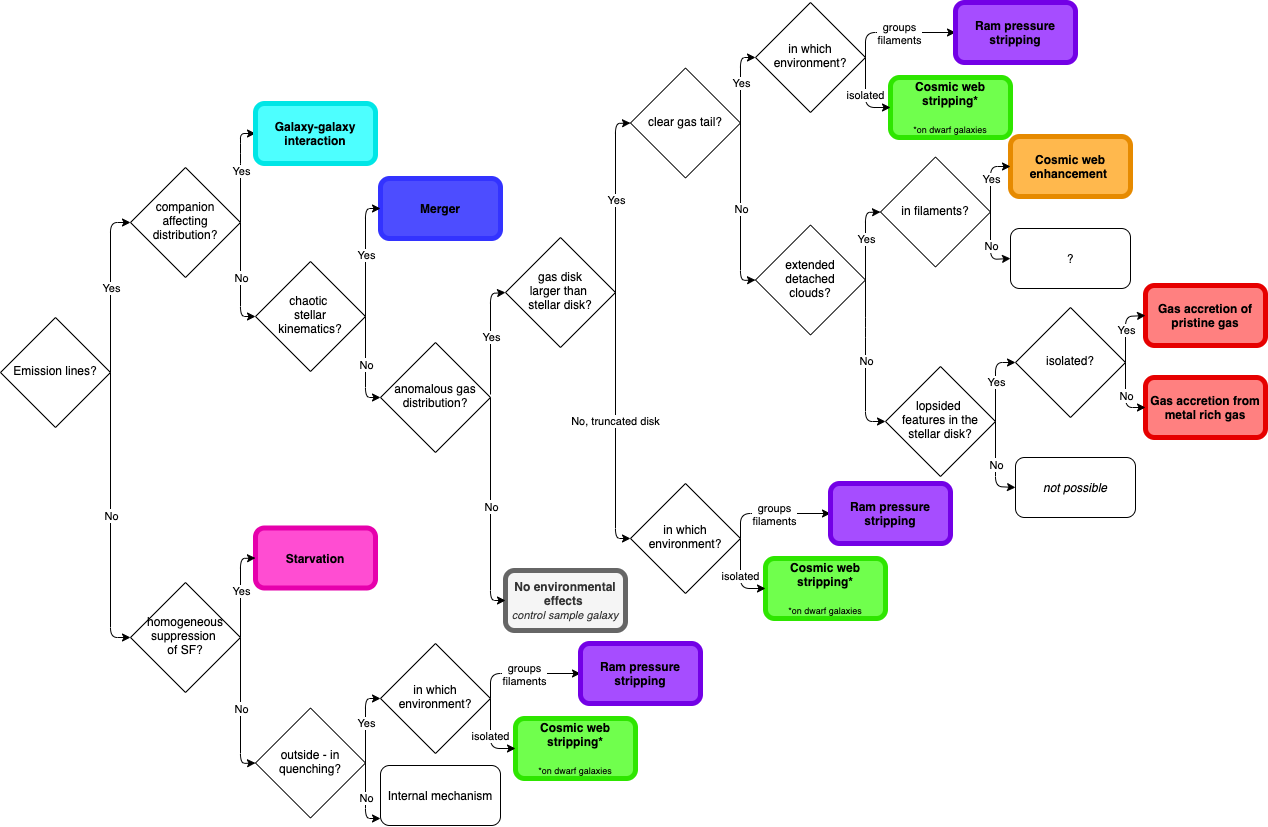}
\caption{Flow chart describing the main criteria adopted to pin point the major mechanism acting on galaxies. Further evidence for each physical process is given the following figures. \label{fig:mech} }
\end{figure*}
In this section we will present the criteria adopted for the classification in sub-classes, describing what are their main expected signatures on the gas and stellar property maps. 

To  assign a physical mechanism to each galaxy in the sample, we adopted the scheme presented in Fig. \ref{fig:mech}. This diagram is as general as possible and can be used to classify also galaxies outside our sample.  However, it includes only the main line of evidence. Our classification is also supported by the analysis of the other galaxy properties, as discussed below. 

{  We note that  our analysis is based on optical observations, and that no information on the distribution and kinematics of neutral gas, which likely dominates the gas mass budget  of our star-forming galaxies, is available for this sample.}

\subsection{The classification criteria}
The first step is to separate galaxies with \Ha in emission throughout the disk - i.e. star-forming galaxies -  from those with no \Ha in emission - i.e. passive systems. 

\subsubsection{Galaxy-Galaxy interactions}

Focusing on galaxies with \Ha in emission, we first select the objects with a close companion (distance between the galaxies $r<50\arcsec$) to identify possible cases of galaxy-galaxy interactions. Indeed, 
even though interactions and mergers were avoided in the target selection based on B-band imaging \citep{Poggianti2016JELLYFISHREDSHIFT, Poggianti2017}, MUSE observations revealed their presence in the sample.

Disk galaxies with a companion can experience gravitational forces which are different from one side of the galaxy to the other. As a result, their material undergoes deforming effects that re-arrange the individual components of the galaxy \citep[e.g.,][]{toomre72}, producing  tails  \citep[e.g.,][]{Mihos1993} and, sometimes,  warps \citep{Semczuk2020}.
Effects of interactions can therefore be seen on the ionized gas and stellar kinematics maps obtained by MUSE. 
Tidal interactions (cyan box in Fig.\ref{fig:mech}, discussed in  Sec. \ref{sec:interactions} and \ref{sec:interactions_a}) are expected to happen when the acceleration $a_{tid}$ produced by the companion on the galaxy of interest is larger than the acceleration from the potential of the galaxy itself, $a_{gal}$. Following \cite{Vollmer2005b}, 
$$
\frac{a_{tid}}{a_{gal}} = \frac{M_{neighbour}}{M_{gal}} \left(\frac{r}{R}-1\right)^{-2}
$$
where R is the distance from the centre of the galaxy, r is the distance between the galaxies, and $\frac{M_{neighbour}}{M_{gal}}$ the stellar mass ratio. This formulation would require the de-projected distance between the two galaxies, which is unfortunately unknown, therefore we  use the projected distance. Typically, if $\frac{a_{tid}}{a_{gal}}>0.15$, meaning that the tidal acceleration is at least 15\% of the galaxy's own acceleration, tidal interaction could be invoked as one of the main mechanisms affecting galaxy properties \citep{Merluzzi2016}. 

\subsubsection{Merging systems}
{  Next, we focus on merging systems.} During an interaction, in some cases galaxies get very close and the exchange of orbital angular momentum (through dynamical friction) becomes very high, so that the mean distance between the progenitors rapidly decreases and they finally merge (blue box in Fig.\ref{fig:mech}, discussed in  Sec. \ref{sec:Mergers} and \ref{sec:Mergers_a}), forming a unique  galaxy \citep{Barnes1992a}. As the gas is redistributed, the regular metallicity gradients typically found in galaxies \citep[e.g.][]{Pilyugin2014} can be destroyed \citep{Kobayashi2004}. Star formation can also be triggered during mergers: gas can infall towards the center, inducing a nuclear starburst \citep{Springel2000, Barnes2002, Naab2006}. At large radii the gravitational torques instead push the material to the outer regions. This outflow enhances the formation of the tails already formed by the tidal field itself \citep{Bournaud2010}.  Therefore, the signs of the mergers on the spatially resolved properties are chaotic gas and stellar kinematics and, in some cases, tidal features \citep[e.g.][]{Mihos1993, Struck1999}; flat or very irregular metallicity gradients, and signs of the merger remnants (i.e. double nuclei). 
Note that simulations by \cite{Hung2016} have shown that the time and duration during which the merger signatures are detectable are typically 0.2 - 0.4 Gyr, except in the case of equal mass merger, when the duration is approximately twice as long.

Interactions and mergers can also produce gas inflows that can feed a central black hole and induce AGN activity \citep{Sanders1988}.
 
If no interactions/mergers occur, galaxies are expected to show rather regular patterns in their stellar kinematics. If they also present a regular ionized gas kinematics, we can consider them as undisturbed galaxies (grey box in Fig.\ref{fig:mech}) and, in our case, they constitute the GASP control sample  \citep{Vulcani2018_L,Vulcani2019b}.

\subsubsection{Ram pressure stripping}
{  In the previous sections we focused on processes that alter the galaxy stellar kinematic, we now discuss hydrodynamical processes, which typically leave the stellar component unaltered, but produce  } disturbed gas kinematics.

Distortions can be limited to the galaxy disk, or can be visible as ionized gas tails. The latter are typically produced by ram pressure stripping \citep{Gunn1972}, one of the most efficient mechanisms to  remove the gas from galaxies, provided the galaxy is embedded in a rather dense medium (clusters and groups).  Indeed the strength of the ram pressure depends on the intracluster gas density and the speed of the galaxy relative to the medium. This pressure can strip gas out of the galaxy where the gas is gravitationally bound to the galaxy less strongly than the force from the intracluster medium wind due to the ram pressure.
Ram pressure stripping (purple boxes in Fig.\ref{fig:mech}, discussed in  Sec. \ref{sec:RPS} and \ref{sec:RPS_a}) can produce tails of ionized gas that is ionized through stellar photoionization due to ongoing star formation within the stripped gas. {  In contrast, the  stellar component is undisturbed.} The gas retains the velocity of the stars at the location of the disc from which it was stripped and continues to rotate coherently with the galaxy until several kiloparsecs away from the main galaxy body.  The stripping proceeds from the outside in, with the outermost regions of the disc stripped first \citep[e.g.][]{Poggianti2017}. As a consequence, the gas properties in the tail are similar to the gas properties in the external regions of the disk (e.g. lower metallicity than in the galaxy core, A. Franchetto et al. in prep). Tails are also very young \citep[few Myr, e.g.][]{Gullieuszik2017}, as stars are born out of the stripped gas. Galaxies undergoing ram pressure stripping can also show a central burst in star formation \cite{Vulcani2018_L}, as expected when the gas is compressed by this mechanism \citep{Roediger2014}.
Ram pressure stripping can eventually remove  the gas, producing first truncated disks \citep[e.g.,][]{Fritz2017} and then quenching star formation and transforming galaxies into passive systems \citep{Vulcani2020}.

\subsubsection{Cosmic web stripping}

Even though many parameters influence the efficiency of
ram pressure stripping \citep[galaxy mass, galaxy orbit within the cluster, cluster properties;][]{Gullieuszik2020}, the density of the medium in which the galaxy is embedded plays a critical role. Galaxies in isolation are not expected to feel ram pressure stripping. These galaxies could instead undergo cosmic web stripping (green box in Fig.\ref{fig:mech},  discussed in  Sec. \ref{sec:cws} and  \ref{sec:cws_a}). This mechanism has been introduced by \cite{Benitez2013} to characterize the effect of the cosmic web on isolated dwarf galaxies. The process is very similar to ram pressure stripping, but exerted by a much less dense environment. {  Similarly to ram pressure stripping, it is expected to give origin to a ionized gas protuberance that resembles a tail, extending in a well defined direction. The comparison between the stellar and gas kinematics is therefore key to distinguish between cosmic web stripping and e.g. irregular galaxies:   in irregular galaxies the stellar kinematics is expected to closely follow the motion of the gas, both in terms of extent and rotation \citep[e.g.,][]{Johnson2012}.}

\begin{table*}
\centering
\small
\begin{tabular}{ccccccc}
\hline
\multicolumn{1}{c}{\multirow{2}{*}{ID}} & presence of   & a$_{tid}$     & asymmetric   &  asymmetric        & inhomogeneous & \multirow{2}{*}{$\rm flag_c$} \\
                    & companion     & $>0.15$       & rgb and \Ha &  $\Delta v_{gas}$ and $\Delta v_{star}$ &$ \sigma v_{star}$  &  \\
\hline
P17048 & \cmark & \cmark & \cmark & \cmark & \cmark  & 1\\
P12823  & \cmark &  {?} & \cmark & \xmark & \cmark & 1\\
\end{tabular}
\caption{{  Summary of the main features investigated to characterize interacting systems. \cmark\ means that the criterion is met, \xmark\ means that the criterion is not met,  {  ?} means that it is not possible to verify the criterion (see text for details). $\rm flag_c$ indicates if the classification is certain (1), or uncertain (0). }
\label{tab:interacting}}
\end{table*}

Cosmic web stripping is also expected to eventually remove the gas, producing first truncated disks  and then quenching star formation and transforming galaxies into passive systems \citep{Benitez2013}. However, no observational studies have observed these phases.

\subsubsection{Cosmic web enhancement}

Distorted gas kinematics can also be produced by processes that rather than removing gas from galaxies either compress the existing gas or feed the galaxies with new gas, aiding star formation. Cosmic Web Enhancement (orange box in Fig.\ref{fig:mech},  discussed in  Sec. \ref{sec:CWE}) is one of these mechanisms. This process has been first proposed by \cite{Vulcani2018_g} to explain the appearance of galaxies found in  filaments. Filaments can indeed assist  gas cooling and increase the extent of the star formation in the densest regions in the circumgalactic gas. \cite{Liao2018} showed that filaments are an environment that particularly favors this gas cooling followed by condensation and star formation enhancement. As the clouds move through the filament, the metal-rich gas of the galaxies mixes with the metal-poor gas constituting the circumgalactic gas. As a consequence, the latter gets enriched and cools more easily. 
Galaxies undergoing Cosmic Web Enhancement show detached \Ha clouds out to large distances from the galaxy center (beyond 4$\times R_e$). The gas kinematics, metallicity map, and the ratios of emission-line fluxes confirm that they do belong to the galaxy gas disc; the analysis of their spectra shows that very weak stellar continuum is associated with them. Similarly, the star formation history and luminosity weighted age maps point to a recent formation of such clouds. 

\subsubsection{Gas accretion}
The last mechanism that can  alter regular gas kinematics is gas accretion (red boxes in Fig.\ref{fig:mech},  discussed in Sec. \ref{sec:accr}) \citep{Sancisi2008, Semelin2005}. An injection of gas can produce a very lopsided morphology. This asymmetry is expected to develop only at late times {and affecting only the gaseous component, leaving the stellar one mostly unaltered}. Galaxies in isolation can be most likely fed by very low-metallicity gas inflow from the cosmic web, and present very steep and asymmetric metallicity gradients, while galaxies in denser environments most likely  underwent substantial gas accretion throught mergers from a metal rich less massive object, whose presence is not visible, in a particular event. Signs of the accretion are again seen on the metallicity distribution, which is different from that observed in undisturbed galaxies \citep{Pilyugin2014}. Gas accretion with angular momentum opposite to that of the host galaxy can also produce a counter-rotating stellar disk \citep[e.g.,][]{Thakar1997, Puerari2001, Algorry2014, Bassett2017}.

\subsubsection{Starvation}

Focusing on galaxies with no \Ha in emission, the analysis of the stellar kinematics can show signatures of past mergers, while the SFH maps can reveal how the quenching proceeded: inside-out, outside-in or homogeneously. An homogeneous suppression of the star formation can be produced by starvation. According to this scenario, star formation in galaxies is quenched because the inflow of gas from the IGM is halted, and as a consequence star formation in these galaxies can continue only for a limited amount of time by using the gas available in the galaxy. However, understanding what is causing the cut off of the gas reservoir is quite hard. 

\section{Results of galaxy classification} \label{sec:results1_1}

In this section we will present an overview of the different mechanisms we identified acting on the galaxies in our sample. We will also characterize the galaxy hosting environments. 
For the sake of clarity, we will discuss in detail only one galaxy per physical process (Sec.\ref{sec:results1_1}), {  showing and discussing only the main galaxy properties used to pin point the acting mechanism. In Appendix \ref{sec:all_fig}, we will instead show the maps of all the properties at our disposal, for completeness. Appendix \ref{sec:all} will show all the other galaxies in the sample, along with their classification.}
We note that in some cases what we propose is just the most probable process, and we can not always  exclude that other not identified mechanisms are responsible for the observed features. Therefore, in our classification, we will distinguish between certain and uncertain cases ($flag_c$ in Tab. \ref{tab:sample}). {  This flag is based on the analysis presented in Tables \ref{tab:interacting}, \ref{tab:merger}, \ref{tab:rps}, \ref{tab:cws}, \ref{tab:cwe}, \ref{tab:accr}, \ref{tab:pass}. If at least two criteria are not met, we assign $flag_c$=0, meaning that the classification is tentative. In the other cases, $flag_c$=1 and  the classification is secure.} Overall, 21 galaxies have a secure classification 
and 6 a tentative one. 

\subsection{Galaxy-Galaxy Interactions} \label{sec:interactions}

\begin{figure*}
\centering
\includegraphics[scale=0.6]{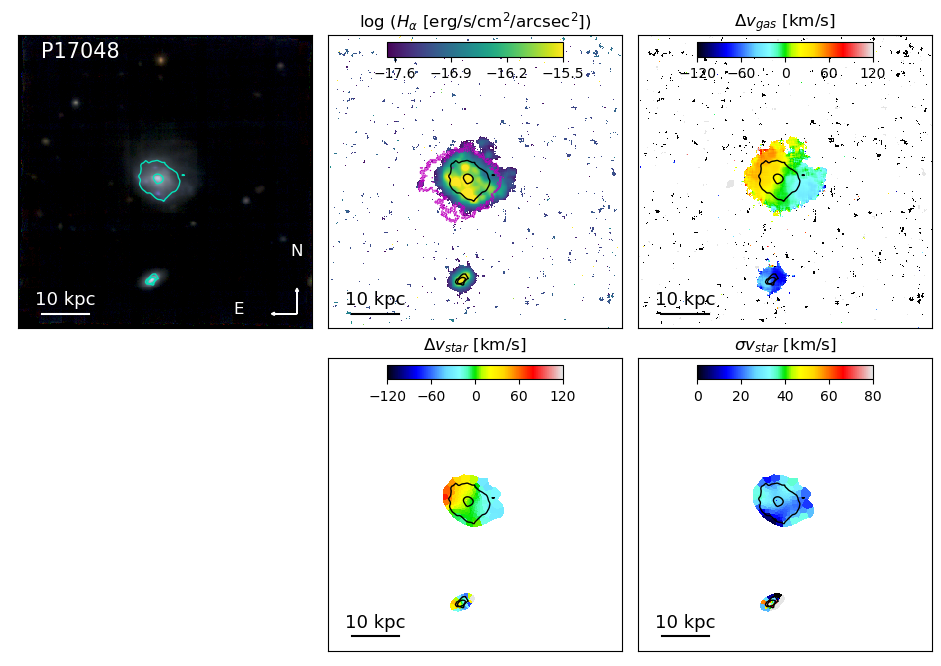}
\caption{P17048: example of interacting candidates in the sample. From left to right: Color composite image obtained by combining the reconstructed g, r, and i filters from the MUSE data cube (rgb), \Ha flux map, \Ha and stellar kinematics maps, stellar velocity dispersion map. The kpc scale is also shown in each panel. North is up, and east is left. Cyan or black contours represent the distribution of the oldest stellar population (from \sinopsis). { The magenta contour on the \Ha flux map represents the galaxy rotated of 180$^\circ$ and is used to quantify the lopsidedness (see text for details). } \label{fig:interacting} }
\end{figure*}

Two galaxies fall in this category, and in both cases the classification is certain. {  Table \ref{tab:interacting} lists them and summarizes the main features investigated to characterize interacting systems. These will be discussed in what follows. Figure  \ref{fig:interacting} shows the useful maps for P17048 and its companion P17044 \citep{Calvi2011TheGalaxies}, used as example to outline the features to look for to characterize interactions. These are:} color composite image obtained by combining the reconstructed g, r, and i filters from the MUSE data cube (from now on called rgb images for brevity), the \Ha flux map and the \Ha and stellar kinematics { (from now on $\Delta v_{gas}$ and $\Delta v_{star}$ for brevity)}\footnote{{ We note that spatially resolved spectroscopy is affected by beam smearing, i.e  the flux profile is spatially blurred in the central regions due to the atmospheric seeing. As a consequence, results in the central 1 arcsec, corresponding to the PSF, are influenced by this effect.}}. {  All the available maps for P17048 (but not useful to determine the main acting mechanisms) are provided in Fig.\ref{fig:P17048_bis}}. The same maps for P12823 and its companion are presented in  Sec.\ref{sec:interactions_a}. 
In these and all the following figures, contours represent the distribution of the oldest stellar population ($t> 5.7 \times 10^9$ yr, from \sinopsis). These contours will help us to identify  the ``original body'' of the galaxy.

\begin{table*}
\centering
\small
\begin{tabular}{ccccccccccc}
\hline
\multicolumn{1}{c}{\multirow{2}{*}{ID}}& no clear    & tidal  & evidence for      & { asymmetric} & { asymmetric}       & inhomogeneous               &  asymmetric   & patchy &\multirow{2}{*}{$\rm flag_c$} \\
                    & companion   &   tail  & merger remnant\tablenotemark{\footnotesize{a}}    &  $\Delta v_s$ & $\Delta v_g$  & $\sigma v_{star}$  &  metallicity  & young regions &  & \\
\hline
JO20 & \cmark & \cmark  & \cmark & \xmark & \cmark &\cmark & \cmark &  \cmark &1\\
JO134 & \cmark & \cmark  & \cmark & \cmark & \cmark &\cmark  & \cmark  & \cmark &1 \\
JO190 & \cmark &  \cmark  & \cmark & \cmark & \xmark &\cmark  & \cmark  & \cmark &1 \\
P3984 & \cmark & \cmark & \xmark & \cmark& \xmark &\cmark& \cmark  & \cmark  &1\\
P877 & \cmark & \xmark & \xmark &  \cmark &\cmark & \cmark   & \xmark  & \xmark &0\\
P63947 & \cmark & \cmark & \xmark & \cmark & \cmark & \cmark &  \cmark & \cmark &1\\
P96949  & \cmark & \cmark & \cmark & \cmark & \cmark &\cmark &  \cmark &\cmark &1 \\
\end{tabular}
\tablenotetext{a}{Note that the evidence for merger remnant is not a absolutely necessary criterion to state if the galaxy underwent a merging event, and is not considered to define $\rm flag_c$.}
\caption{{  Summary of the main features investigated to characterize merging systems. The meaning of the symbols is as in Tab. \ref{tab:interacting}. }
\label{tab:merger}}
\end{table*}

As shown in Fig.\ref{fig:interacting},  P17048 is characterized by a lopsided morphology: the rgb image highlights that it extends much more towards North-West than towards South-East. The \Ha map is similarly asymmetric and shows a tattered distribution towards North. { To quantify  the degree to which the light is lopsided, we measure the overlap between the original image and the image of the galaxy rotated by 180$^\circ$ around the galaxy center. Specifically, we take the ratio of the area that overlap in the two images to the area of the original image.  Symmetric galaxies have a level of  overlap ($A_o$) of $>85\%$. The rotated map of P17048 is overlaid as pink contour to the \Ha flux map in Fig.\ref{fig:interacting}. P17048 has an overlap of 73\%, suggesting that about a quarter of the light 
is lopsided.} Such lopsidedness is seen also in the motion of the gas: the gas kinematics is distorted. The velocity field spans the range -50$<v [km/s]<$ 50. Also P17044 shows rotation and its gas kinematics is distorted. The galaxy presents a warp in position angle in the direction of the companion. 
The stellar kinematics is similarly disturbed and spans the same $\Delta(v)$ range. The southern part has very low rotational velocity. The velocity dispersion map of the stellar component is rather inhomogenous, suggesting that the motion of the stars has been altered. The velocity dispersion is higher towards North, reaching values of $\sigma v_{star}$= 35 km/s.  

{To quantify the asymmetries in the velocity fields (both for the stellar and the gas components, separately), we consider only the stellar disk and therefore the region defined by the stellar velocity map.} We adapt the definition of asymmetry first introduced by \cite{Conselice1997}. As done for the \Ha flux image, we rotate the image of the galaxy around the galaxy center and  perform a comparison of the resultant rotated image with the original image. We then measure the asymmetry $A$ as follows:
$$
A = \sqrt{\frac{\sum \left(|V_o|-|V_{180}|\right)^2}{\sum V_o^2}}
$$
with $V_o$ either stellar or ionized gas velocity in the original image and $V_{180}$ stellar or ionized gas velocity in the  image rotated by 180$^\circ$. The sum is performed over all pixels within the matching region of the original and rotated images. The lowest possible value for the asymmetry parameter is 0, corresponding to a completely symmetric velocity field,   while the highest is 1 and corresponds to a galaxy that is completely asymmetric. We set a threshold of 0.3 above which we define a velocity field as asymmetric.  For P17048 the asymmetry of the stellar kinematics ($A_v$) is 0.58, for the gas kinematics ($A_g$) is 0.65.

\begin{figure*}
\centering
\includegraphics[scale=0.6]{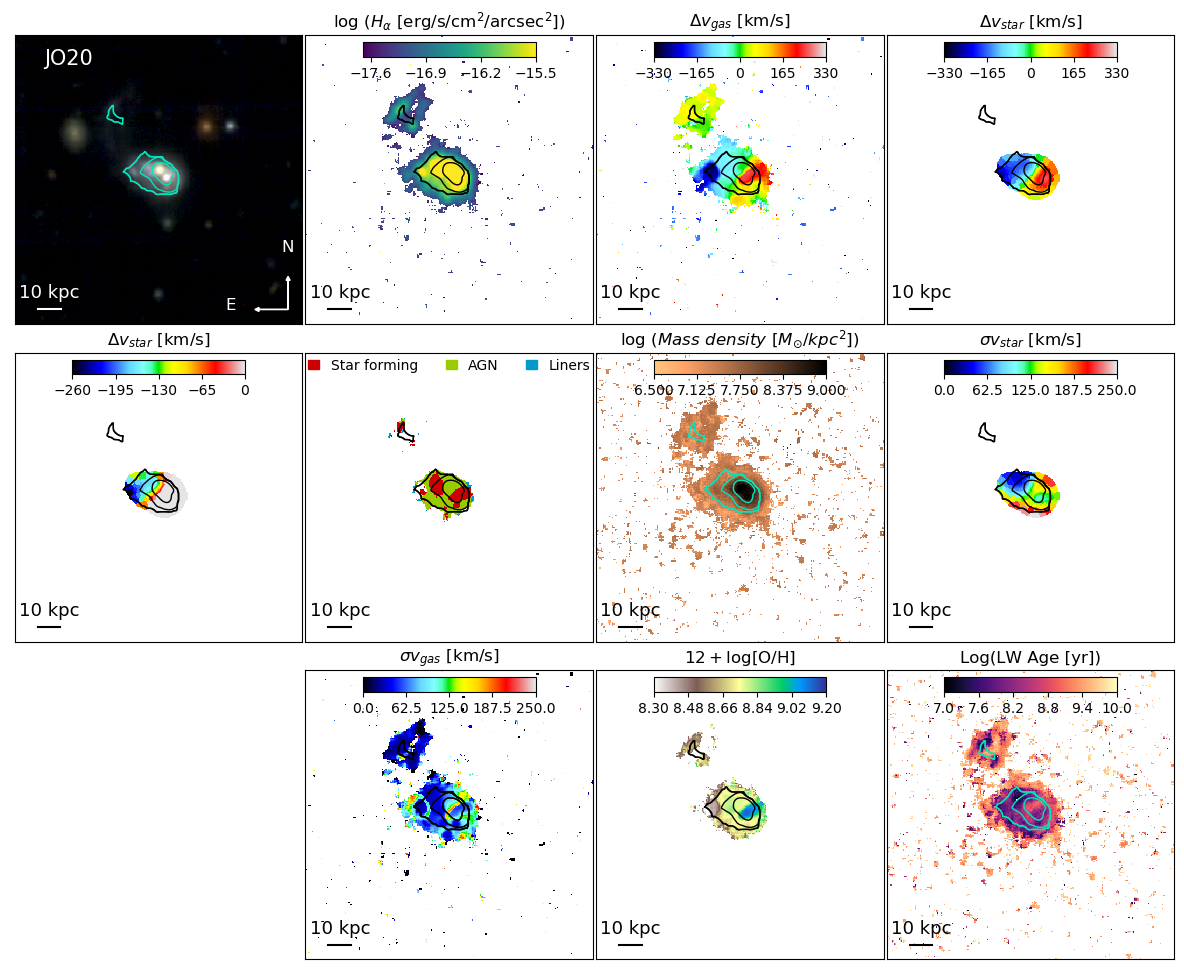}
\caption{JO20: example of merging system. From left to right: rgb image, \Ha flux map,   \Ha and stellar kinematics maps, gas and stellar velocity dispersion maps,  BPT map obtained using the OI line and the division by \cite{Kauffmann2003},  metallicity map, mass density map and luminosity weighted age map. {  The stellar velocity map is shown twice with a different range of velocities, to better highlight the rotation of the eastern object.} The kpc scale is also shown in each panel. North is up, and east is left.  Cyan or black contours represent the distribution of the oldest stellar population. \label{fig:merger} }
\end{figure*}

{  To understand whether P17048 and P17044 are indeed interacting, we also inspect their surrounding environment.}  P17048 is  part of a small four-member group for \cite{Tempel2014_g}, while both \cite{Calvi2011TheGalaxies} and \cite{Saulder2016} classify it as binary system.  P17044 is at 22\arcsec\ from P17048, towards South, and it is visible in rgb image shown in the first panel of Figure \ref{fig:interacting}.  P17044 is at z= 0.04938, its mass is $\log (M_\ast/M_\sun) \sim 8.8$. 
Following the \cite{Vollmer2005b} prescription, $\frac{a_{tid}}{a_{gal}}>0.15$ is reached at r$\sim$10\arcsec (=$\sim$ 10 kpc), implying  that at larger distances tidal interaction effects indeed play a role.

P17048 and P17044 are therefore likely interacting, and, given that the disturbance is more pronounced in the opposite side with respect to the position of the companion, they could have already had a closer approach, and they might be orbiting around each other.

\subsection{Merging systems}\label{sec:Mergers}

Seven galaxies fall in this category: JO20, JO134, JO190, P3984, P877, P63947 and P96949.  In all the cases but P877 the classification can be considered certain. {  Table \ref{tab:merger} summarizes the main features used to state they are undergoing mergers. As it can be seen, some galaxies do not meet all the criteria, but as explained in the text, these do not challenge the proposed classification.} 

We note that for these galaxies no companions were detected either from MUSE images or from the available redshifts in the literature, but we can not exclude the presence of very faint objects.  

 The details for the major merger P96949 can be found in \cite{Vulcani2017c} {  and are summarized in Tab. \ref{tab:merger}}, those for JO134, which is an example of multiple physical processes at play, can be found in Sec.\ref{sec:multiple},  while the description of the other galaxies is deferred to Sec.\ref{sec:Mergers_a}. 

In this section we will focus on JO20, which is a very clear case of a major 1:1 merger {  and use this galaxy to outline the typical features of galaxies undergoing mergers.}

JO20 is found in the field of view of the cluster Abell 147 for which there is no spectroscopic coverage available. Indeed, JO20  had no redshift before MUSE observations and it is the farthest away object in our sample (z=0.1471). It is quite isolated in the universe: the closest galaxy at similar measured redshift is at 18\arcmin. Nonetheless, we can not exclude the presence of some unidentified small structure. 

\begin{figure*}
\centering
\includegraphics[scale=0.6]{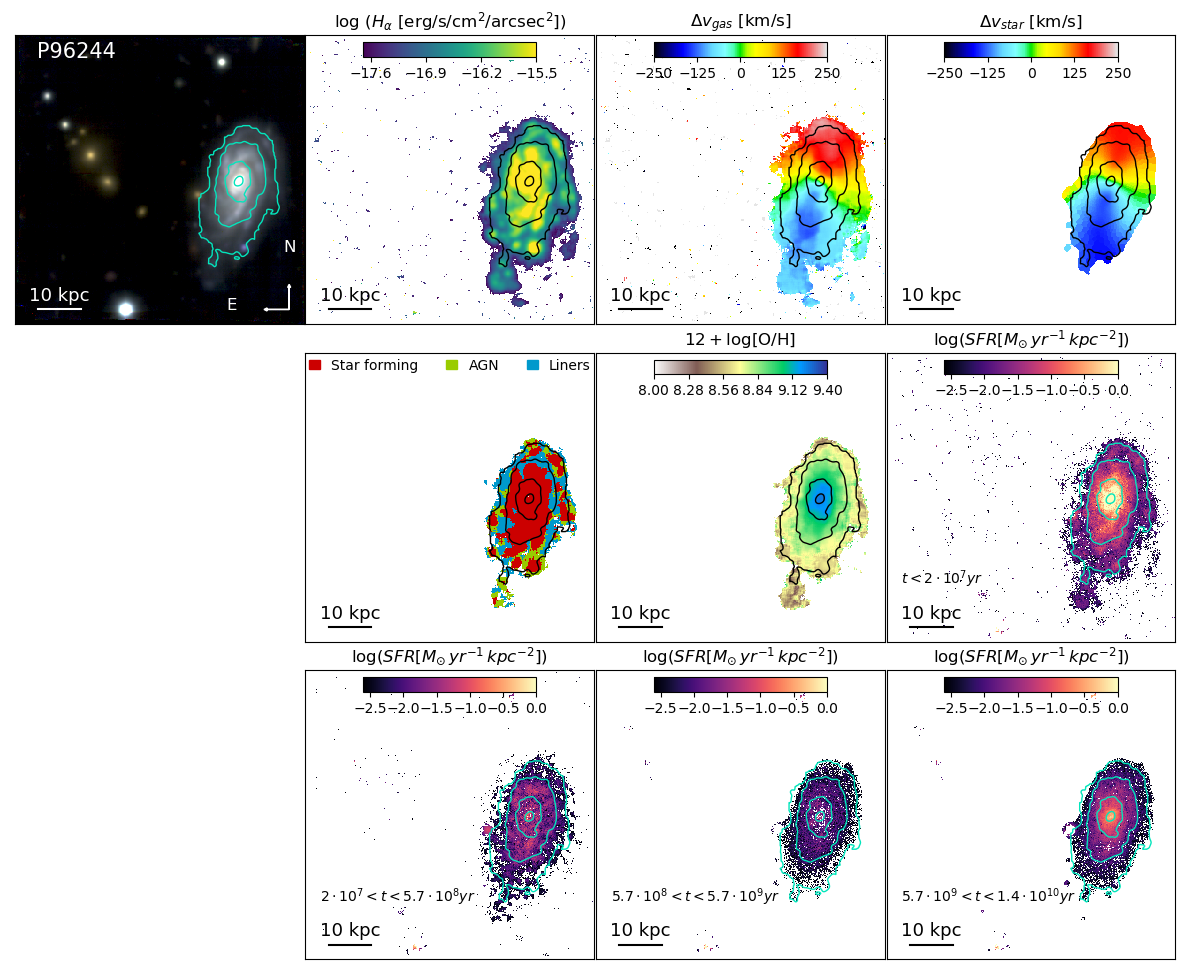}
\caption{P96244: example of ram pressure stripped candidate. From left to right: rgb images, \Ha flux maps,   \Ha and stellar kinematics maps, metallicity maps,  BPT maps obtained using the OI line and the division by \cite{Kauffmann2003}, and star formation histories maps in 4 age bins.  The kpc scale is also shown in each panel. North is up, and east is left.  Cyan or black contours represent the distribution of the oldest stellar population.\label{fig:rps} }
\end{figure*}

\begin{table*}
\centering
\small
\begin{tabular}{lccccccccc}
\hline
\multicolumn{1}{c}{\multirow{2}{*}{ID}} & member of & presence of  & { asymmetric}       & { symmetric} & evidence & stretched  & young & central & \multirow{2}{*}{$\rm flag_c$}   \\
                    & a group   & gas tail         & $\Delta v_{gas}$   &   $\Delta v_{star}$      & for shocks  &      metallicity                & tail  & burst & \\
\hline
P96244 & \cmark & \cmark  & \cmark & \cmark & \cmark & \cmark & \cmark & \cmark &1\\
JO134 & {  ? } & \cmark  & \cmark  & \xmark  & \cmark  & \xmark & \xmark & \cmark & 1\tablenotemark{\footnotesize{a}} \\
P20159  & {  ? } & \cmark & \cmark & \cmark & \cmark & \cmark & \cmark & \cmark & 0  \\
P59597 & \cmark & \cmark & \cmark & \cmark & \cmark & \cmark & \cmark & \cmark & 1 \\
P5055 & \cmark & \xmark\tablenotemark{\footnotesize{b}} & \cmark  & \cmark  &\cmark & \cmark &\xmark\tablenotemark{\footnotesize{b}} & \cmark & 1\\
P5125 & \cmark & {  ?}  & \cmark  & \cmark &\xmark  &\xmark & \cmark &\cmark & 0 \\
\end{tabular}
\tablenotetext{a}{This galaxy also underwent a merger event, so flag$_c$=1 has not been determined using the usual criterion.}
\tablenotetext{b}{This galaxy is a truncated disk, which is another clear signature for RPS at a very advanced stage.}
\caption{{  Summary of the main features investigated to characterize ram pressure stripped galaxies. The meaning of the symbols is as in Tab. \ref{tab:merger}.}
\label{tab:rps}}
\end{table*}

Figure \ref{fig:merger} shows the rgb image, the gas and stellar kinematics and their velocity dispersions,  the BPT,  the ionized gas metallicity,\footnote{
In this case metallicities have been derived neglecting the information from the BPT classification.
} the mass density and luminosity weighted age maps for JO20. {  Other galaxy properties can be found in Fig. \ref{fig:JO20_bis}.}
The rgb image highlights the presence of a clear tail extending towards North. 
Two bright nuclei are visible and these are most likely the remnants of the two merging galaxies. These have similar brightness and size, suggesting the two objects had similar mass. In the \Ha flux map the tail is also visible. Part of it is detached from the main galaxy body, but ionized gas is still visible between the galaxy and the tail. The velocity field of the ionized gas is disturbed, spanning the range -300$<v [km/s]<$300 { ($A_g$=0.55)}. The locus of zero velocity is bent and there are hints for an inner disk which is fast rotating. 

The velocity field of the stellar component reveals even more clearly the existence of the two objects: the Eastern one has a typical velocity with respect to the center of -180km/s and rotates with a speed of about 60 km/s, the Western one has the same velocity as the gas, but it has almost no rotation, suggesting it could be face on.  The presence of the two components in the velocity field is evidence of the presence of merger remnants. In this case, $A_v$=0.21.\footnote{{We note that in this case $A_v$ might be biased by the very high relative velocity of the two objects ($\sim600$ km/s). This number is used as denominator in the formula to obtain $A_v$.}}
Also the stellar velocity dispersion map is very chaotic and inhomogenous: {
 its median value is 135 km/s, but reaches very high values (up to 250 \kms)}. For comparison, control sample galaxies have median velocity dispersion of 37$\pm$7 km/s, highlighting the chaotic motions of the stars in the aftermath of the collision. 
The BPT map shows the presence of an AGN in the core, along with a bi-conical structure. Very little star formation in the disk is ionizing the gas; the tail is instead most likely ionized by young stars  formed as a consequence of the merger. 

The metallicity map is very asymmetric and highlights again the presence of two systems: the eastern component - roughly corresponding to the peak of negative gas velocity -  is metal poor ($12 + \log(O/H)$ = 8.5), while the western one - corresponding to the peak of positive gas velocity - is much richer ($12 + \log(O/H)$ = 9.1). Most of the mass of the galaxy is confined in the core, where the two merger remnants are visible as a double peak in the mass distribution.  Finally, the luminosity weighted age map shows many young regions (LWA$<10^{7.5}$ yr), suggesting that the merger has produced a burst in star formation.  An arc characterized by young ages (LWA$\sim10^{7}$ yr) is also clearly visible towards the East, where the metallicity is the lowest.

\subsection{Galaxies undergoing ram pressure stripping} \label{sec:RPS}

\begin{figure}
\centering
\includegraphics[scale=0.3]{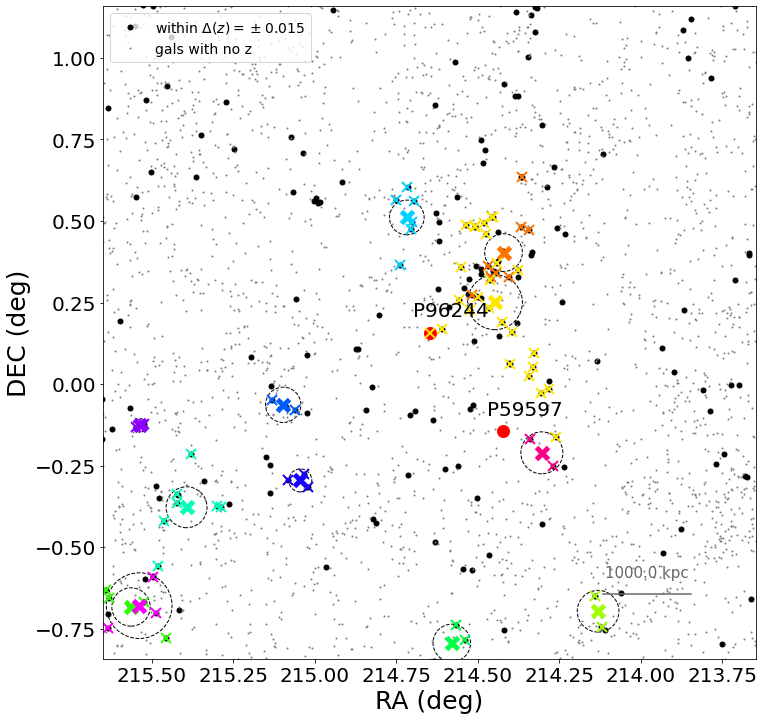}
\caption{Spatial distribution of galaxies around P96244. The group definition is taken from \cite{Tempel2014_g}. P96244 is represented with thick red symbol. Colored crosses represent galaxies in groups (each color identifies a different group) whose redshift is within $\pm0.01$ from the galaxy redshift. Black dots represent galaxies whose redshift is within  $\pm0.015$ the galaxy redshift, grey small dots represent galaxies with no redshift available (from Hyperleda). Note that P96244 and P59597 (Sec.\ref{sec:RPS_a}) are relatively close in space and are in the same plot, therefore also P59597 is indicated here. \label{fig:rps_env} }
\end{figure}

\begin{figure*}
\centering
\includegraphics[scale=0.3]{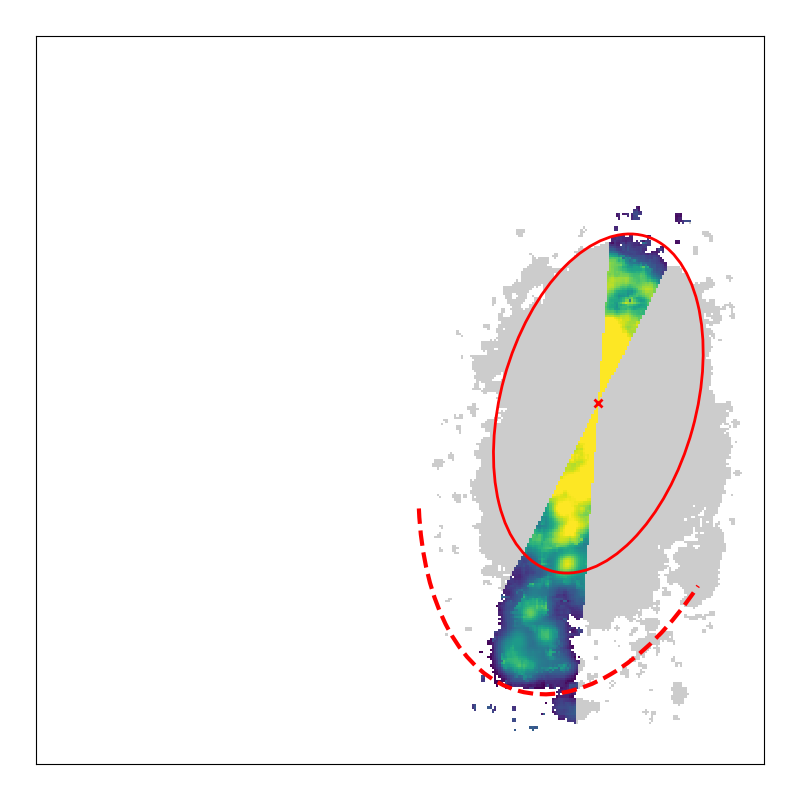}
\includegraphics[scale=0.37]{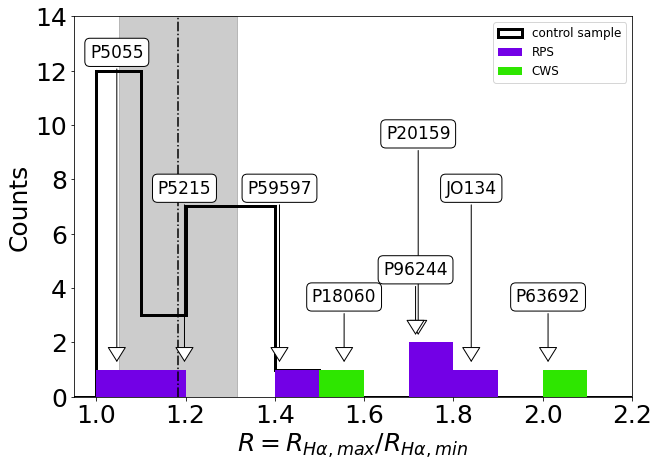}
\caption{{  Left: Example of the procedure adopted to compute the asymmetric extent of \Ha in galaxies undergoing stripping, with the intent of identifying the tail. The grey region identifies the extent of \Ha in P96244. Colored spaxels are the ones used to compute the maximum \Ha extent ($R_{H\alpha_{max}}$), identified by the red dashed line,  and the minimum  \Ha ($R_{H\alpha_{min}}$) extent, identified by the red solid ellipse (see text for details. Right: distribution of $R=R_{H\alpha_{max}}/R_{H\alpha_{min}}$ for control sample galaxies (black histogram), ram pressure stripping galaxies (purple histogram) and cosmic web stripping galaxies (green histogram). Vertical dashed line represents the median control sample value, grey region indicate 1$\sigma$ uncertainty. Stripping galaxies have systematically larger R than control sample galaxies, highlighting the presence of a ionized gas tail. } \label{fig:tail} }
\end{figure*}

\begin{figure*}
\centering
\includegraphics[scale=0.6]{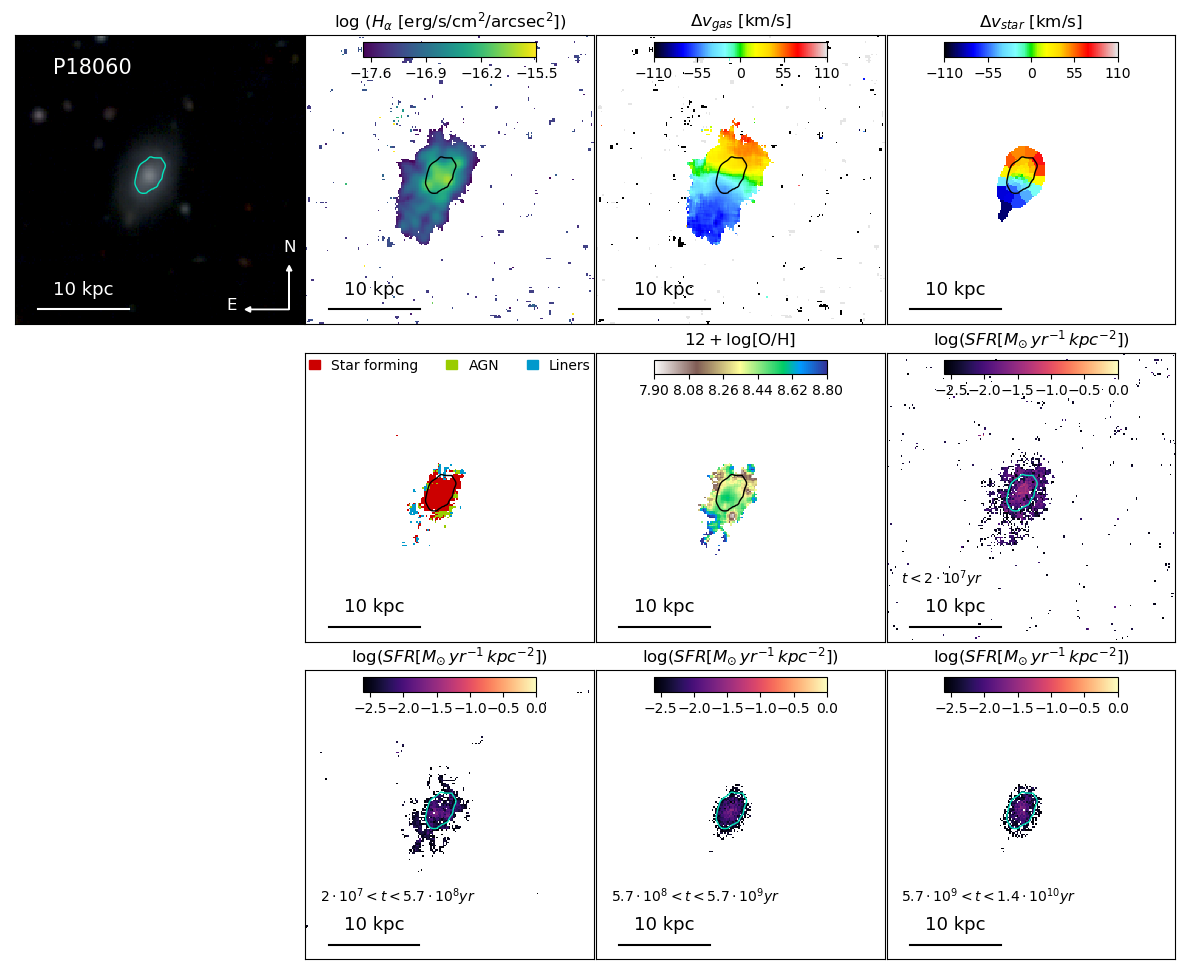}
\caption{P18060: example of cosmic web stripping candidate. From left to right: rgb image, \Ha flux map,  \Ha and stellar kinematics maps, metallicity map,  BPT map obtained using the OI line and the division by \cite{Kauffmann2003}, and star formation histories maps in 4 age bins.  The kpc scale is also shown in each panel. North is up, and east is left.  Cyan or black contours represent the distribution of the oldest stellar population.\label{fig:cws} }
\end{figure*}

We identify {  six} galaxies in our sample whose characteristics are consistent with ram pressure stripping. { All of these galaxies but JO134 have $A_v<0.3$, suggesting  symmetric stellar kinematics.} The best candidate is P96244 {  and will be used to outline the main features of galaxies falling in this class. Its main} properties are shown in Fig. \ref{fig:rps}: the rgb image, the BPT, \Ha flux, \Ha gas, and stellar kinematics, metallicity and the SFH  maps. {  Additional maps are shown in Fig.\ref{fig:P96244_bis}.} As the hosting environment is a  key feature for understanding if ram pressure stripping can be advocated as the main mechanism, Fig.\ref{fig:rps_env} shows the surrounding region of P96244. The same plots for the other candidates (P20159 and P59597) are shown in Sec. \ref{sec:RPS_a}. P5055 and P5215 are the other two cases of possible ram pressure stripping in groups and have been characterized in \cite{Vulcani2018_g}. {  The properties of all galaxies falling in this class are summarized in Tab.\ref{tab:rps}.} {  As already mentioned, JO134  is an example of multiple physical processes (merger+ram pressure stripping) at play, and will be discussed in Sec.\ref{sec:multiple}.}

Its rgb image shows that P96244 is a spiral galaxy with a clear tail extending towards South. The signatures of unwinding arms (i.e. dislodged material that appears to retain the original structure of the spiral arms) are also observed, a feature that can be induced by ram pressure stripping \citep{Bellhouse2020}. The tail is even more visible in the \Ha flux map, where small detached clouds are also detected. {  The presence of the tail for this and for all the other ram pressure and cosmic web stripping candidates has been confirmed by comparing the extent of \Ha along two opposite directions with the extent of \Ha in undisturbed galaxies (see Sec.\ref{sec:results2} for details on the control sample). Specifically, for each galaxy showing a possible tail, we selected a wedge centered on the galaxy center and with an aperture of 40 deg enclosing the tail. We then measured the maximum extent of \Ha ($R_{H\alpha_{max}}$), i.e. the semi-major axis of an ellipse with the same PA, inclination and ellipticity of the galaxy, enclosing 95\% of the data points. Next we took the symmetrical wedge with respect to the galaxy center and measured again the extent of \Ha ($R_{H\alpha_{min}}$). We then took the ratio of the two quantities ($R$) to obtain a measurement of how much \Ha is extended on one side (tail) with respect to the other. An example of the procedure is shown in the left panel of Fig.\ref{fig:tail} for P96244. We did the same analysis for the control sample galaxies. As these ones do not have a tail by definition, we selected two symmetric wedges along the semi-major axis.\footnote{Choosing a random position did not alter the control sample results.} The right panel of Fig.\ref{fig:tail} shows the distribution of $R$ for the stripping candidates and for the control sample. The mean $R$ for the control sample is $\bar{R} =1.18\pm0.12$. No control sample galaxies has $R>\bar{R}+2\sigma$. In contrast,  stripping candidates typically exceed this value, allowing us to conclude that they are characterized by a tail. The only exceptions are P5055, which is a truncated disk, and P5215, whose tail is only hinted, as extensively discussed by \cite{Vulcani2018_g}.}

The \Ha flux map of P96244 also indicates that the objects seen towards East in the rgb image are actually background objects. {  The stellar velocity field is overall regular. The  bending of the locus of zero velocity is due to the presence of a bar (O. Sanchez et al. in prep.). The same bending is observed also in the ionised gas velocity field. We also observe that the gas in the tail maintains a coherent rotation with the galaxy.} { The velocity field though shows  asymmetries, as $A_g=0.68$}. 
The metallicity gradient in the stellar disk is overall regular, but the metallicity in the tail is stretched and  the lowest (A. Franchetto et al. in prep.). 
The line ratios in the tail are indicative of processes other than star formation as producers of the ionized gas \citep{Poggianti2019b}. Finally, the SFH maps point to a recent formation of the tail and shows a strong burst of star formation in the galaxy core for $t<2\times 10^7$yr. 

The tail is also aligned along the direction that connects the galaxy and the center of its hosting group:  P96244 is indeed part of a 26 member group (Fig.\ref{fig:rps_env}, \citealt{Tempel2014_g, Saulder2016}) at z=0.053, with $\sigma=201.9 km/s$, $R_{vir}=0.31$ Mpc and halo mass of 10$^{13.65}$\ms. P96244 is at 2.6$R_{vir}$ and 1.2 $\sigma$ from the group center and might be at first infall.  However, Fig. \ref{fig:rps_env}  shows that the  approach adopted by \citealt{Tempel2014_g} might have broken the same structure into many smaller systems, so the galaxy could actually be part of a bigger group whose center is closer to the galaxy.

\begin{table}
\centering
\small
\begin{tabular}{lccccccc}
\hline
\multicolumn{1}{c}{\multirow{2}{*}{ID}} & \multirow{2}{*}{D/T} & $r_d$   & r$_{extent}$ &  \multirow{2}{*}{r/R$_{200}$} & \multirow{2}{*}{$v/\sigma$} & $\Pi_{gal}$  \\
                    &    & (kpc)        & (kpc)   &     &    &   (N m$^2$)   \\
\hline
P96244 & $\sim 1$ & 3.9  &12.8 &2.6 &1.2& 1.0$\times 10^{-13}$ \\
P20159  &   $\sim 1$ & 1.6 & 5.3 & 2.2 & -3.7 & 8.4$\times 10^{-14}$ \\
P59597 &  $\sim 1$ & 1.8 & 6.8 & 4.3 & -6.5 &4.0$\times 10^{-14}$ \\
P5055\tablenotemark{\footnotesize{a}} &  $\sim 1$ &2.3 &  12.6 & 2.4 & -0.5 & 5.4$\times 10^{-15}$\\
P5125\tablenotemark{\footnotesize{a}} &  $\sim 1$ & 4.2 &  15.0 & 0.3 & -2.1 & 1.4$\times 10^{-14}$ \\
JO134 &  $\sim 1$ & 1.1 &2.6 & ? & ? &2.4$\times 10^{-13}$\\
P18060  & $\sim 1$ & 1.1 &3.5 & 5.6 & -2.7 &8.3$\times 10^{-14}$ \\
P63692 &  $\sim 1$ &  1.0  & 3.6 & 6 & 0 &3.2$\times 10^{-14}$ \\
\end{tabular}
\tablenotetext{a}{Values taken from \citet{Vulcani2018_g}}
\caption{{  Structural parameters and other physical properties of the galaxies undergoing stripping: disk to total light ratio (D/T)  disk scale-length ($r_d$),  extent of the \Ha emission within the mayor axis of the disk (r$_{extent}$), projected phase-space coordinates, and $\Pi_{gal}$ at r$_{extent}$.}
\label{tab:rps_math}}
\end{table}

{  To better understand whether ram pressure stripping can be invoked to explain the features observed in P96244 an in the other stripping galaxies, we follow the approach presented in \cite{Vulcani2018_g}. As we do not have at our disposal X-ray observations of the environment surrounding the galaxies, we can not directly infer the intensity of ram- pressure from the density of the intra-group medium ($\rho_{IGM}$). We instead must compare the expected ram- pressure with the self-gravity or anchoring pressure across the galaxy ($\Pi_{gal}(r_{gal})$, where $r_{gal}$ is the radial distance from the galaxy centre), which reflects their ability to retain gas. Gas stripping will occur when $P_{ram} > \Pi_{gal}(r_{gal})$. 
We compute $\Pi_{gal}$ using a pure disk model as described in \cite{Jaffe2018}, with the parameters listed in Tab. \ref{tab:rps_math}. We assume that all galaxies are disk-dominated (D/T$\sim1$), measure the extent of the \Ha emission disk along the semi major axis ($r_{extent}$) of the galaxies and compute the stellar disk scale length $r_d$ from the stellar mass of galaxies, as in \cite{Wu2018}. We assume a gas fraction of 0.25 (corresponding to $\sim 10^{10} M_\odot$ late-types, \citealt{Popping2014}) and a disk scale-length for the gas 1.7 times that of the stars \citep{Cayatte1994}. We  compute $\Pi_{gal}$ at $r_{extent}$ assuming this is the maximum radius at which ram- pressure has been able to strip gas.  $\Pi_{gal}(r_{gal} = r_{extent})$ can be considered an upper limit to the $P_{ram}$ experienced by the galaxies. We report in Tab. \ref{tab:rps_math} the maximum ram-pressure experienced by the analyzed galaxies. These numbers must be compared to the predicted $P_{ram}$ from hydrodynamical simulations of groups and filaments: according to \cite{Bahe2013}, for groups with $\rm M_{host} = 1 \, to \, 3 \times 10^{13} M_\odot$ the estimated $P_{ram}$ ranges from $\sim 3\times 10^{-14}$ to $\sim 10^{-15}$ N m$^2$ in the region 1-3$\times R_{200}$. Filaments around low-mass structures have a $P_{ram}$ ranging from  $\sim 2\times 10^{-13}$ to $\sim 6\times 10^{-15}$ N m$^2$   in the region 1-5$\times R_{200}$ from the centre of the group. In voids, $P_{ram}$ can range from  $\sim 10^{-14}$ to $\sim 10^{-16}$ N m$^2$  in in the region 1-5$\times R_{200}$. This analysis confirms that the galaxies analyzed in this Section can indeed be undergoing a ram pressure sufficient to produced the observed features. P96244 is the galaxy most likely feeling the largest $P_{ram}$ of the sample. Similarly, also cosmic web stripping galaxies can indeed be feeling some stripping, but it is not clear if that is produced by the closest groups, indeed found too far in both space and velocity to exert a significant pressure, or by the voids. } 

Finally, we can exclude that P96244 is feeling tidal interaction: its closest galaxy is P59391 at 145\arcsec\, (=150 kpc), whose stellar mass is $10^{9.9}$\ms. According to \cite{Vollmer2005}, P96244 would feel the effect of the galaxy galaxy interaction only 74\arcsec\ away from the galaxy center. Note, though, that P59391 is present in the \cite{Poggianti2016JELLYFISHREDSHIFT} catalog of stripping candidates, included in the mildest stripping category (JClass=1). This galaxy could therefore be another case of ram pressure stripping exerted by the group.

\subsection{Galaxies undergoing cosmic web stripping} \label{sec:cws}
\begin{table*}
\centering
\small
\begin{tabular}{cccccccc}
\hline
\multicolumn{1}{c}{\multirow{2}{*}{ID}} & group only in   & presence of  &  { asymmetric}   &  { symmetric}     & stretched & young  & \multirow{2}{*}{$\rm flag_c$}   \\
                    & the surrounding   & gas tail         & $\Delta v_{gas}$      & $\Delta v_{star}$       &    metallicity                & tail & \\
\hline
P18060 & \cmark & \cmark  & \xmark & \cmark & \cmark & \cmark & 1\\
P63692 & \cmark & \cmark & \cmark & {  ?} & {  ?} & \cmark & 0\\
\end{tabular}
\caption{{  Summary of the main features investigated to characterize cosmic web stripping candidates. The meaning of the symbols is as in Tab. \ref{tab:merger}.}
\label{tab:cws}}
\end{table*}

Within the sample, we detected two cosmic web stripping candidates: P18060 and P63692. They both have low values of stellar mass ($\log (M_\ast/M_\odot)<9.1$). 
Note the the galaxies analyzed by \cite{Benitez2013}  are passive, while our galaxies are  still star forming, so we hypothesize we are witnessing the early phases of this phenomenon.

\begin{figure}
\centering
\includegraphics[scale=0.3]{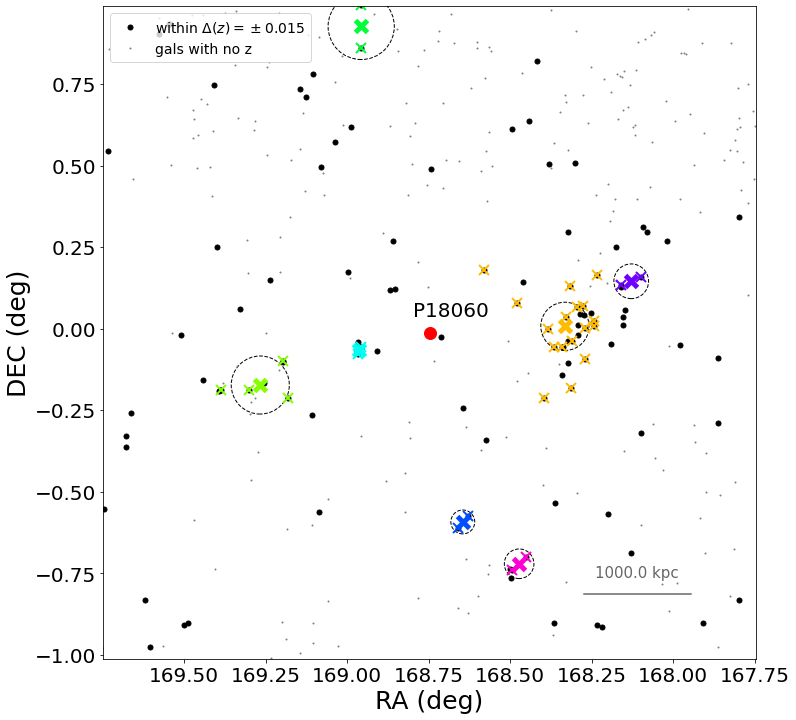}
\caption{Spatial distribution of galaxies around P18060. Colors  and symbols have the same meaning as in Fig. \ref{fig:rps_env}. \label{fig:cws_env} }
\end{figure}

{  Table \ref{tab:cws} summarizes the main features investigated to assess cosmic web stripping. These are also shown in  Fig. \ref{fig:cws}  for P18060. These are  the rgb image, the BPT, \Ha flux, \Ha gas, and stellar kinematics, metallicity, SFHs maps. Additional maps can be found in Fig. \ref{fig:P18060_bis}}. Figure \ref{fig:cws_env} shows the environment where P18060 is found. The description of  P63692 is deferred to Sec.\ref{sec:cws_a}. 

Its rgb image shows that P18060 seems to have a rather regular morphology.  The clouds seen in the rgb image are background objects.
However, the \Ha map highlights the presence of a tail towards the South. {  The presence of the tail is supported by the analysis of the \Ha extent (Fig.\ref{fig:tail}).} The { extent of the} gas distribution is asymmetric, { but the rotation in the disk is symmetric ($A_g=0.22$).} The gas rotates slower than the stars (-60$<v_{gas} [km/s]<$60, -100$<v_{star} [km/s]<$100). 
The galaxy has an inverted metallicity gradient, with the tail having a metallicity of $12+\log[O/H]\sim 8.8$  the core a metallicity of $12+\log[O/H]\sim 8.4$, the side opposite to the tail a metallicity of  $12+\log[O/H]\sim 8.0$. This inverted gradient is often found in low mass galaxies \citep[e.g.,][]{Wang2019}. This piece of evidence could suggest that either the gas was displaced during the stripping, or that rather than stripping we are witnessing  accretion of high metallicity gas from the North. However, there are no other indications for gas accretion. The BPT maps shows that the gas is mostly powered by star formation. The SFH maps point to a rather recent formation of the tail. 

While all its properties are overall consistent with ram pressure stripping, P18060 does not belong to any structure (Fig.\ref{fig:cws_env}). A binary system is at 677 kpc. According to \cite{Tempel2014_g, Saulder2016}, the closest group  is at 1268 kpc from P18060, at a redshift of 0.04062. Given the estimated radius of the group (0.2259 Mpc), the galaxy is at 5.6 $R_{vir}$. The group velocity dispersion is $\sigma= $267.1 km/s, meaning that the velocity difference  between the galaxy and the group is $\sim 720 km/s$, corresponding to $\sim 2.7\times \sigma$. {   Tab.\ref{tab:rps_math} reports the expected $\Pi_{ram}$ on the galaxy, which is consistent with stripping exerted by filaments/voids. }
Even though the group could hardly be responsible for the galaxy properties, it is worth mentioning  that the position of the tail of P18060 is roughly aligned along the direction of the group.

The closest galaxy to P18060 is P18079 at z = 0.03951, found at 137.89\arcsec. This galaxy has no mass estimate available from the literature, but it is too far to exert a significant tidal influence on P18060, so we conclude that the most likely mechanism is cosmic web stripping.

\begin{table*}
\centering
\small
\begin{tabular}{ccccccccc}
\hline
\multicolumn{1}{c}{\multirow{2}{*}{ID}} & member of & \Ha beyond  & symmetric & properties of \Ha clouds      & \Ha clouds      & \multirow{2}{*}{$\rm flag_c$}    \\
                    & a filament   & 4R$_e$   & $\Delta v_{star}$   & similar to main body   &  young  &  \\
\hline
P14672  & {  ?}  & \cmark  & \cmark & \cmark & \cmark & 0\\
P95080 & \cmark  & \cmark  & \cmark  & \cmark  & \cmark & 1\\
P63661  & \cmark  & \cmark  & \cmark  & \cmark  & \cmark & 1\\
P8721  & \cmark  & \cmark  & \cmark  & \cmark  & \cmark & 1\\
P19482  & \cmark  & \cmark  & \cmark  & \cmark  & \cmark & 1\\
\end{tabular}
\caption{{  Summary of the main features investigated to characterize cosmic web enhancement galaxies. The meaning of the symbols is as in Tab. \ref{tab:merger}. }
\label{tab:cwe}}
\end{table*}

{  We note that at first sight, cosmic web stripping galaxies could also simply be irregular galaxies:  the two populations cover a similar mass and luminosity range \citep{Hunter1986, Hunter2004}. Nonetheless, the main difference between the two populations lies in the comparison between their stellar and gas kinematics.  In irregular galaxies the stellar kinematics is expected to closely follow the motion of the gas \citep[e.g.,][]{Johnson2012}, while in cosmic web stripping galaxies the gas extends much further away than the stellar component, especially in a preferential direction that gives origin to a tail. This is  suggestive of stripping.}

\subsection{Galaxies undergoing cosmic web enhancement} \label{sec:CWE}

\begin{figure*}
\centering
\includegraphics[scale=0.6]{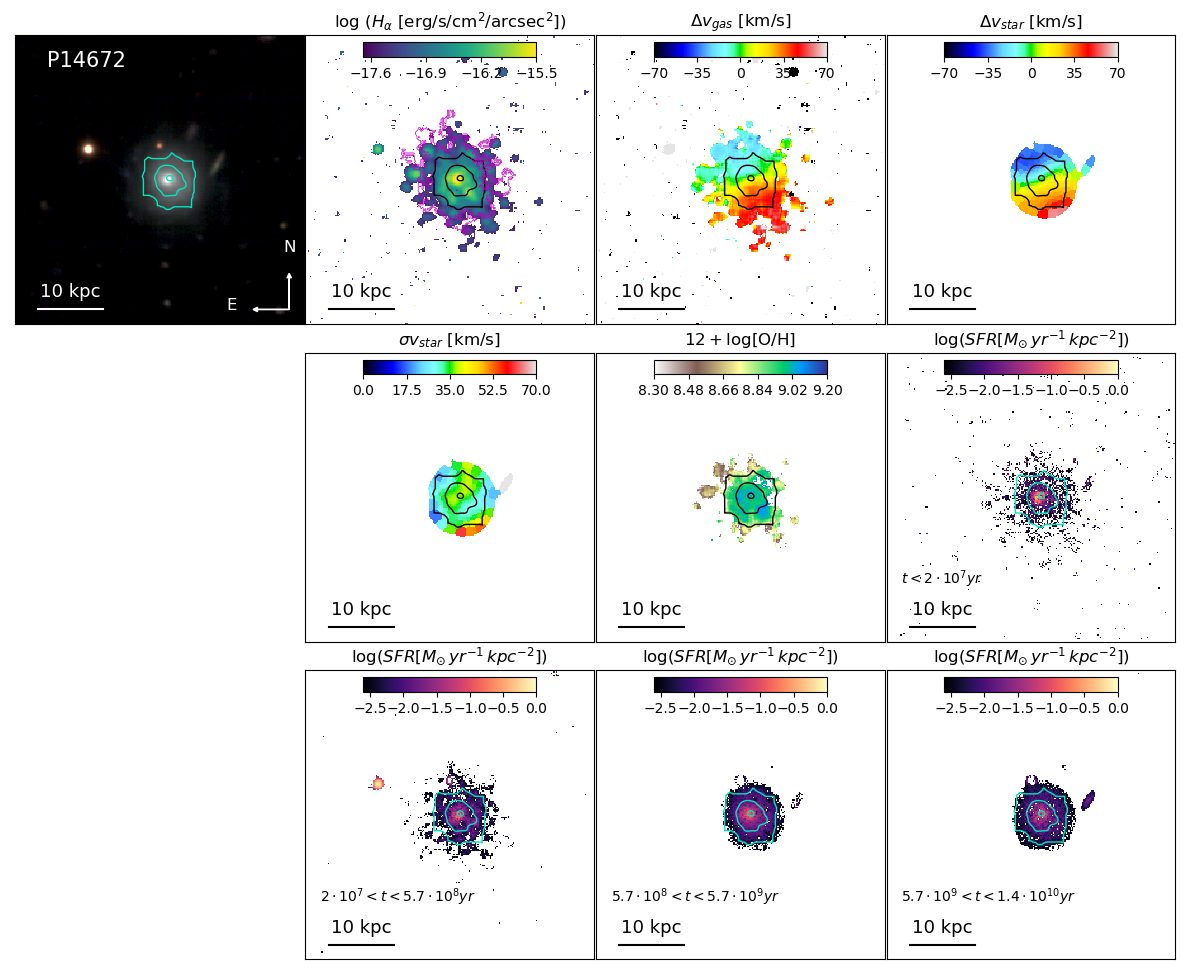}
\caption{P14672: example of cosmic web enhancement candidate. From left to right: rgb image, \Ha flux map,  \Ha and stellar kinematics maps, stellar velocity dispersion map, metallicity map,  BPT map obtained using the OI line and the division by \cite{Kauffmann2003}, and star formation histories maps in 4 age bins.  The kpc scale is also shown in each panel. North is up, and east is left.  Cyan or black contours represent the distribution of the oldest stellar population.  { The magenta contour on the \Ha flux map represents the galaxy rotated of 180$^\circ$ and is used to quantify the asymmetry (see text for details). } \label{fig:cwe} }
\end{figure*}
\begin{figure}
\centering
\includegraphics[scale=0.3]{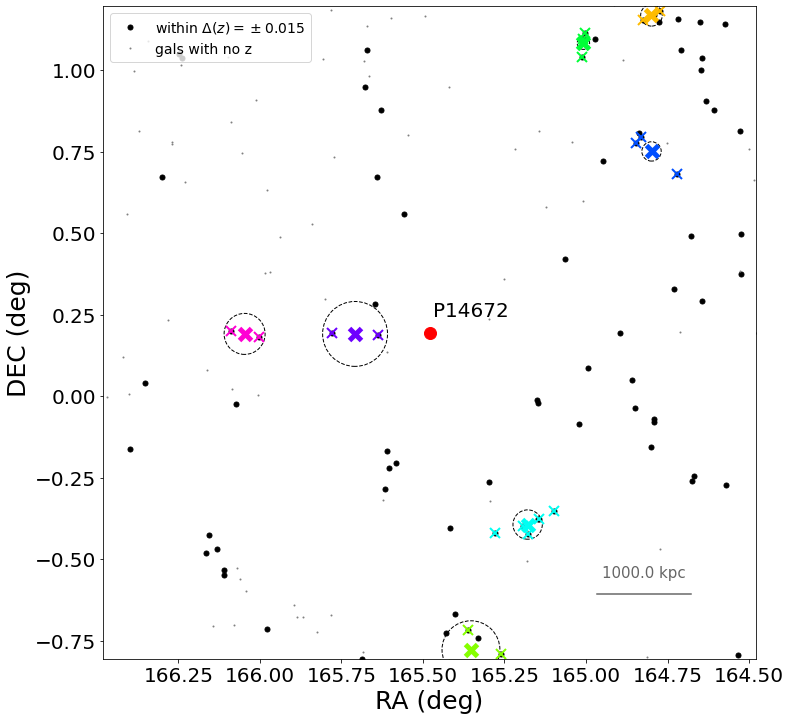}
\caption{Spatial distribution of galaxies around P14672.  Colors  and symbols have the same meaning as in Fig. \ref{fig:rps_env}. 
\label{fig:P14672_env} }
\end{figure}

In \cite{Vulcani2019_fil} we identified four galaxies feeling Cosmic Web Enhancement (P95080, P19482, P8721 and P63661) and in  \cite{Vulcani2018_g} we proposed that P5215 might also be feeling the same mechanism, instead of ram pressure stripping.

Within the GASP sample, there is another candidate that might be undergoing cosmic web enhancement, as it shows properties similar to those of the galaxies discussed by \cite{Vulcani2019_fil} {  and summarized in Tab.\ref{tab:cwe}}: P14672, whose {  main} properties are shown in Fig. \ref{fig:cwe}.\footnote{{  Additional maps are shown in Fig.\ref{fig:P14672_bis}.}} Nonetheless, this classification is highly uncertain, as the redshift coverage is very poor in the part of the sky surrounding the galaxy and we can not properly characterize its  environment in order to detect the possible presence of a filament (Fig. \ref{fig:P14672_env}). The closest group at similar redshift (z=0.051) to the galaxy is at 800 kpc. The velocity difference between the galaxy and the group center is 370 km/s. {  In this case, the environmental information is the key, therefore even though all the other criteria adopted to define the mechanism are met, we assign flag$_c$=0.}

The rgb  shows no peculiar features, but it is evident that the galaxy is lopsided {( $A_o=65\%$).} We note that the elongated source towards the West, at the edge of the galaxy, is a background star-forming galaxy. The \Ha flux map shows a large number of knots, many detached clouds that are at the same velocity as the galaxy, so they belong to them. These clouds have no preferential direction. In \cite{Vulcani2019_fil} we selected galaxies with a maximum extent of \Ha in units of $r_e$ (R(\Ha)$_{max}$) $>4$. For P14672 R(\Ha)$_{max}$=3.97, therefore it just barely did not pass the selection threshold. As for the galaxies in \cite{Vulcani2019_fil}, both the gas and stellar kinematics are quite regular, { the stellar velocity field is symmetric ($A_v$=0.17)} and the stellar velocity dispersion is relatively low. The ratios of emission line fluxes confirm that they do belong to the galaxy gas disk, the metallicity map shows a rather regular gradient, with the metallicity in the clouds consistent with that of the galaxy outskirts. The SFH  maps point to a recent formation of the clouds.

There is no other mechanism able to explain the morphology of the galaxy, so alternatively it could simply be an irregular galaxy. However, in the literature we have found no object with similar properties. Narrow band image surveys like  the \Ha Galaxy Survey \citep[\Ha GS,][]{Shane2001}, the \Ha galaxy survey \citep{James2004}, the H-alpha Galaxy Groups Imaging Survey (H$\alpha$ggis, PI. Erwin), An \Ha Imaging Survey of Galaxies in the Local 11 Mpc Volume \cite[11Hugs][]{Kennicutt2008}, Dynamo \citep{Green2014}, or  Fabry-Perot observations like the Gassendi \Ha survey of SPirals \cite[GHASP][]{Epinat2008} have no field galaxies with such extended and luminous ($\rm{\log (H\alpha [erg/s/cm^2/arcsec^2]>-17.5}$) \Ha regions located well beyond R25. 

\subsection{Galaxies experiencing gas accretion} \label{sec:accr}

P11695 is the best example in our sample of a galaxy fed by low-metallicity gas inflow from the cosmic web, as it is found in isolation and has one of the steepest and most asymmetric metallicity gradients observed so far \citep[e.g.][]{Pilyugin2014}. All its properties are described at length in \cite{Vulcani2018_b}.

P40457 and P4946 are instead good candidates for the galaxies that underwent substantial gas accretion from a metal rich less massive object, whose presence is not visible, even though in both cases the classification is uncertain. P4946 has been already discussed in \cite{Vulcani2018_g} and it is also characterized by a gas disk counter-rotating with respect to the stellar disk, suggesting that the accretion happened with a retrograde motion.  

\begin{table*}
\centering
\small
\begin{tabular}{cccccccccc}
\hline
\multicolumn{1}{c}{\multirow{2}{*}{ID}} & \multirow{2}{*}{isolated}  &  asymmetric & regular           & asymmetric & asymmetric & counter-rotating & \multirow{2}{*}{$\rm flag_c$}   \\
                     &                              & rgb and \Ha     &$\Delta v_{star}$ & metallicity & growth &disk\tablenotemark{\footnotesize{a}} & \\
\hline
P40457 &{  ?} &\cmark & {  ?} &\cmark &\cmark & \xmark  & 0 \\
P11695 &\cmark &\cmark &\cmark &\cmark &\cmark  & \xmark  & 1\\
P4946 &\xmark &\xmark &\cmark  & {  ?} & \cmark & \cmark &0\\
\end{tabular}
\tablenotetext{a}{This is not a critical feature, but a clear evidence for gas accretion.}
\caption{{  Summary of the main features investigated to characterize galaxies undergoing gas accretion. The meaning of the symbols is as in Tab. \ref{tab:merger}. }
\label{tab:accr}}
\end{table*}

\begin{figure*}
\centering
\includegraphics[scale=0.6]{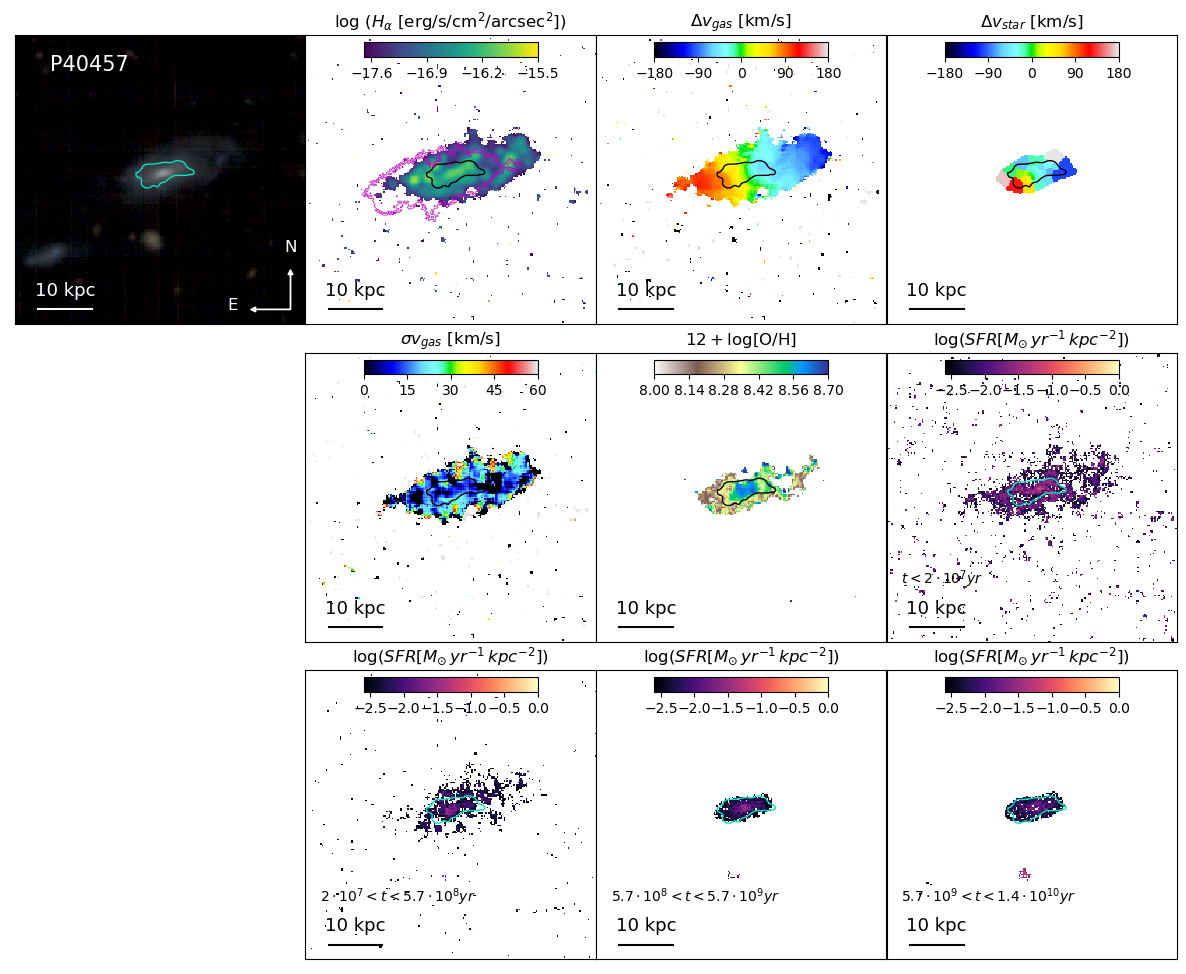}
\caption{P40457: example of gas accretion from a minor object.From left to right: rgb image, \Ha flux map,   \Ha and stellar kinematics maps, gas velocity dispersion map,  metallicity map, and star formation histories maps in 4 age bins.   The kpc scale is also shown in each panel. North is up, and east is left.  Cyan or black contours represent the distribution of the oldest stellar population.  { The magenta contour on the \Ha flux map represents the galaxy rotated of 180$^\circ$ and is used to quantify the asymmetry (see text for details). }
\label{fig:accretion_min} }
\end{figure*}

{  Table \ref{tab:accr} summarizes some  features that can point to accretion, even though it is important to keep in mind that galaxy properties can change depending on the properties of the accreted material. }

Figure \ref{fig:accretion_min} shows the rgb image, \Ha flux, \Ha gas, and stellar kinematics, metallicity, SFHs maps for P40457, {  while additional maps are shown in Fig.\ref{fig:P40457_bis}}. The rgb  shows that the stellar disk, when compared to the contours representing the original body, is lopsided and extends towards the West. Note that the object with similar color detected 20\arcsec\ East from the galaxy is a foreground object (z=0.046).  The lopsidedness is also very accentuated in \Ha { $A_o=59\%$}. It is  quite astonishing how extended the gas disk is compared to the stellar one, especially when we consider the original body.  The stellar kinematics is quite regular ($-180<v [km/s]<180$, { $A_v=0.17$}), even though it needs to be taken with caution due to the low number of bins used, due to the low S/N. { The gas kinematics is instead asymmetric: $A_g=0.44$}.  The gas velocity dispersion is also overall quite low ($<15 km/s$) and only slightly higher in the Northern side of the galaxy. The metallicity map is very asymmetric, with different gradients in the different parts of the galaxy (A. Franchetto et al. in prep.). The northern part is much richer than the southern one. The SFH maps unveil that the galaxy grew much in size in the last two age bins, during which it developed the lopsidedness. 

Regarding the environment, the galaxy is in a portion of the sky with quite low spectroscopic coverage, so the environment is difficult to identify. According to the available data,  and the closest galaxy with the same redshift is 250\arcsec\ away, the closest group is about 1 Mpc away \citep{Tempel2014_g}. 

All these properties are consistent with gas accretion, most likely of some metal rich gas, proceeding from the North West.

\subsection{Galaxies experiencing starvation} \label{sec:passive}

\begin{table}
\centering
\small
\begin{tabular}{cccccccc}
\hline
\multicolumn{1}{c}{\multirow{2}{*}{ID}} & no \Ha   & regular &  homogeneous& \multirow{2}{*}{$\rm flag_c$} \\
                     & emission &  $\Delta v_{star}$ & SF suppression & \\
\hline
P16762 & \cmark   & \cmark  & \cmark  &1 \\
P443 & \cmark   & \cmark  & \cmark &1 \\
P5169  & \cmark   & \cmark  & \cmark &1 \\
\end{tabular}
\caption{{  Summary of the main features investigated to characterize galaxies underwent a starvation event. The meaning of the symbols is as in Tab. \ref{tab:merger}. }
\label{tab:pass}}
\end{table}

\begin{figure*}
\centering
\includegraphics[scale=0.6]{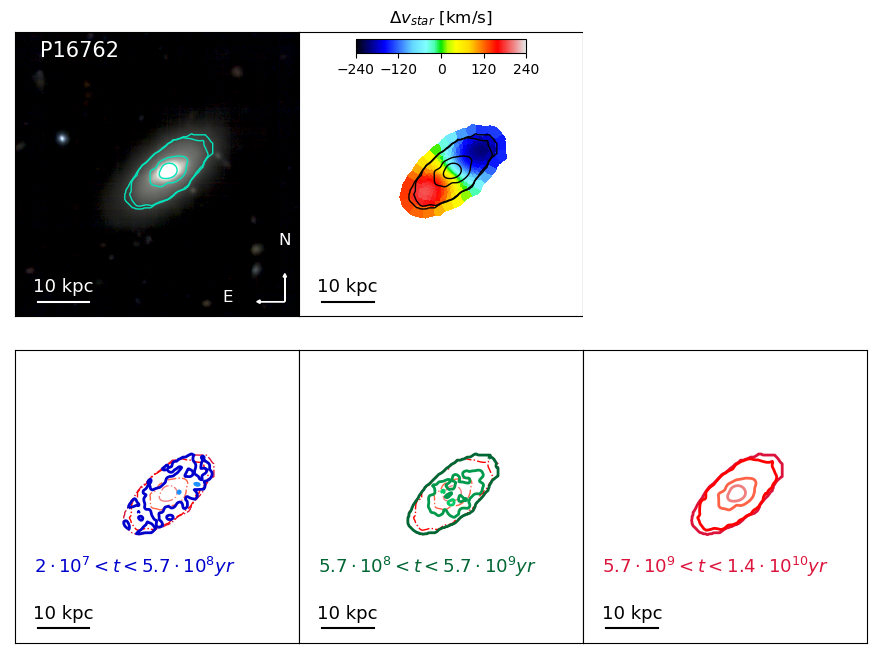}
\includegraphics[scale=0.33]{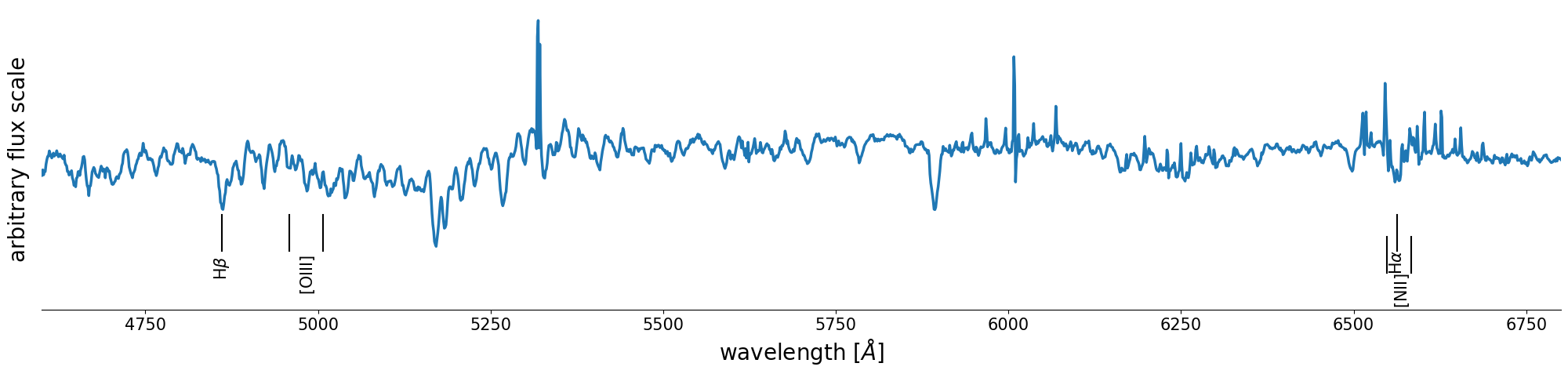}
\caption{P16762: example of starvation candidate. The rgb, stellar velocity map, the comparison between the different age maps and the integrated spectrum are shown.  Cyan or black contours represent the distribution of the oldest stellar population (from \sinopsis). In the comparison between the stellar maps of different ages. The old star formation bin ($t > 5.7\times 10^9$ yr; red lines) is shown in all the panels, for reference. Left: recent star formation bin ($2 \times 10^7$ yr $< t < 5.7 \times 10^8$ yr; blue lines). Middle: intermediate star formation bin ($5.7 \times 10^8$ yr $< t < 5.7 \times 10^9$ yr; green lines). Right: old star formation bin. Contours are logarithmically spaced between SFR = 0.0001 and 0.1 $M_\sun$/yr.\label{fig:passive} }
\end{figure*}
The three passive galaxies in the sample were inserted among the targets to study the final stages of galaxy evolution in the field. 

P5169 is a group galaxy that has been already studied in  \cite{Vulcani2018_g}. The other two galaxies in this category are P443 and P16762. {  The main properties investigated are summarized in Tab.\ref{tab:pass}. These are also shown in Fig. \ref{fig:passive} for P16762. These are: the rgb image, the stellar kinematics map,  the comparison between the different age maps and the integrated spectrum. All the other properties are shown in Fig.\ref{fig:P16762_bis}.} 
To ease the comparisons between the SFR at the different epochs, we only plot the contours logarithmically spaced between SFR = 0.0001 and 0.1 $M_\sun$/yr/kpc$^2$ and use as reference the oldest age bin. Note that the youngest age bin is not shown because no ongoing star formation is detected.  P443 is instead discussed in Sec.\ref{sec:passive_a}.

\begin{figure*}
\centering
\includegraphics[scale=0.6]{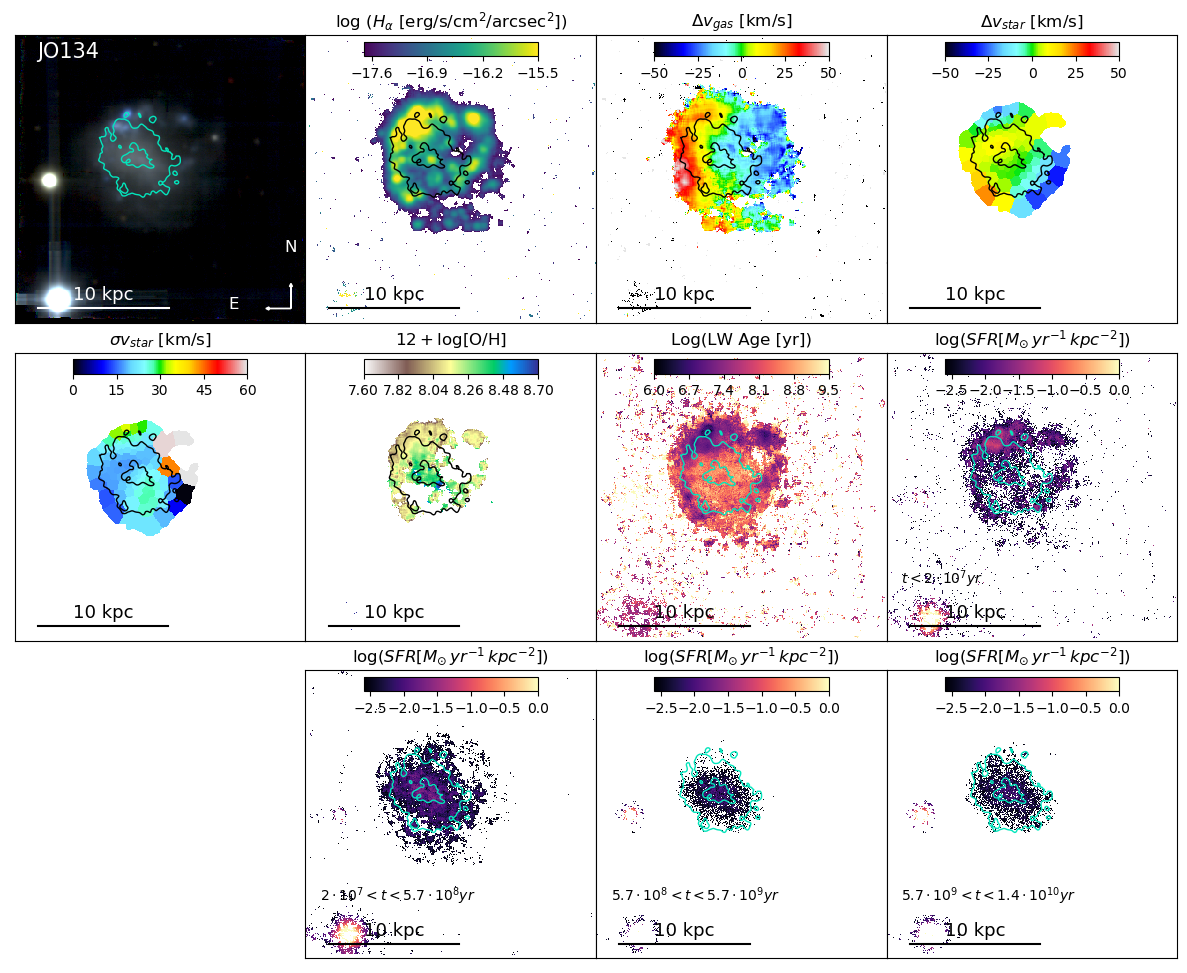}
\caption{JO134: example of a combination of merger and ram pressure stripping.  From left to right, top to bottom: rgb image, \Ha flux map,   \Ha and stellar kinematics maps, metallicity map, luminosity weighted age map and star formation histories maps in 4 age bins.  The kpc scale is also shown in each panel. North is up, and east is left.  Cyan or black contours represent the distribution of the oldest stellar population.
\label{fig:mul} }
\end{figure*}

\begin{figure*}
\centering
\includegraphics[scale=0.5]{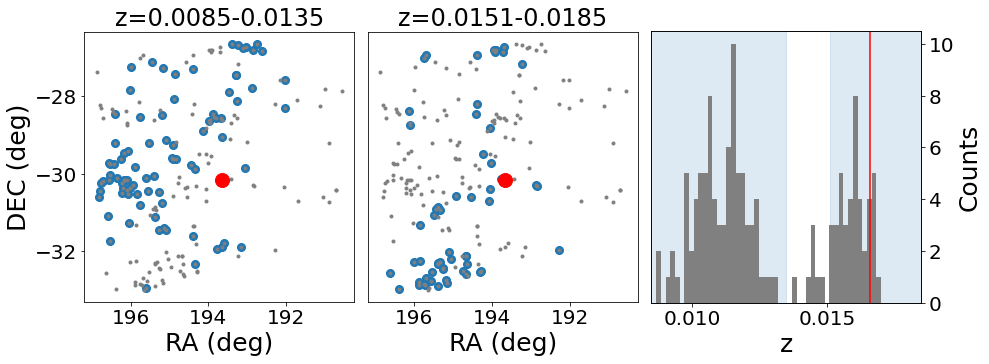}
\caption{Environment around JO134. As the galaxy is outside the SDSS footprint, no group definition is available. Grey dots represent galaxies in the redshift range $0.05<z<0.019$, blue filled circles are the galaxies in the redshift ranges indicated in the titles. JO134 is represented as red circle. The histogram represents the redshift distribution of galaxies. Vertical line is the redshift of JO134. Shaded areas are the redshift intervals indicated in the spatial distributions. \label{fig:JO134_env} }
\end{figure*}

P16762 is a typical example of passive disks \citep{bundy10, Bamford2009}: it is classified as an S0 galaxy \citep {Calvi2018MorphologyUniverse} and the rgb image reveals the presence of  a disc, but there is no star formation throughout the disk, as indicated by the absence of emission lines in the integrated spectrum. The stellar velocity field is very regular (-240$< v[km/s]<$240), { and symmetric ($A_v=0.09$),} excluding the possibility that interactions or a merger affected the galaxy, at least recently. P16762 also has a very high central velocity dispersion ($\sigma v_{star}=$170 km/s), which indicates the presence of a prominent bulge and excludes the existence of a bar \citep{Das2008, Aguerri2009}.
The comparison of the extent of the SFR at different epochs suggests the same suppression happened throughout the disk: the extension of the maps is very similar at all epochs. No outside-in quenching, {  often used as a proxy for ram pressure stripping \citep[e.g.][]{Vulcani2020},} is observed. This result points to  starvation as the main mechanism. The galaxy is part of a binary system: the companion is a massive galaxy (10$^{11.1}$\ms) located at 70\arcsec. According to the \cite{Vollmer2005} formulation, tidal interaction could have an effect on the galaxy for $r>13\arcsec$, but no signs are visible. 

\subsection{JO134: a case of multiple processes} \label{sec:multiple}
To conclude this overview, we focus on a galaxy that might be simultaneously affected by two mechanisms. This is actually the only clear example of coexistence of two mechanisms in the field sample, suggesting that typically the presence of a dominant mechanism washes out the effect of the secondary one \citep[but see][for an example in GASP clusters]{Fritz2017}. 

Figure \ref{fig:mul} shows the rgb image, the \Ha flux, the \Ha and stellar kinematics, the metallicity, the luminosity weighted age maps and star formation histories maps in 4 age bins for JO134, while Fig.\ref{fig:JO134_env} shows its environment.  {  Tables \ref{tab:merger} and \ref{tab:rps} summarize the most important features used to pin point the mechanisms. Fig.\ref{fig:JO134_bis} summarizes all the maps we have at our disposal for this galaxy.}

The rgb image unveils a low luminosity central part, and a very clear bright region towards North East, composed by many blue knots. Overall the galaxy morphology is very irregular, especially in the North-West side of the galaxy. The \Ha maps shows very bright regions in the Northern part, and a clear low luminosity tail towards the South is visible. Some detached clouds are also observed. The ionized gas has an overall { asymmetric ($A_g = 0.71$)} and low rotation, in the range -30$<v_{gas} [km/s]<$ 30. The rotation axis goes from South to North and clear distortions are seen in the Northern side of the galaxy, where the locus of zero velocity has a meander. The gas kinematics in the tail has a coherent rotation with the rest of the galaxy, though it is stretched. The stellar kinematics is  much more chaotic and no regular rotation is detected. { The measured asymmetry is  $A_s = 0.45$.} A clear trail of constant velocity (v$\sim 20$ km/s) crosses the galaxy from East to West. The stellar velocity dispersion is the highest along this trail (plot not shown) reaching a $\sigma_{star}>100$ km/s. The same region also has very low metallicity. The LWA map shows very young ages in all of the northern part of the galaxy (LWA$\sim10^6$ yr), distributed along an arc,  and also at the  end of the tail. Similarly, the SFH maps show that the galaxy was born as an overall regular object, but from t$<6\times10^9$ yr an asymmetry started to develop. In the youngest age bin, both the Northern side and the tail developed.

JO134 had no redshift measurement prior to MUSE observations and  is also outside the SDSS footprint, so no group definition is available. Extracting all the available redshifts in the surrounding area, there are two main structures, one at the redshift of the galaxy and one at slightly lower redshift (Fig. \ref{fig:JO134_env}). Both structures have a velocity dispersion of $\sim 250$km/s and are still in formation, with a no clear center, so it is not possible to measure the distance of JO134 from the groups' centers.  

The most probable scenario that could explain the characteristics of JO134 is that on one side the galaxy is falling towards a group and therefore feeling ram pressure stripping that is producing the bow shock in the Northern part and the tail developing on the opposite side, and on the other hand it is also undergoing a minor merger event. 
The bright blue clump seen towards North-West in rgb image could indeed be a bullet { (i.e. a compact galaxy moving at very high speed}) hitting the galaxy with a different velocity and affecting  the orbits of the galaxies in the Northern part. The chaotic stellar velocity field and the high velocity dispersion in that region are consistent with this possibility.  Alternatively  to the bullet, it could be an old gas-rich merger from which the disk regrew before being stripped.

\section{General trends} \label{sec:results2}

\begin{figure}
\centering
\includegraphics[scale=0.47]{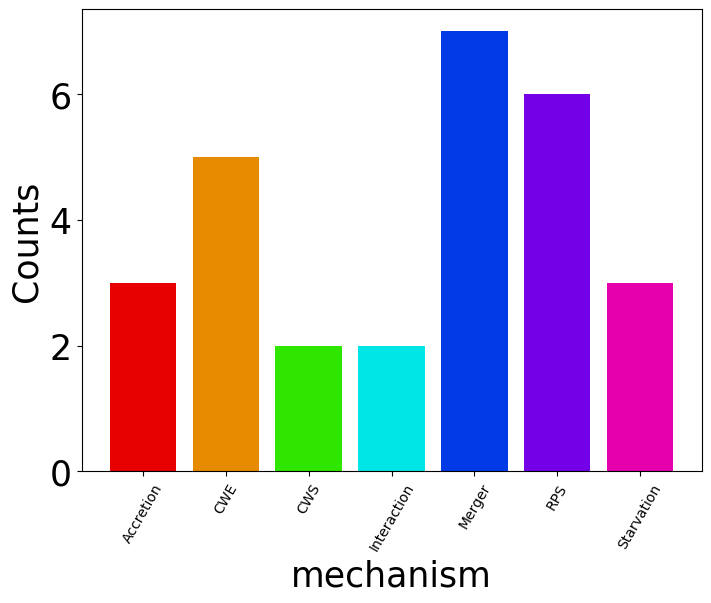}
\caption{Incidence of the different categories in the sample: Accretion (either gas accretion or from a small object), Cosmic web enhancement (CWE), Cosmic web stripping (CWS), Interaction, Merger, Ram pressure stripping (RPS), Starvation. \label{fig:class} }
\end{figure}

In the previous section we have combined the information obtained from the spatially resolved properties of the galaxies and the characterization of the environment in which they are embedded to infer the most probable mechanism acting on them. 
Figure \ref{fig:class} presents a summary of the different categories and shows the distribution of galaxies within each class.\footnote{{  We note that as for JO134 we identified two coexisting main mechanisms, this galaxy will be counted twice in the following distributions. As for P5215, where we can not firmly distinguish between ram pressure stripping and cosmic web enhancement \citep{Vulcani2018_g}, we assume ram pressure stripping as the dominant process.} }

Overall, the most populated category is that of merger (either minor or major) with 7 galaxies, followed by  ram pressure stripping  (6) and cosmic web enhancement (5). We remind the reader that GASP started with the aim of characterizing gas removal processes, paying particular attention to ram pressure stripping. It is therefore interesting to note that even though merging events and interactions were purposely excluded, they still represent more than 35\% of the non-cluster star-forming sample. 
This result indicates that a visual selection based on optical images can easily mismatch processes affecting both the stellar and gas component and processes that leave the stellar component unaltered. The presence of interacting galaxies also suggests that recognizing companions on the basis of optical images is not trivial, even for expert inspectors. 

\begin{figure}
\centering
\includegraphics[scale=0.44]{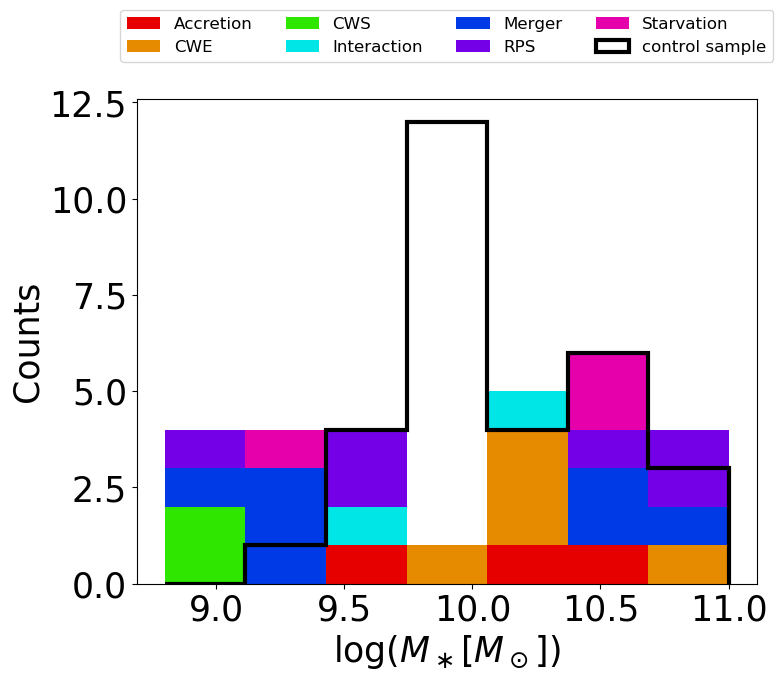}
\caption{Stacked mass distribution for galaxies in the sample, as indicated by the label. The black line shows the mass distribution of the undisturbed sample from \citet{Vulcani2019b}, for reference. \label{fig:mass_distr} }
\end{figure}

Taking into account the original selection performed by \cite{Poggianti2016JELLYFISHREDSHIFT}, three galaxies were 
classified as undisturbed (P12823, P14672 and P877), but instead they are not, while the other candidates were labelled with a different degree of stripping. There is however no clear correlation between the classification obtained from the optical selection (``JClass'', with numbers ranging from 0 to 5 with 5 representing the galaxies with the clearest evidence for stripping) and the proposed mechanism resulting from the spatially resolved analysis (e.g.  it is not the case that all galaxies with JClass = 4 or 5 turning out as ram pressure stripping galaxies, while galaxies with JClass=1 being cosmic web enhancement). 
The three galaxies included as passive systems (P443, P16762, P6169) based on fiber spectroscopy turned out to be passive across their entire galaxy disk and no galaxy selected as star forming  from fiber spectroscopy turned out to be passive instead. 

\begin{figure}
\centering
\includegraphics[scale=0.47]{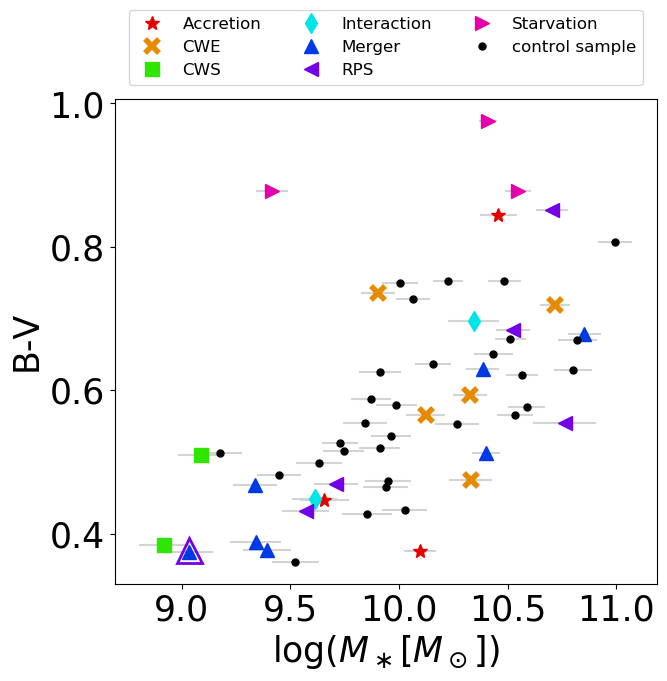}
\caption{Reconstructed Absolute B-V -\ma relation of galaxies in the sample, compared to that of the undisturbed galaxies. Colors and symbols refer to galaxies of the different classes, as indicated in the label. The empty purple triangle represents JO134, which is simultaneously undergoing both a merging event and ram pressure stripping.  \label{fig:color_mass} }
\end{figure}

We can now investigate some general trends of the sample, to inspect whether some category stands out in any of the main scaling relations, obtained using integrated values. The aim is to understand whether, in  the absence of spatially resolved data, we can use 
one of these relations to select galaxies undergoing a specific process. 
In what follows, we will use for comparison the GASP undisturbed sample, already exploited e.g. in \cite{Vulcani2019b}. {  Though small, this sample allows us to avoid systematics that would be introduced by using external samples, where different approaches to measure stellar masses, star formation rates, metallicities and sizes have been adopted.} This sample includes both field and cluster galaxies, for a total of 30 objects. 

\begin{figure*}
\centering
\includegraphics[scale=0.47]{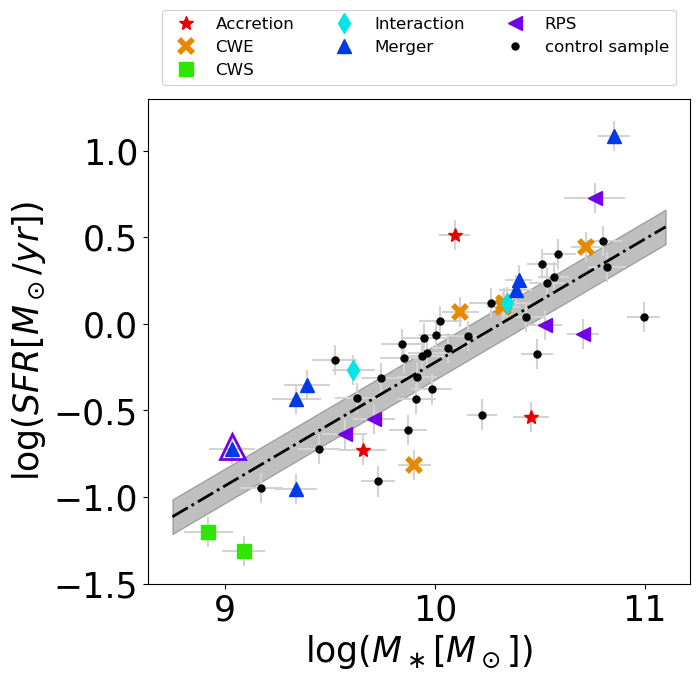}
\includegraphics[scale=0.47]{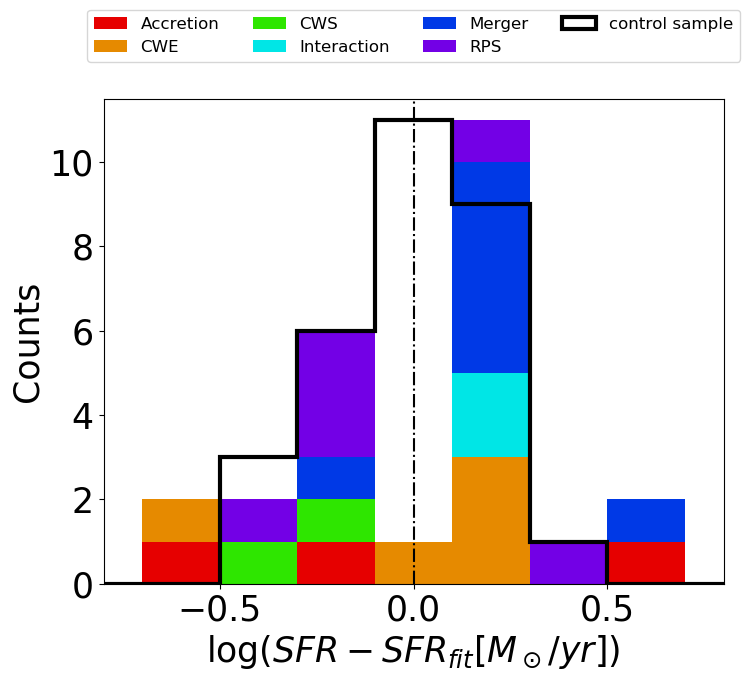}
\caption{Left: SFR-\ma relation of galaxies in the sample, compared to the undisturbed relation. Colors and symbols refer to galaxies of the different classes, as indicated in the label. The empty purple triangle represents JO134, which is simultaneously undergoing both a merging event and ram pressure stripping.  The dashed line represents the fit to the control sample from \cite{Vulcani2018_g}. Right: Distributions of the differences between the galaxy SFRs and their expected value according to the fit to the control sample, given their mass. Colored histograms represent galaxies of the different classes, black empty histogram represent the control sample.  The dashed vertical line is centered at 0. 
\label{fig:sfr_mass} }
\end{figure*}

Figure \ref{fig:mass_distr} shows the mass distribution of the galaxies in the different categories. Cosmic web stripping galaxies populate the lowest mass bin. 
In contrast,  cosmic web enhancement tends to populate the high mass end of the distribution. This might be due to the fact that given the existence of mass segregation inside the filaments \citep{Malavasi2017}, more massive galaxies are found in the inner parts of the filaments where densities are higher, and therefore more easily develop signs of cosmic web  enhancement. Galaxies undergoing all the other processes seem not to have a clear dependence on stellar mass. The lack of trends could also be due to the low number statistics, even though \cite{Gullieuszik2020}  showed that for ram pressure stripping stellar mass is not the driving parameter. 
For comparison, the undisturbed sample is also plotted. 
Though the control sample lacks very low mass galaxies ($<10^{9.2}$\ms),
both median and mean values are very similar ($\sim 10^{10}$\ms) and the Kolmogorov-Smirnov test  is not able to detect statistically significant  differences between the samples.  

\subsection{Scaling relations}
\subsubsection{(B-V)-\ma relation}
Figure \ref{fig:color_mass} presents the (B-V)-\ma relation. As expected, the passive galaxies are the reddest objects in the sample, followed by the ram pressure stripped galaxy P5055, which is a truncated disk and will most likely soon become completely passive \citep{Vulcani2018_g}. Compared to the control sample, none of the classes stand out in a particular part of the diagram, although there are hints that the  other ram pressure stripping galaxies and the mergers might be slightly bluer, suggesting a large contribution of young stars. 

\subsubsection{SFR-\ma relation}

We next inspect the SFR-\ma relation, which can probe whether a sub-population is currently forming stars at higher rate than the others. 
The same relation for cluster ram pressure stripping galaxies was presented in \cite{Vulcani2018_L} where we showed
that galaxies undergoing ram pressure stripping occupy the upper envelope of the control sample SFR-\ma relation, showing a systematic enhancement of the SFR at any given mass. The star formation enhancement occurs in the disk (0.2 dex), and additional SF takes place in the tails. \citeauthor{Vulcani2018_L}'s results suggest that strong ram pressure stripping events can moderately enhance the star formation in the disk prior to gas removal. 

Figure \ref{fig:sfr_mass} shows the SFR-\ma relation and the distribution of the difference between the SFR of each galaxy and the value derived from the control sample fit given the galaxy mass \citep[from][]{Vulcani2018_L}.
While the sample size is small to perform definitive statistical tests, some qualitative trends  emerge. First of all, we note that overall the plane spanned by the sample analysed in this paper and the control sample are very similar, even though control sample galaxies have a narrower distribution in the difference between the measured and estimated SFRs.

\begin{figure*}
\centering
\includegraphics[scale=0.47]{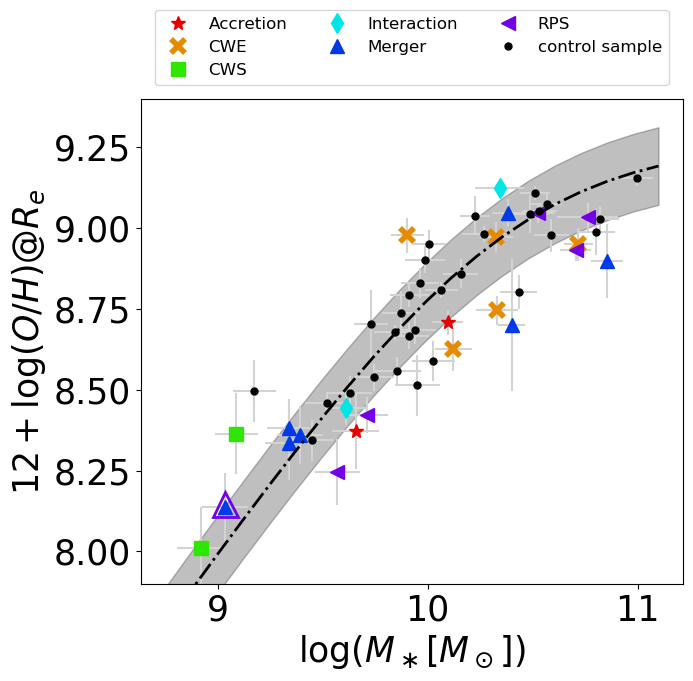}
\includegraphics[scale=0.47]{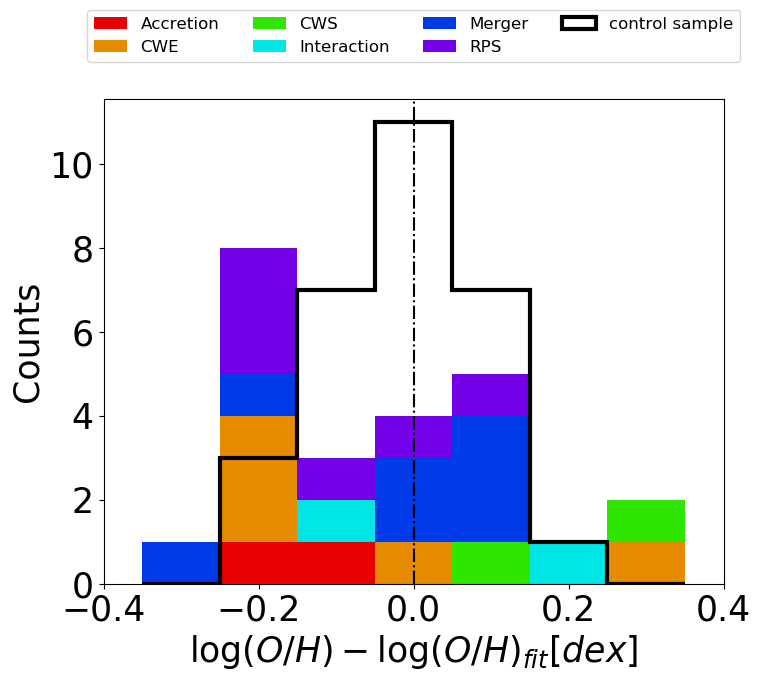}
\caption{Left: \ma-ionized gas metallicity relation of galaxies in the sample, compared to the undisturbed relation.  Colors and symbols refer to galaxies of the different classes, as indicated in the label. The empty purple triangle represents JO134, which is simultaneously undergoing both a merging event ans ram pressure stripping. The dashed line represents the fit to the control sample from \cite{Franchetto2020}. Right: Distributions of the differences between the galaxy ionized metallicity and their expected value according to the fit to the control sample, given their mass. Colored histograms represent galaxies of the different classes, black empty histogram represent the control sample.  The dashed vertical line is centered at 0.\label{fig:mass_met} }
\end{figure*}

Focusing on the different sub-populations, all but one (P63947) of the mergers lie well above the control sample fit. Their mean SFR excess is $\sim 0.2$ dex { (corresponding to a 2$\sigma$ deviation)}. The two interactions also lie just above the relation. This result is consistent with the notion
that mergers and interactions can indeed enhance star formation \citep[e.g.,][]{Barnes2004, Kim2009, Saitoh2009, Schweizer2009, Davies2015, lin08, Perez2011,  Athanassoula2016}. 
Only two ram pressure stripping candidates (P96244 and JO134) lie above the control sample relation, while all the others actually tend to occupy the lower envelope, suggesting they could  be transitioning toward the passive sequence. We remind the reader that P96244 (see Sec. \ref{sec:RPS}) is our best candidate for ram pressure stripping in groups and is most likely found at the peak of the stripping. Its excess is of $\sim 0.4$ dex { (= 4$\sigma$ deviation)} and the enhancement was also visible in the SFH maps shown in Fig. \ref{fig:rps}. 
P5055, the ram pressure stripping candidate with the largest negative excess (0.3 dex, {  i.e. 3$\sigma$ deviation}), is indeed a truncated disk.  

Cosmic web enhancement galaxies also lie along the upper envelope of the relation 
suggesting that  the star formation induced by this mechanism alters the global star-forming properties of the galaxies. The only exception is P14672, which is the only galaxy in this group with an uncertain classification. 

Among the accretion candidates, P11695, which accreted gas with low metallicity, is well above the control sample SFR-\ma relation, and it excess is of 0.7 dex. In contrast, the two galaxies undergoing a metal rich inflow show a suppression of their SFR. Both cosmic web stripping candidates are  below the control sample fit.  

Overall, all the trends described above are very mild, suggesting that none of the considered physical mechanisms dramatically affect the  properties of star-forming galaxies.

\subsubsection{Ionized gas metallicity-\ma relation}

We next inspect the ionized gas metallicity-\ma relation, to test whether the different physical mechanisms can impact the metallicity of galaxies  measured at  R$_e$. 
The same relation for cluster galaxies undergoing ram pressure stripping  was presented in \cite{Franchetto2020} where it was shown  that the chemical properties of these galaxies are similar to those of the control sample galaxies.

Figure \ref{fig:mass_met} shows the ionized gas metallicity -\ma relation and the distribution of the difference between the metallicity of each galaxy and the value derived from the control sample fit given the galaxy mass \citep[from][]{Franchetto2020}.\footnote{{  We note that for P4946 we can not obtain a reliable estimate of the ionized gas metallicity. The presence of the central AGN does not allow for a sufficient number of star-forming spaxels  to estimate the metallicity at the effective radius, therefore this galaxy is not shown  in these plots.}}

As in the case of the SFR-\ma relation, some trends are worthwhile to mention, even though the small number of galaxies in the sample prevent us from confirming the results on a statistical ground.

\begin{figure*}
\centering
\includegraphics[scale=0.47]{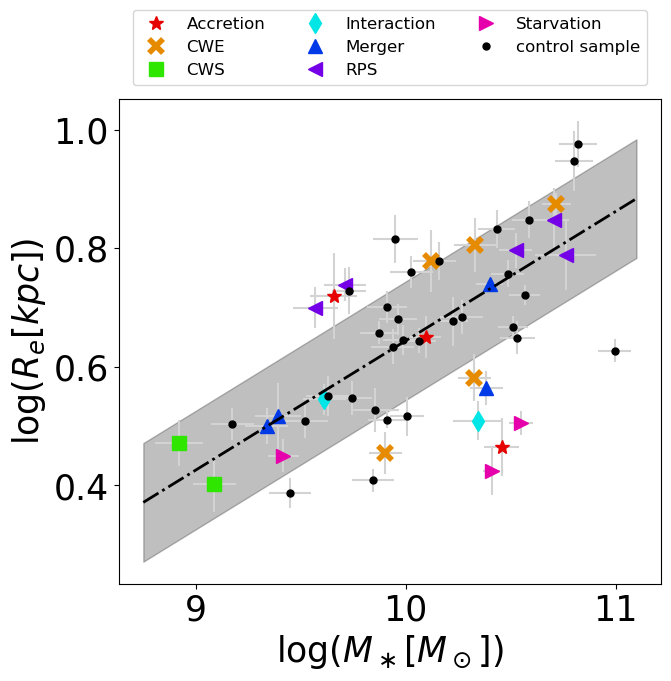}
\includegraphics[scale=0.47]{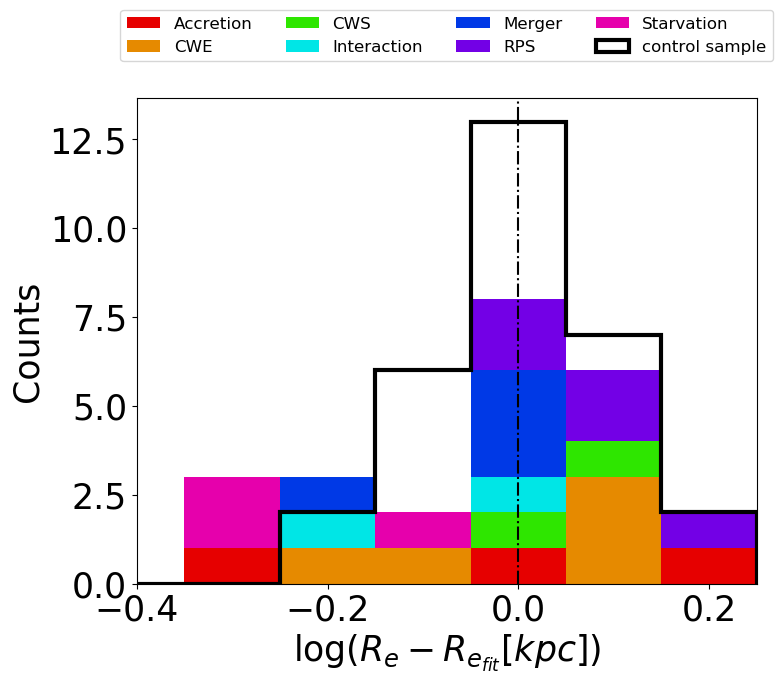}
\caption{Left: R$_e$-\ma relation of galaxies in the sample, compared to the undisturbed relation.  Colors and symbols refer to galaxies of the different classes, as indicated in the label. The dashed line represents the fit to the control sample from \cite{Franchetto2020}. Right: Distributions of the differences between the galaxy size and their expected value according to the fit to the control sample, given their mass. Colored histograms represent galaxies of the different classes, black empty histogram represent the control sample.  The dashed vertical line is centered at 0. \label{fig:mass_size} }
\end{figure*}

All ram pressure stripped galaxies but JO134\footnote{{  We recall that JO134 is also affected by a merger. Moreover, since it has no estimate of R$_e$, the expected R$_e$ was computed from the size-mass relation and therefore its metallicity at R$_e$ has an additional uncertainty.}} have a lower metallicity than expected given their mass. Excluding JO134, their mean $\log(O/H)-\log(O/H)_{fit}$ is -0.13 dex ({  $\sim 1\sigma$}). This suggests that for these galaxies the stripping is at a relatively advanced stage, as ram pressure have been able to alter the metallicity of the gas at R$_e$. 
Most mergers have a positive metallicity excess and two of them (P96949 and JO20) a negative one. The metallicity of mergers is clearly influenced by the metallicity of the merging galaxies, thus a large spread is indeed expected. 

Among the interacting systems, one is slightly richer than expected given the fit of the control sample, the other is metal poorer. Cosmic web stripping galaxies are both  metal richer, with a mean excess of 0.2 dex ({  1.6$\sigma$}).  

Cosmic web enhancement galaxies have typically a negative excess of metallicity. The only outlier is again P14672 which has instead an excess of metallicity of almost 0.3 dex ({  2.5$\sigma$}). This is another indication that this galaxy could actually be undergoing some other process. The mean negative excess of the other four cosmic web enhancement galaxies is -0.15 dex ({  1.25$\sigma$}). 

Finally, the two candidates that might have experienced gas accretion and  for which we have a  metallicity measurement show a negative excess. While this was expected in the case of P11695, whose properties are explained in terms of an inflow of low metallicity gas \citep{Vulcani2018_b} that is indeed affecting the properties of the gas at R$_e$, this is surprising for P40457 (Sec.\ref{sec:accr}), for which we proposed  an infall of enriched gas. It could be that such infall has not yet affected the gas properties at R$_e$.

\subsubsection{Size-\ma relation}

Figure \ref{fig:mass_size} shows the size-\ma relation and the distribution of the difference between the size of each galaxy and the value derived from the control sample fit given the galaxy mass \citep[from][]{Franchetto2020}.\footnote{{  We stress again that for JO134, JO190 and JO20 structural parameters could not be determined, given the irregularities of the galaxies, therefore they will be excluded from this analysis.}} Similarly to the  stripping galaxies in clusters, also the group galaxies undergoing ram pressure stripping have larger sizes (of $\sim 0.1$dex = 1$\sigma$) than control sample galaxies of similar mass. Galaxies that suffered from starvation have instead systematically smaller sizes (by 0.2 dex = 2$\sigma$), in agreement with many literature results that found that passive/early-type systems are much smaller than their star-forming counterparts \citep[e.g.,][]{shen03}.
No other clear trend is visible.

To conclude, none of the scaling relations studied above could give us irrefutable evidence for one acting mechanisms rather than another, even used in combination. This means that 
integrated values for galaxies with optical signs of unilateral debris can not firmly distinguish among the different processes. 

\section{Discussion and conclusions} \label{sec:summary}

In this paper we have analyzed the spatially resolved properties of the GASP galaxies targeted for showing  unilateral debris and gas tails in the optical imaging \citep{Poggianti2016JELLYFISHREDSHIFT} and located in the field (groups, pairs, filaments, isolated), for a total of 24 objects. In addition, we have also studied three passive galaxies in the same environments, that were included in the GASP sample aiming at characterizing the final stages of galaxy evolution. Considering also the characteristics of their hosting environment, we have therefore identified the most probable mechanism occurring for each galaxy. The intent was to test whether a visual inspection of optical images is suitable to select and classify the processes that remove exclusively the gas from galaxies, and probe the role of spatially resolved data in distinguishing among the different mechanisms.

To classify galaxies, we have strictly followed a  scheme (Fig.\ref{fig:mech}) that, being very general, could be used to classify all field galaxies in the local universe, provided the availability of spatially resolved observations that  cover the whole extent of the disk, including the galaxy outskirts, and some knowledge on the galaxy environment. 
We note that in  previous  papers \citep{Vulcani2017c, Vulcani2018_b, Vulcani2018_g, Vulcani2019_fil}, we had already characterized 10/27 galaxies, but without putting them in a  general context. 

To perform the classification, we mainly inspected the rgb images, the \Ha flux maps, the BPT maps, the stellar and gas kinematics, the metallicity maps and ages maps (either the luminosity weighted ages or the SFHs in four age bins), and the mass density maps. 

Having classified all galaxies (21/27 with a secure classification),  we have inspected the position of such galaxies on  scaling relations (SFR-\ma, R$_e$-\ma, ionized gas metallicity - \ma) obtained studying undisturbed GASP galaxies, to identify possible deviations and test whether these relations could be used to identify the acting physical mechanism.  

Thanks to the exquisite quality of the MUSE data, which allows us to study the gas and stellar properties on the kpc scale well beyond the stellar disk extent, we have identified the following mechanisms:

\begin{itemize}

    \item {\it Galaxy interactions:} The two interacting galaxies show an enhancement of the SFR, but are located on opposite sides in all the other scaling relations. Differences are most likely due to the different properties of the interactions: in one case the interacting galaxies are quite similar in mass and size and could be at the first approach, in the other case the companion is much smaller and could have been orbiting around the main galaxy for longer.
    
    \item {\it Mergers:} Seven galaxies enter this category. Almost all of them have an excess of SFR given their mass, suggesting that new stars were born after the impact. They cover a rather wide range of sizes and metallicities, most likely due to the diversity in ages, orbits of the mergers and in properties of the progenitors.
    
    \item {\it Ram pressure stripping:} Six galaxies are undergoing ram pressure stripping, one in combination with a merger event. They are clearly at different stages of stripping: one of them
    has already lost most of its gaseous disk \citep{Vulcani2018_g},  is very red and has a low SFR, another one is instead at the peak of its stripping and shows an enhancement of the star formation (especially in the core). Except for the galaxy also undergoing a merger, all these galaxies lie below the metallicity-mass relation, differently from what is found for ram pressure stripped galaxies in clusters by \cite{Franchetto2020}. Similarly to cluster galaxies \citep{Franchetto2020}, the ram pressure stripping galaxies in groups have slightly larger sizes than undisturbed galaxies. 
    
    \item {\it Cosmic web stripping:} The two galaxies undergoing this process are the lowest mass galaxies in the sample, therefore they are in the most favourable position to detect even mild environmental effects. These galaxies are quite blue, have a suppressed SFR with respect the control sample SFR-Mass relation, but a much higher metallicity. 

    \item {\it Cosmic web enhancement:} Five galaxies belong to this class, even though for one the classification is quite uncertain.
    This galaxy has also a very different position on the scaling relations with respect to the other four objects. In general, cosmic web enhancement galaxies show an excess of SFR and lie below the metalllicity-mass relation. The outlier is the reddest of this class, and has suppressed SFR and enhanced metallicity.

   \item {\it Gas accretion:} Three galaxies enter this class. One of them is most likely undergoing an inflow of low metallicity gas, as largely discussed in \cite{Vulcani2018_b}, while the others are most likely acquiring enriched gas. The former galaxy shows an enhancement of the SFR with respect to the SFR-mass relation, suggesting that this new  gas prompted the formation of new stars. In contrast, the other two galaxies are well below the control sample SFR-mass relation. 
    
    \item {\it Starvation:} All three passive galaxies can be explained in terms of starvation   having suppressed the SFR homogeneously throughout the disk. As expected, the galaxies are redder and have smaller sizes compared to the star-forming population. They are clear examples of passive disks \citep{bundy10, Bamford2009, Rizzo2018}
\end{itemize}

To summarize,  mechanisms that affect only the gas component leaving the stellar properties unaltered (ram pressure stripping, cosmic web enhancement, cosmic web stripping, accretion) constitute 65\% of the mechanisms affecting the star-forming sample. 
We stress again that this sample  was selected using deep optical imaging to characterize hydrodynamical  processes, especially  ram pressure stripping, and great care was taken to exclude mergers and interactions \citep{Poggianti2016JELLYFISHREDSHIFT}. Our results therefore imply that at least in the non-cluster sample almost 2/5 times a visual inspection of B band images is not able to identify processes affecting both the stellar and gas component, nor detect companions, highlighting how difficult it is to identify these processes from optical images. 
Also, three galaxies were observed with MUSE as part of the control sample, but the spatially resolved observations unveiled signs of disturbance.

On the other hand, we also remind the reader that passive and star-forming galaxies were selected on the basis of fiber spectroscopy \citep{Calvi2011TheGalaxies, Moretti2014WINGSClusters} and all galaxies maintained their classification after spatially resolved observations. This result implies that  galaxies with highly star-forming cores but passive outskirts or galaxies with passive cores but still star forming in the outskirts are overall quite rare \citep{Tuttle2020}.   

The analysis of the scaling relations does not give definitive results, although some general trends are observed. In general,
different physical processes considered in this work produce similar signatures on  global galaxy properties, and integrated values can not firmly distinguish among the different processes.

To conclude, spatially resolved data are very powerful to robustly identify the different mechanisms, when used in combination with accurate definition of the environment. The latter is very important and in some of the  uncertain cases a better characterization of the surrounding environment would help. 

Secondly, some uncertain cases could also be solved using observations at other wavelengths.  This paper indeed focused only on ionised gas, while the signature of environmental processes are visible also for other gas phases (molecular gas, atomic gas). 
For example, the presence of tails, streams or bridges between galaxies of atomic gas could help to better study the cases of ram pressure stripping, gas accretion and interactions. As the cold gas discs are typically more extended than ionised gas discs, it is possible that in some cases  stripping of the cold gas occurs while the  ionised gas is still undisturbed.

Similarly, also theoretical predictions on the spatially resolved properties of galaxies in the different environments and identifying the different mechanisms would be very important to support our findings. 

\acknowledgments
We thank the anonymous referee whose comments helped us to improve the manuscript.
We thank Stephanie Tonnesen for useful discussions and for providing comments on the manuscript. Based on observations collected at the European Organization for Astronomical Research in the Southern Hemisphere under ESO programme 196.B-0578. This project has received funding from the European Research Council (ERC) under the Horizon 2020 research and innovation programme (grant agreement N. 833824).  We acknowledge financial contribution from the contract ASI-INAF n.2017-14-H.0, from the grant PRIN MIUR 2017 n.20173ML3WW\_001 (PI Cimatti) and from the INAF main-stream funding programme (PI Vulcani).  Y.~J. acknowledges support from CONICYT PAI (Concurso Nacional de Inserci\'on en la Academia 2017) No. 79170132 and FONDECYT Iniciaci\'on 2018 No. 11180558. 
J.F. acknowledges financial support from the UNAM-DGAPA-PAPIIT IN111620 grant, M\'exico.

\bibliography{references}{}
\bibliographystyle{aasjournal}



\appendix

\section{All galaxy properties}\label{sec:all_fig}
{  In this section we show all the maps we have at our disposal for the galaxies discussed in the main text. Maps surrounded by a thicker contours are the ones discussed in detail in the text.}

\begin{figure*}
\centering
\includegraphics[scale=0.6]{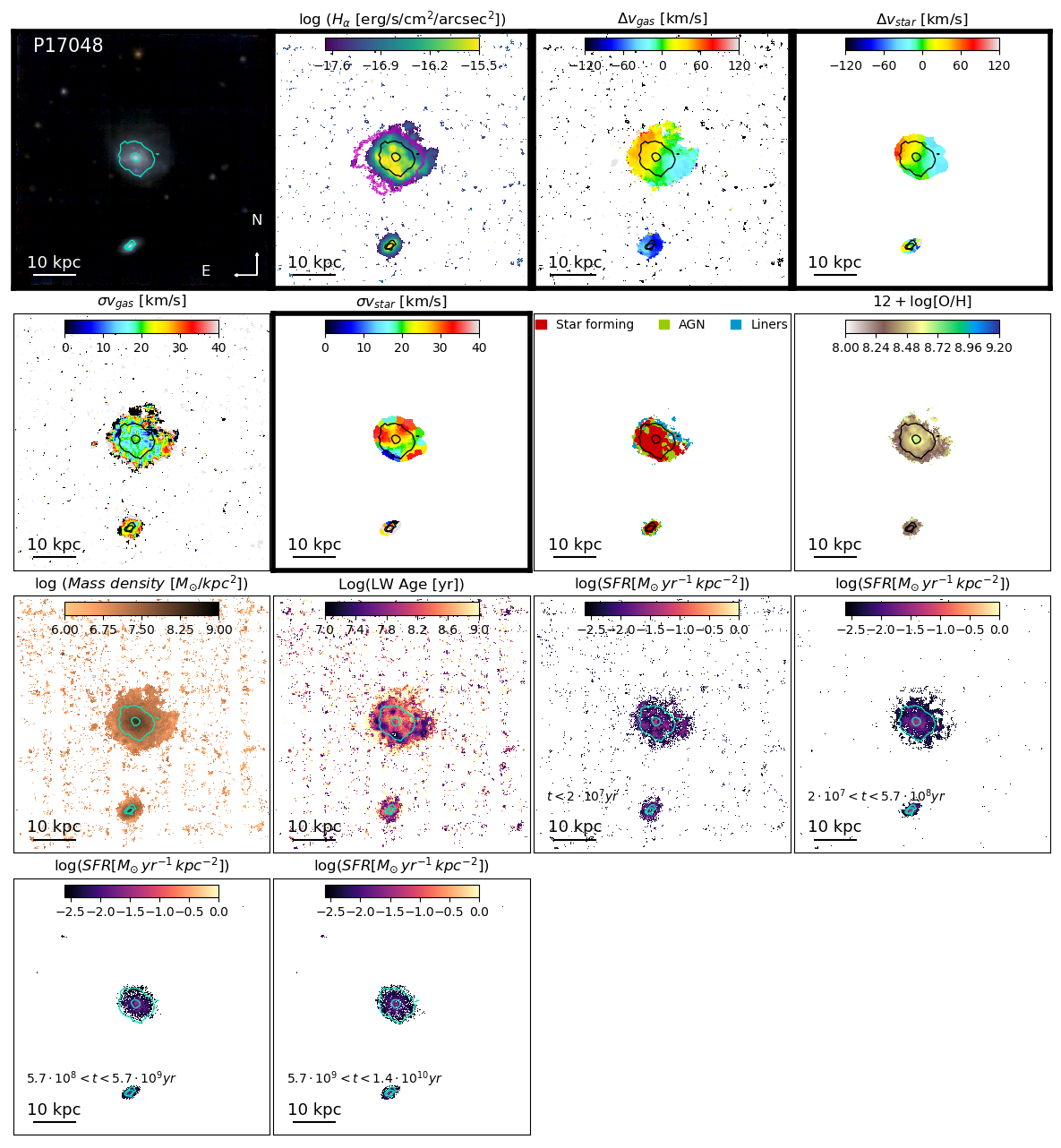}
\caption{P17048. From left to right: Color composite image obtained by combining the reconstructed g, r, and i filters from the MUSE data cube (rgb), \Ha flux map, \Ha and stellar kinematics maps, \Ha and stellar velocity dispersion map, bpt, ionized metallicity, mass density, luminosity weighted age map and SFR map in four age bins. The kpc scale is also shown in each panel. North is up, and east is left. Cyan or black contours represent the distribution of the oldest stellar population (from \sinopsis). Maps surrounded by a thicker contours are the ones discussed in detail in the text. \label{fig:P17048_bis} }
\end{figure*}

\begin{figure*}
\centering
\includegraphics[scale=0.6]{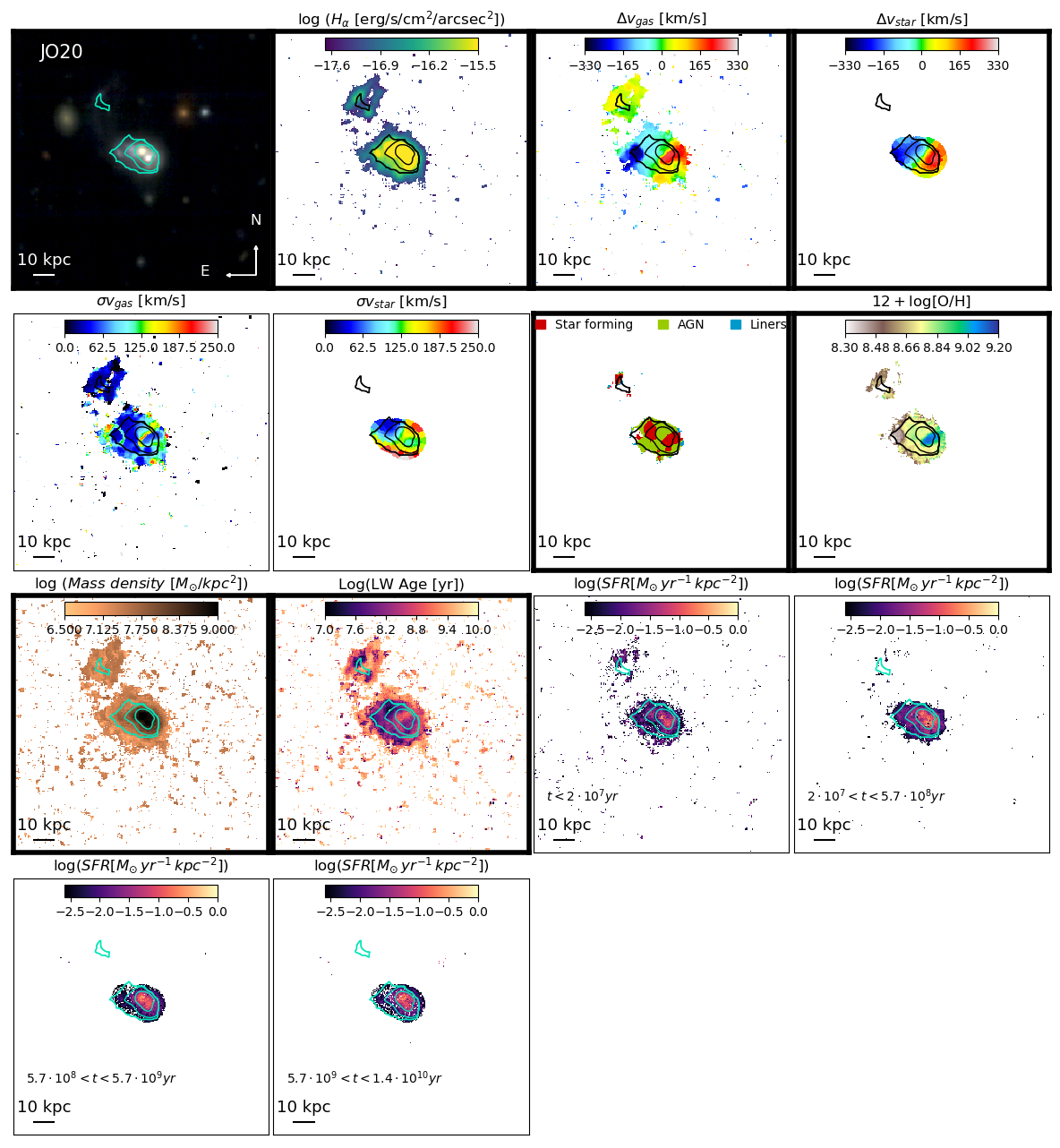}
\caption{JO20.  Panels, lines and colors are as in Fig. \ref{fig:P17048_bis}. \label{fig:JO20_bis} }
\end{figure*}

\begin{figure*}
\centering
\includegraphics[scale=0.6]{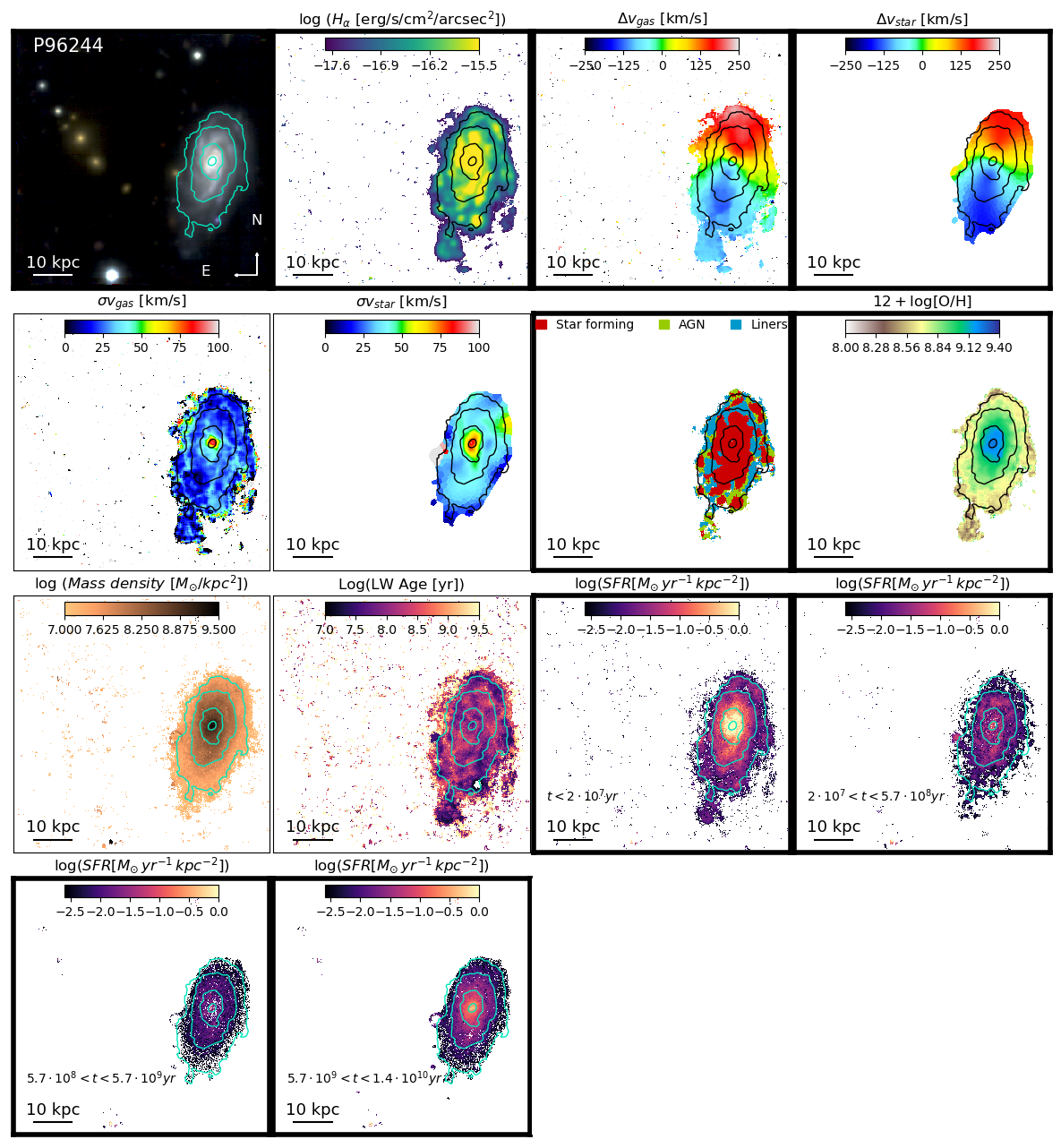}
\caption{P96244.  Panels, lines and colors are as in Fig. \ref{fig:P17048_bis}. \label{fig:P96244_bis} }
\end{figure*}

\begin{figure*}
\centering
\includegraphics[scale=0.6]{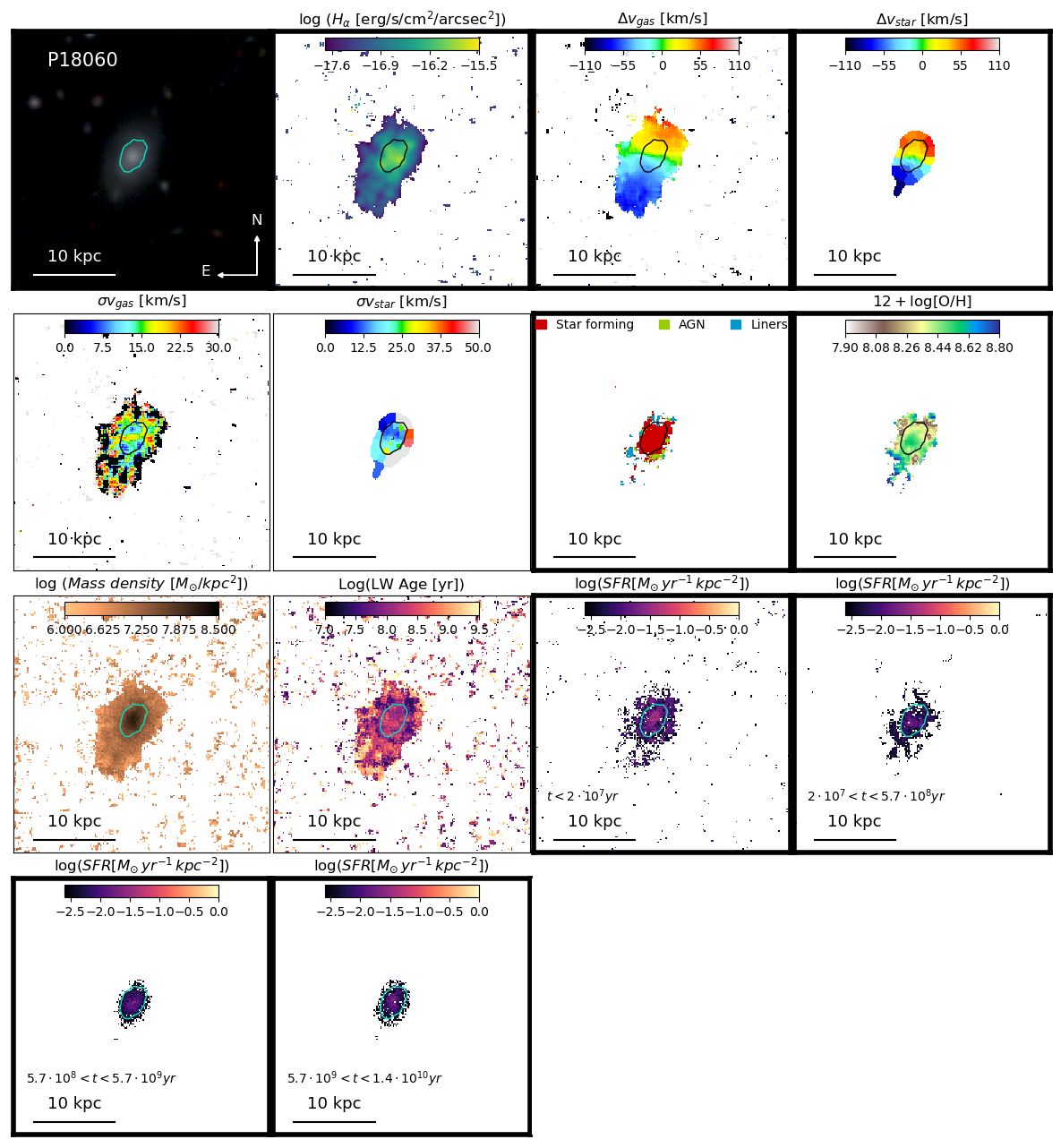}
\caption{P18060.  Panels, lines and colors are as in Fig. \ref{fig:P17048_bis}. \label{fig:P18060_bis} }
\end{figure*}

\begin{figure*}
\centering
\includegraphics[scale=0.6]{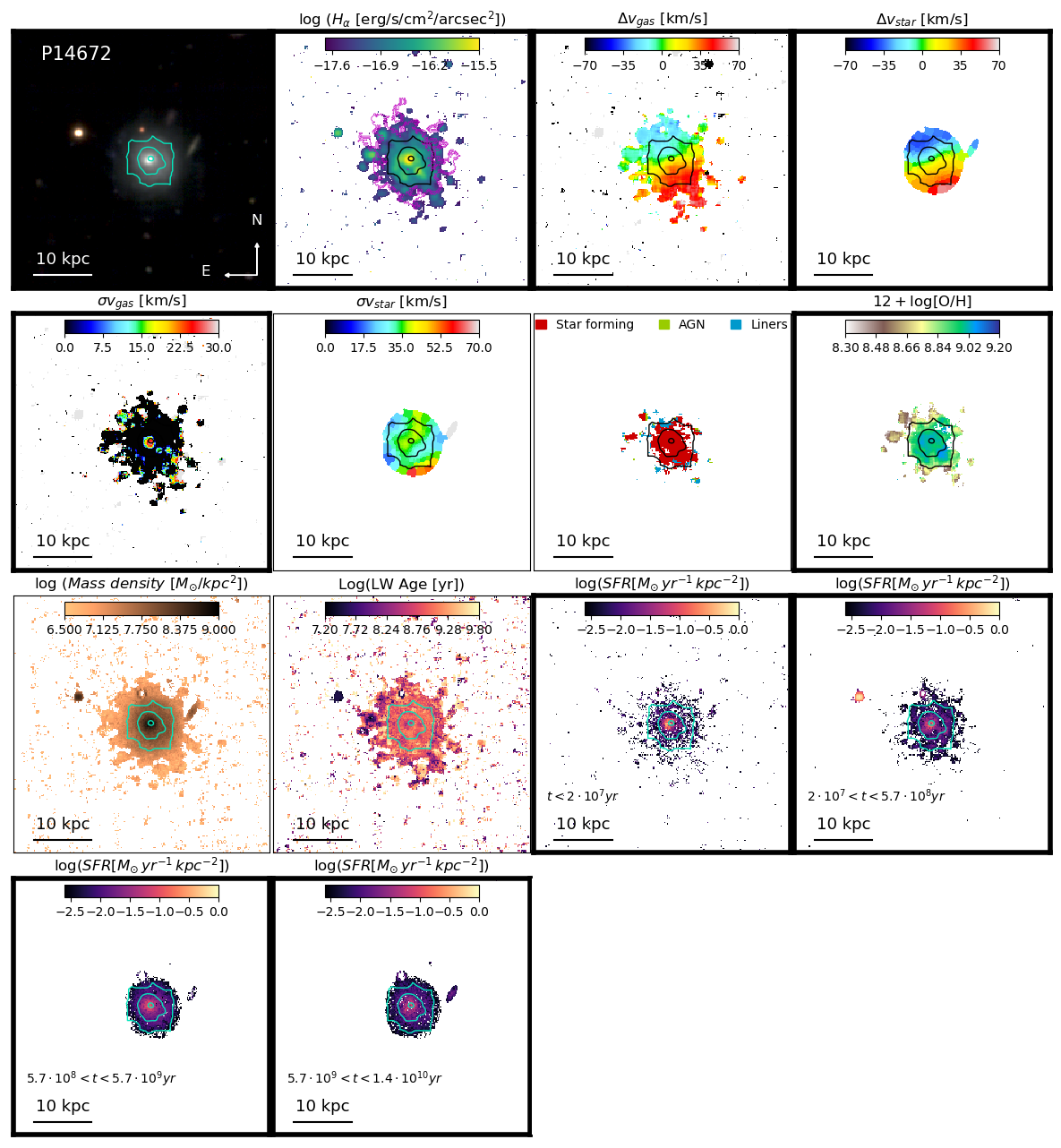}
\caption{P14672.  Panels, lines and colors are as in Fig. \ref{fig:P17048_bis}. \label{fig:P14672_bis} }
\end{figure*}

\begin{figure*}
\centering
\includegraphics[scale=0.6]{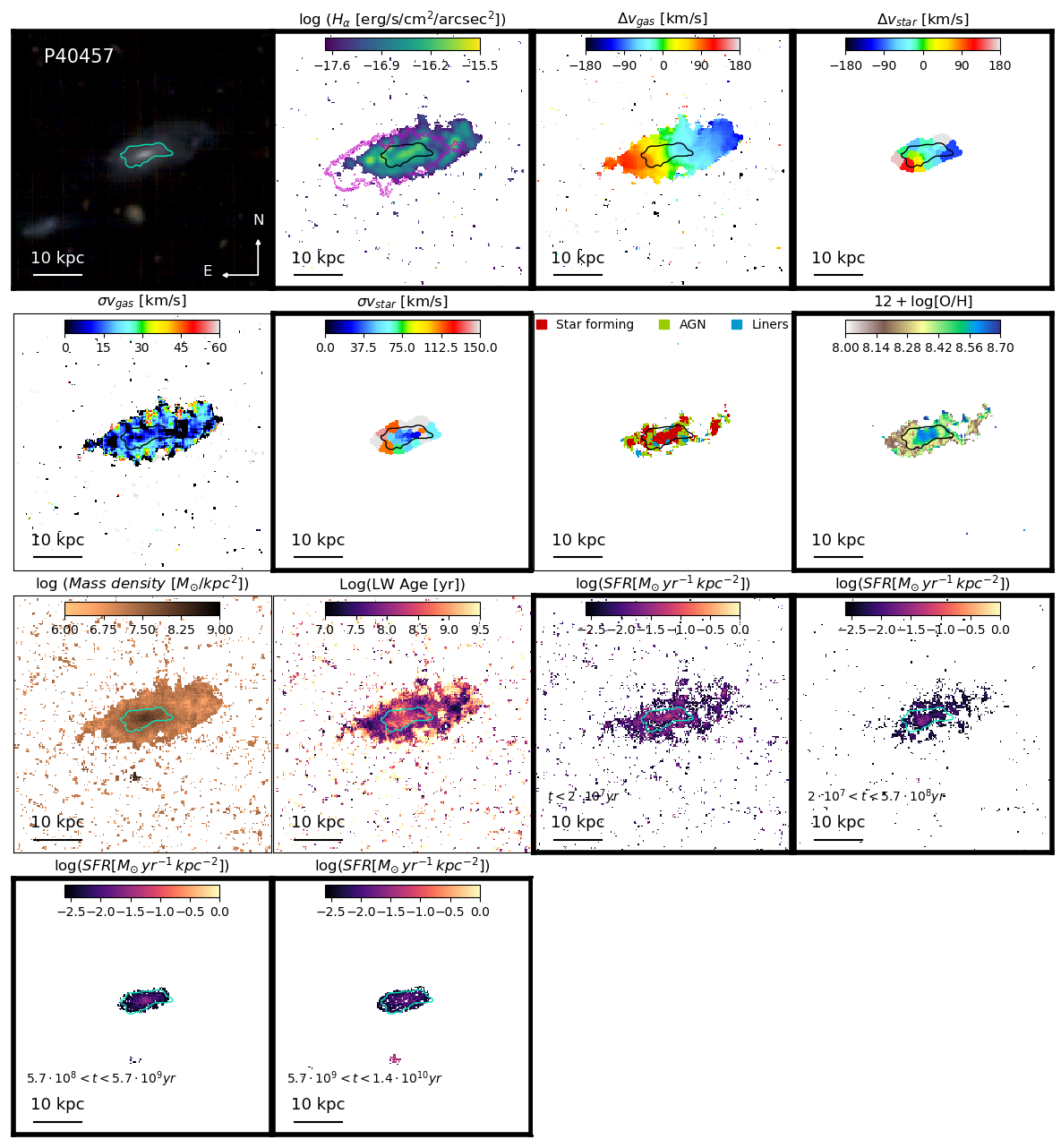}
\caption{P40457.  Panels, lines and colors are as in Fig. \ref{fig:P17048_bis}. \label{fig:P40457_bis} }
\end{figure*}

\begin{figure*}
\centering
\includegraphics[scale=0.6]{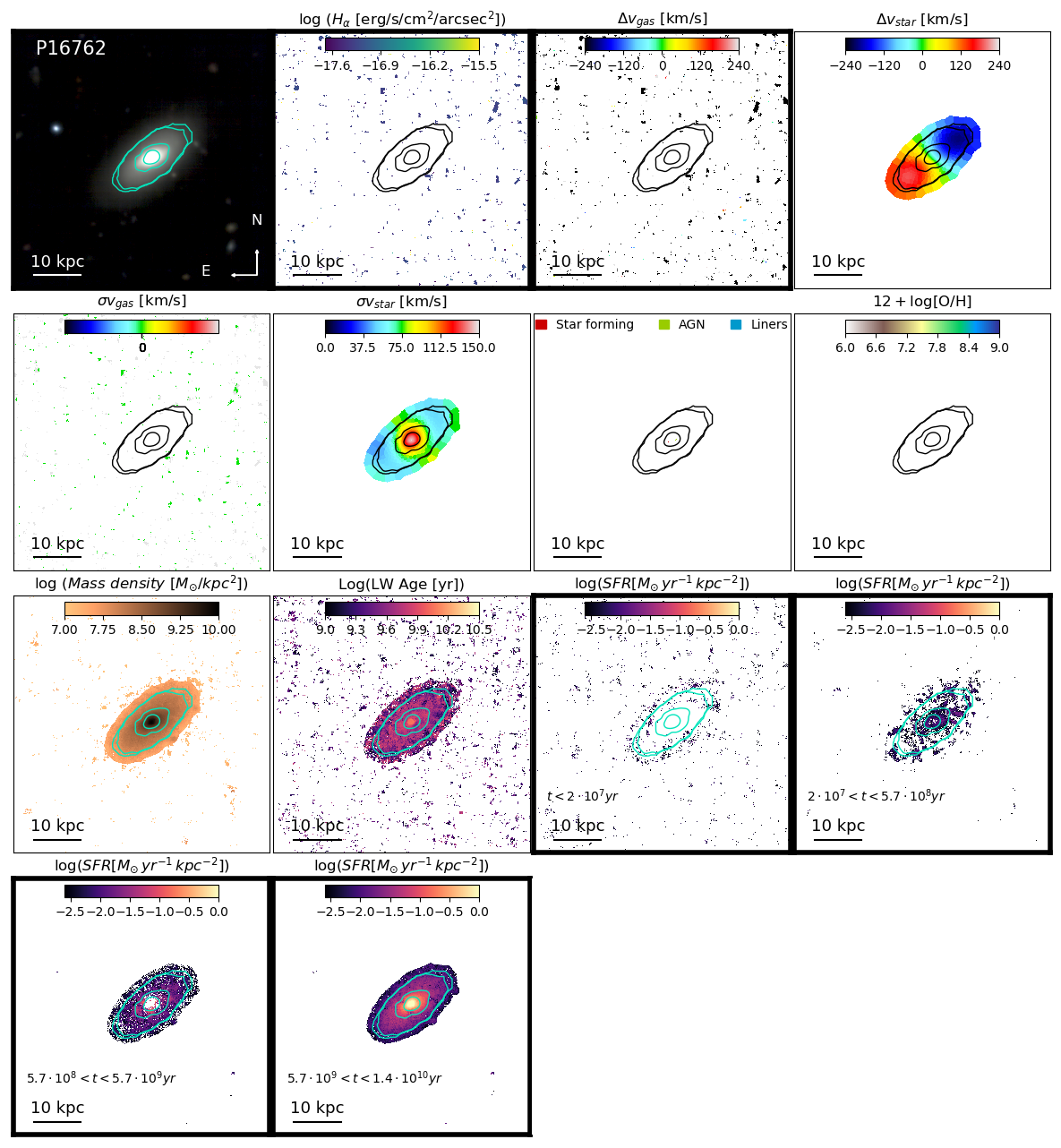}
\caption{P16762.  Panels, lines and colors are as in Fig. \ref{fig:P17048_bis}. \label{fig:P16762_bis} }
\end{figure*}

\begin{figure*}
\centering
\includegraphics[scale=0.6]{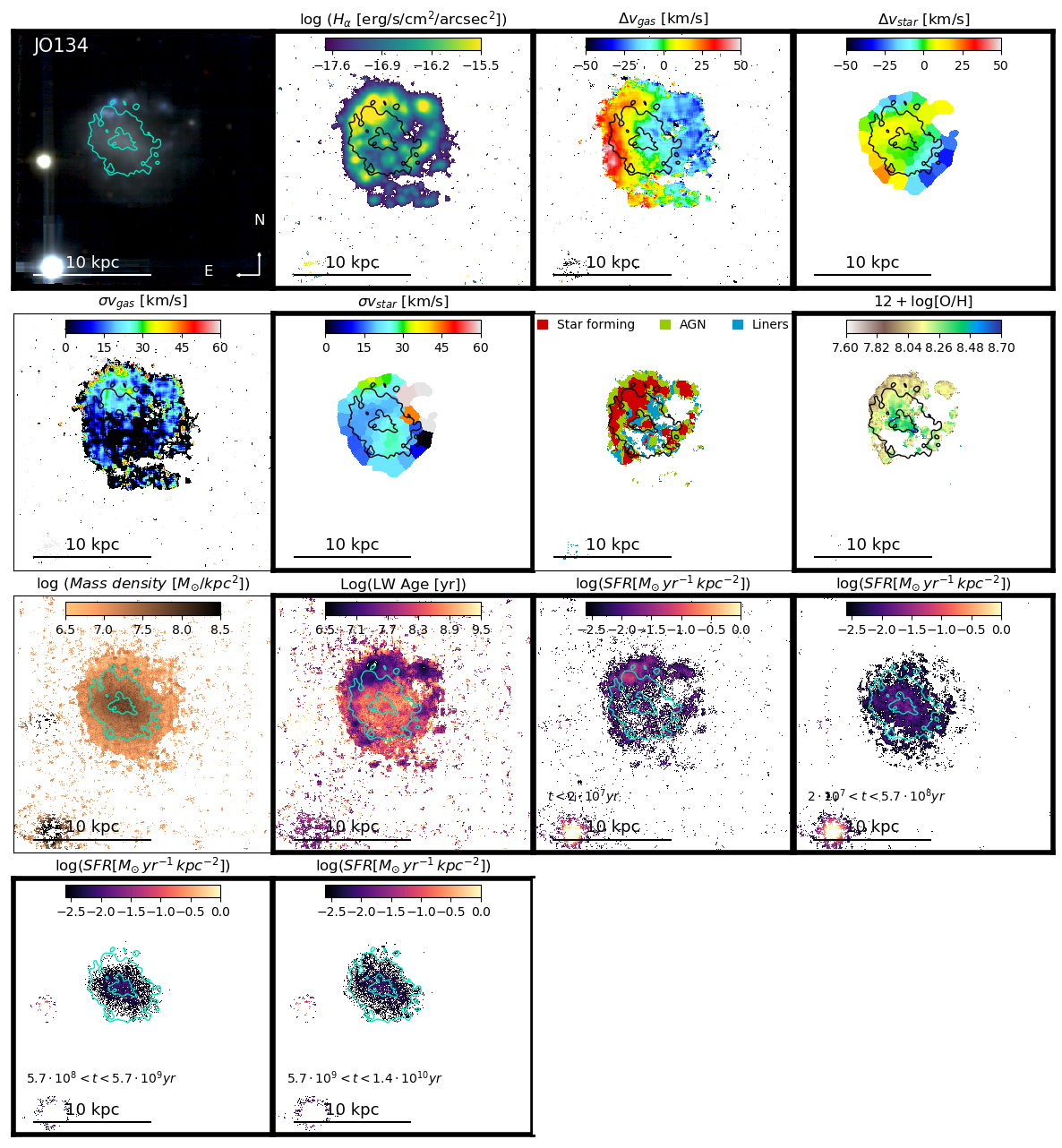}
\caption{JO134.  Panels, lines and colors are as in Fig. \ref{fig:P17048_bis}. \label{fig:JO134_bis} }
\end{figure*}

\section{All galaxies}\label{sec:all}
In this section we present an overview of all the other galaxies in the sample, grouping them by the most probable physical process at play. {  We will discuss in detail  only the maps that are relevant to determine the ongoing physical process (marked by thicker contours in Figures 28-35), but for the sake of completeness we show all the maps.}

\subsection{Galaxy-Galaxy Interactions} \label{sec:interactions_a}

\begin{figure*}
\centering
\includegraphics[scale=0.6]{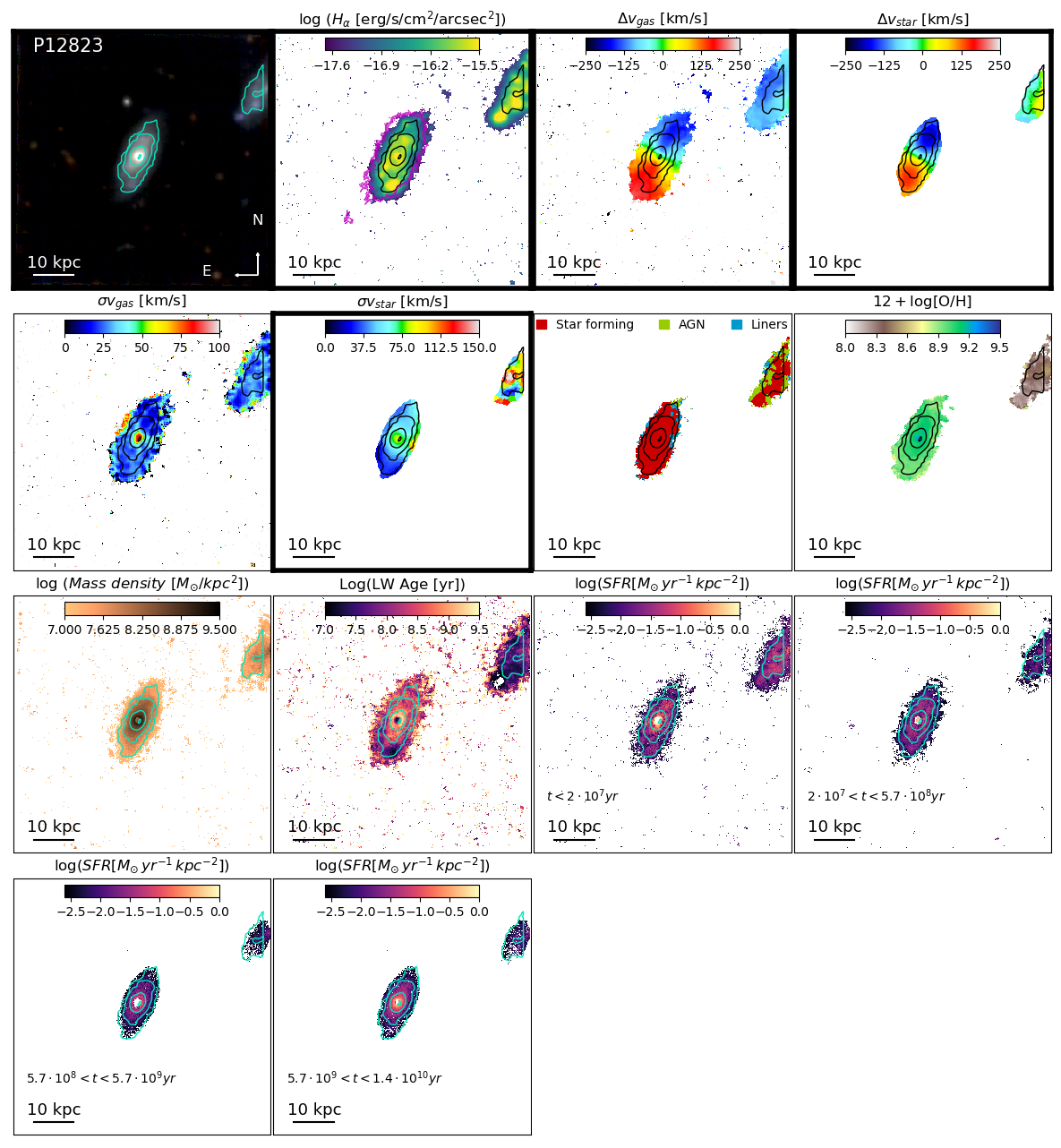}
\caption{P12823: interacting candidate. Panels, lines and colors are as in Fig. \ref{fig:P17048_bis}. \label{fig:interacting_a} }
\end{figure*}

\subsubsection{P12823}
Figure \ref{fig:interacting_a} shows that P12823 is a spiral galaxy with quite high inclination ($i=65^o$). {  Table \ref{tab:interacting} summarizes the main features investigate to characterize the system.} The galaxy is part of binary system \citep{Tempel2014_g} and its companion is visible in the rgb, 30\arcsec\ away from P12823. This galaxy \citep[P12813,][]{Calvi2011TheGalaxies} is at z= 0.05023 and no stellar mass estimate is available from the literature; we therefore can not compute the strength of the tidal interaction. The rgb image highlights some distortions in the external parts of the disk. Such distortions are more evident from the \Ha flux map, which shows that the portion of the disk facing the companion is shredded. { However, the galaxy is only marginally lopsided (the level of overlap is 80$\%$)}. Some \Ha clouds are also visible in between the two objects, suggesting that stripped gas may have triggered the formation of new stars. 
Both the gas and the stellar kinematics are overall regular ({  $A_v$=0.15 and $A_g$=0.17)}, showing significant distortions only on the side closer to the companion, where the stellar velocity dispersion  is also much higher.  This might suggest that the galaxies are approaching their first pericenter passage. Velocities span the range -200$<v [km/s]<$ 200.  It worth mentioning though that the stars and gas have different extents: taking the  contour of the original body as reference, it is evident that the gas disk is more extended, while the stellar disk did not grow much in the last few Gyr. Note that P12823 has also a very  high central gas velocity dispersion, with $\sigma (v_{gas})\sim 80$km/s. The { high value extends { beyond} the region dominated by beam smearing.} This measurement is an indication of a very hot center that can result  when a bar dissolves \citep{Friedli1993, Hasan1993} or weakens \citep{Athanassoula2005}.


\subsection{Mergers}
\label{sec:Mergers_a}
\begin{figure*}
\centering
\includegraphics[scale=0.6]{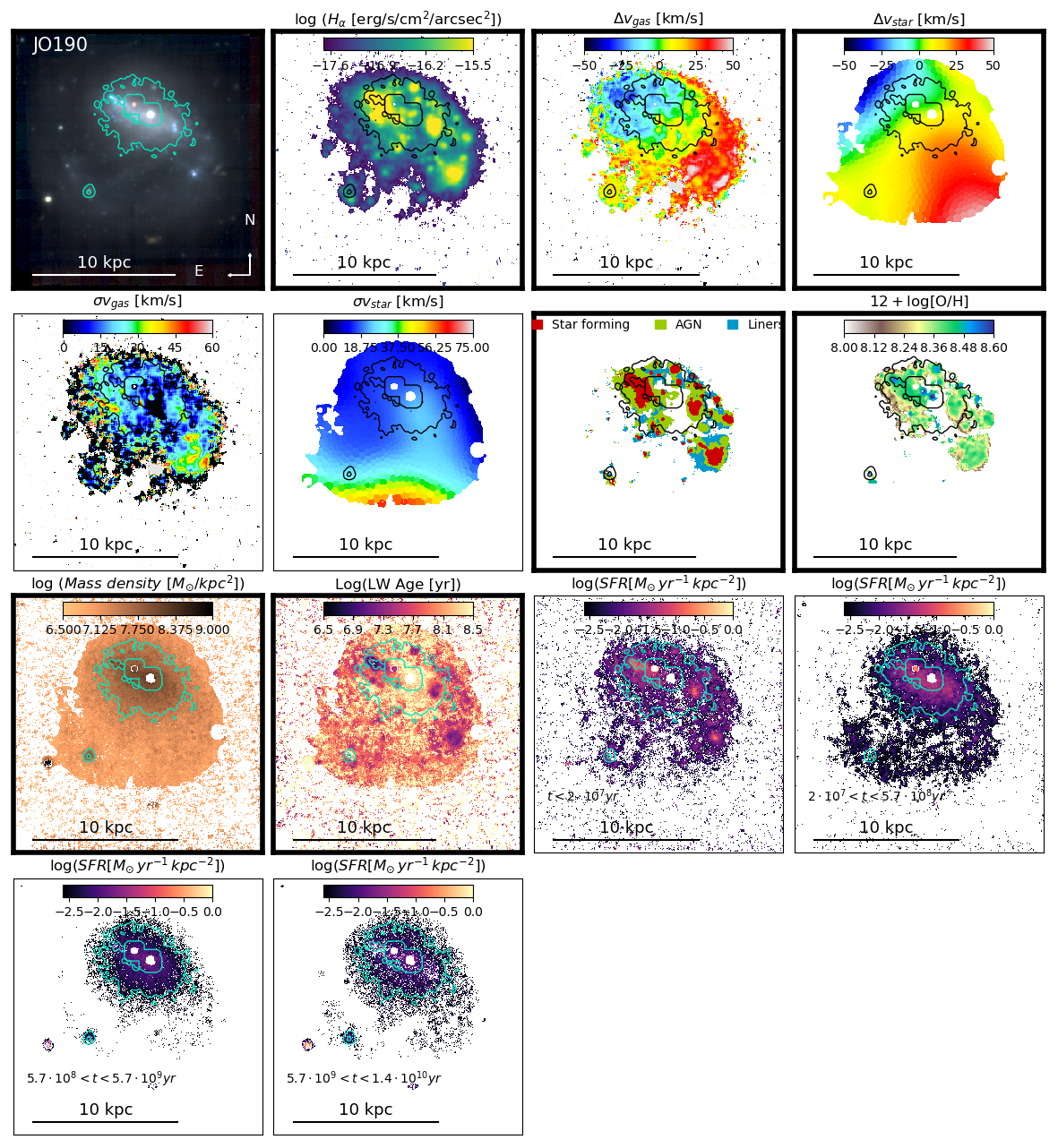}
\caption{JO190: merger candidates.  Panels, lines and colors are as in Fig. \ref{fig:P17048_bis}. \label{fig:merger_JO190} }
\end{figure*}

\begin{figure*}
\centering
\includegraphics[scale=0.6]{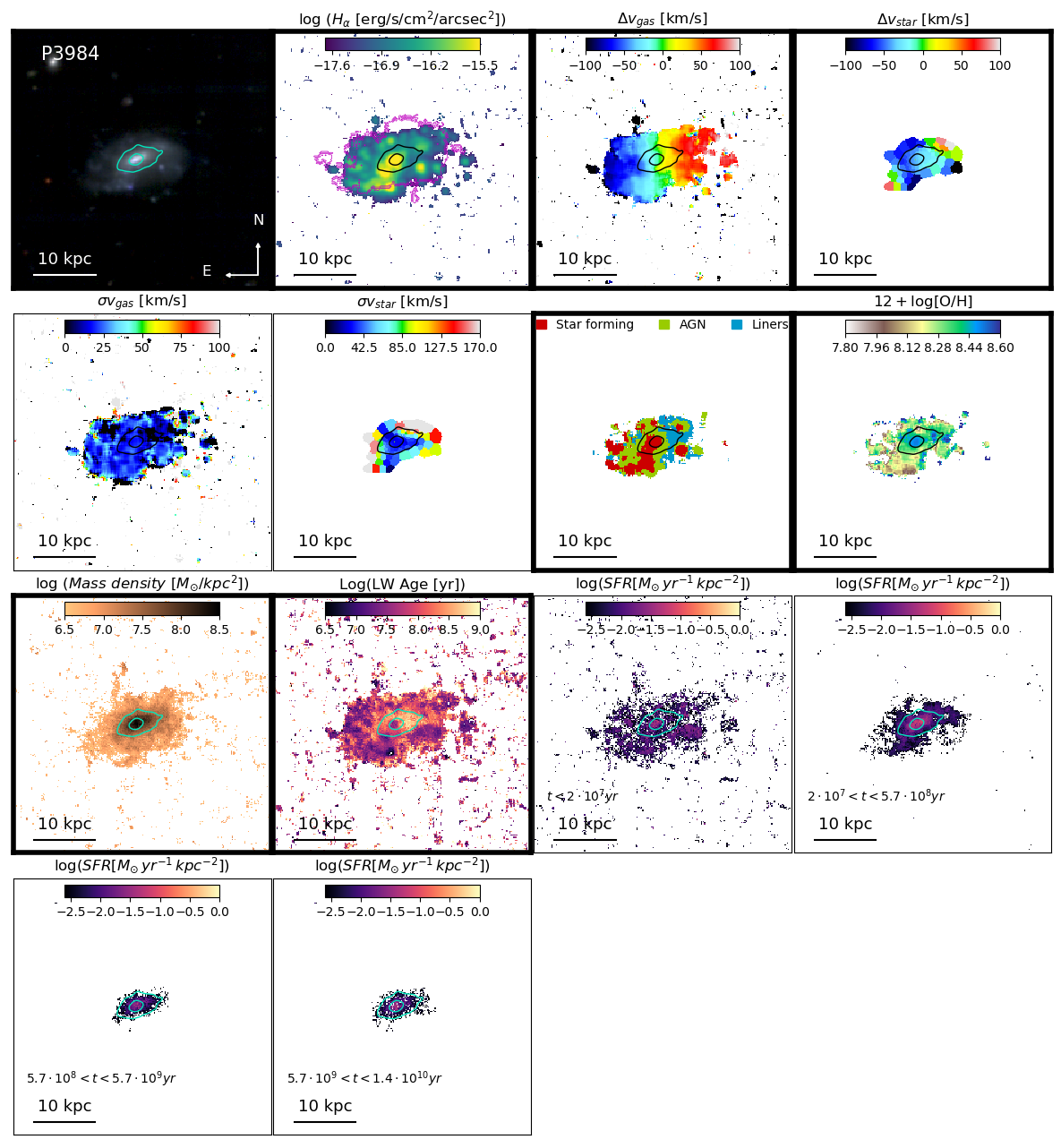}
\caption{P3984: merger candidates.  Panels, lines and colors are as in Fig. \ref{fig:P17048_bis}. \label{fig:merger_P3984} }
\end{figure*}

\subsubsection{JO190}

JO190 is the closest galaxy in the sample, at z= 0.013218. Prior to MUSE observations, no redshift was available. The galaxy is part of an overdensity: within 1 degree from JO190, about 40 galaxies have similar redshift (0.013$<z<$0.015). Nonetheless, this structure has no characterization. The closest galaxy at similar redshift (ESO 467-63) is at 572.83\arcsec. 

The galaxy has been included in the merger category even though it does present a rather regular stellar kinematics { ($A_v = 0.3$)}. However, as shown in what follows, all the other pieces of evidence point to a merging event. The regular kinematics can be an indication that the merger occurred more than 0.5 Gyr ago \citep{Hung2016}.

The rgb image in Fig. \ref{fig:merger_JO190} shows the peculiar morphology of JO190: the main body of the galaxy is in the northern part and the galaxy is quite edge on. Then a long extraplanar tail extends towards the South, with a shell shape typical of tidal interactions \citep[e.g.,][]{Peirani2010}.
Many bright blue knots are visible in the image: these might correspond to star-forming regions that could eventually end up as tidal dwarf galaxies \citep[as in][]{Vulcani2017c}.  Note however that the two bright knots (one redder) on the galaxy disk are background objects and will be masked in the forthcoming analysis (producing the empty circles visible in the maps).  A redder clump in the tail is particularly outstanding and it could be the remnant of a minor merger, being at the same redshift of JO190. The same clump is also quite old (LWA $\sim 10^{8.5}$ yr), and was already in place in the oldest age bin (plot not shown). The \Ha flux map shows that the ionized gas has a peculiar morphology as well: the northern part is quite well defined, while the southern one is tattered. Many bright regions, some of them also very extended, are visible. The BPT map highlights how this gas is not ionized by star formation, but shocks are present throughout the map. The gas and star velocity fields highlight that both components have rather small rotation (-50$<v [km/s]<$50), tilted with respect 
to the major axis of the galaxy as seen from the rgb image. The gas velocity dispersion is overall quite low ($\sigma v_{gas}<35 km/s$), indicating the medium is rather cold, while the stellar velocity dispersion map shows an increase of the velocity dispersion from the center towards roughly the location of the merger remnants. \cite{Nipoti2020} discuss how off axis mergers can produce increases in the velocity dispersions from the central regions of the main progenitor towards the outskirts \citep[see also][]{Bendo2000}.  The gas metallicity map has no clear gradient, indicating that there has been a mixture of the gas during the merger event \citep{Kobayashi2004}.  In the mass density map the merger remnant is clearly visible as a massive clump in the South-East part of the galaxy. Finally, the LWA map is also rather flat, suggesting that  stars have been redistributed. \cite{Peirani2010} have indeed showed that the  shell structure within the host galaxy progressively produced by the satellite remnant is composed of old stars at large radii because of the stellar accretion and ensuing relaxation occurred via dissipationless mechanisms.
However, some younger regions are present, corresponding to the bright \Ha regions and the bright clumps observed in the rgb. These regions have also higher metallicity.

To conclude, JO190 is most likely an example of an old minor merger: both the stars and gas have had the time to settle down and again acquire a rather regular rotation.

\subsubsection{P3984}
P3984 is most likely a case of post gas rich dwarf–dwarf merger. The rgb image shown in Fig.\ref{fig:merger_P3984}  highlights the presence of a bent shape, similar to a stream, with a very blue compact core. The latter  is usually used as evidence for the formation via dwarf–dwarf merging \citep{Bekki2008}. The larger extent of the current galaxy compared to the original is clearly visible, indication that some event puffed up the galaxy disk. The \Ha flux map shows a tattered distribution, especially in the western part of the galaxy, where a sort of tidal is evident. { The overlap between the original and rotated image is of 75\%}. Detached clouds are also visible. The gas velocity field indicates that these clouds do belong to the galaxy, as they have velocities similar to that of the galaxy. The gas velocity field (ranging from -100 to 100 km/s) is quite regular { ($A_g$=0.19)}, suggesting that the gas  had the time to be redistributed after the merger. It has indeed been demonstrated in both simulations and observations that gaseous disks are able to survive during the interaction between gas-rich systems or reform through accreting gas after two nuclei merge \citep[e.g.,][]{Downes1998,  Barnes2002,  Springel2005, Hopkins2009, Ueda2014, Hung2016}. 
The gas velocity dispersion is overall very low ($\sigma v_{gas}\sim 30$ km/s). In contrast, the stellar kinematics is chaotic and shows little rotation { ($A_s$=0.39)}. The stellar velocity dispersion is much higher in the outskirts compared to the galaxy core and chaotic, emphasizing that some event altered the stellar orbits. { The median stellar velocity dispersion is 80 km/s. For reference, control sample galaxies have a median value of 37$\pm$7 km/s.} The low gas and stellar velocity dispersion in the core suggests again that the merger occurred long ago and the nuclei have had time to reach coalescence \citep{Stickley2014}. The different kinematics between stars and gas is another indication that the galaxy is the remnant of a dwarf-dwarf merger \citep{BekkiChiba2008}. 

The BPT map highlights that the  star formation is the main mechanism ionizing the gas only in the eastern part of the galaxy, while towards West other mechanisms become dominant. The metallicity map is irregular: in the Eastern part it reaches $12 + \log(O/H)$ = 8.0, in the center and Western part it peaks at $12 + \log(O/H)$ = 8.6. We might be observing the two components of the merger. Most of the mass is confined within the original galaxy body. Finally, the luminosity weighted age map shows younger ages (LWA$<10^{7.5}$ yr) in the southern part of the galaxy and older ages close to the center.  Numerical simulations show that the remnants of mergers between two gas-rich dwarf irregulars with initially extended gas disks can have both extended spheroids composed of older stellar populations and disks composed mostly of gas and young stars \citep{Bekki2008}. 

As a final note, P3984 is isolated in all our environmental catalogs.

\begin{figure*}
\centering
\includegraphics[scale=0.6]{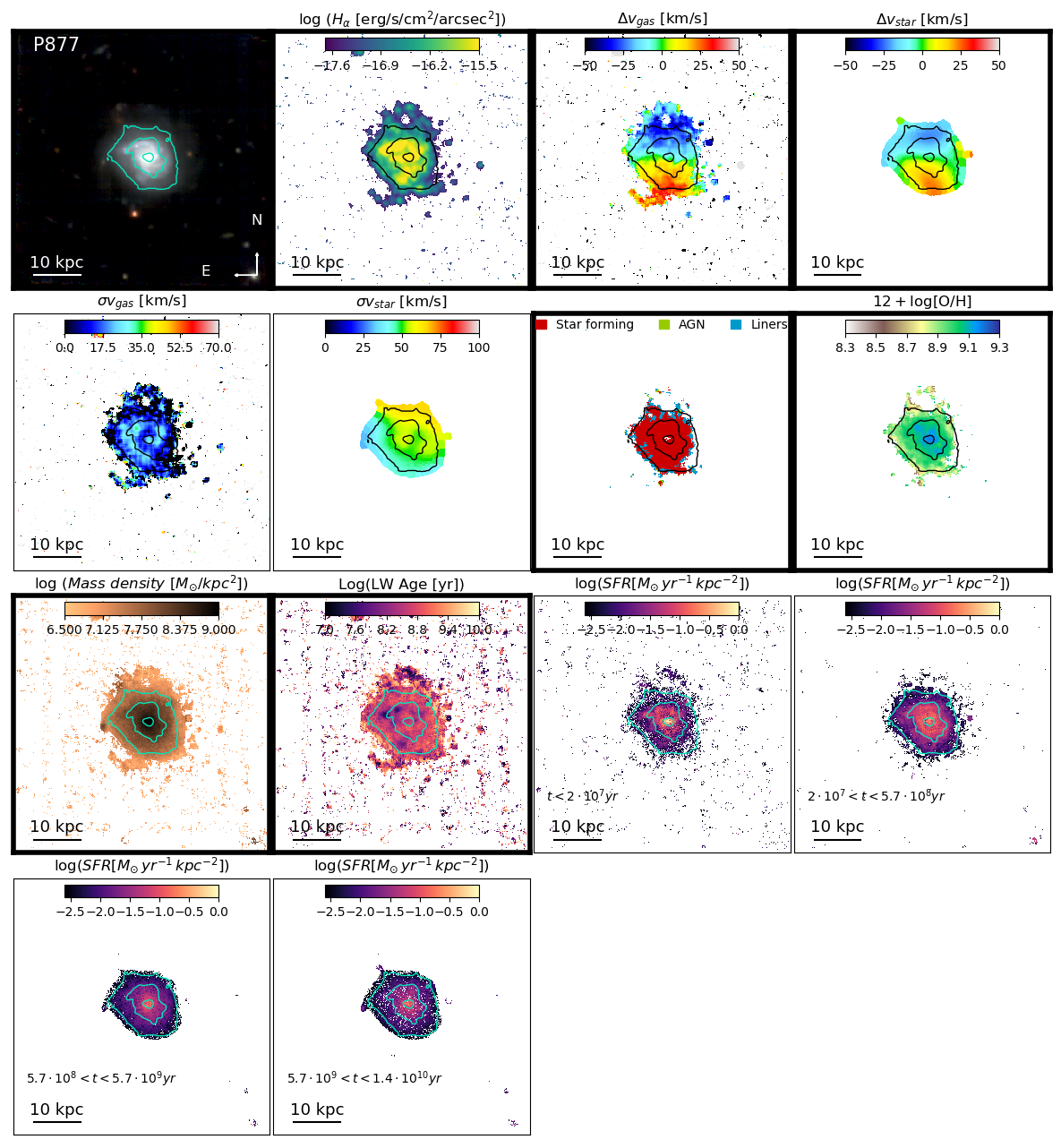}
\caption{P877: merger candidates.  Panels, lines and colors are as in Fig. \ref{fig:P17048_bis}. \label{fig:merger_P877} }
\end{figure*}

\begin{figure*}
\centering
\includegraphics[scale=0.6]{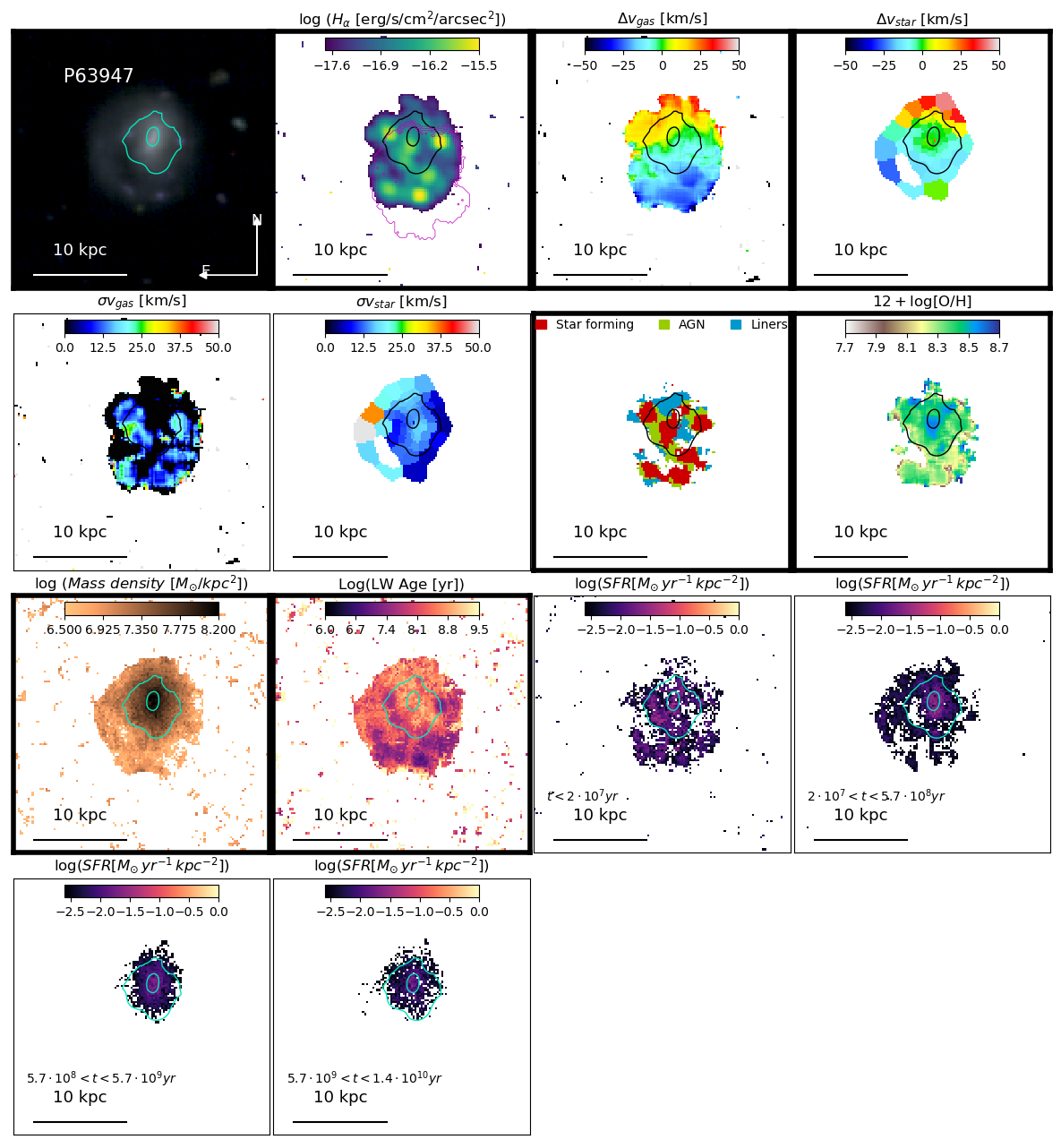}
\caption{P63947: merger candidates.  Panels, lines and colors are as in Fig. \ref{fig:P17048_bis}. \label{fig:merger_P63947} }
\end{figure*}

\subsubsection{P877}
Among the merging candidates, P877 is the least certain. The rgb image in Fig. \ref{fig:merger_P877} shows a rather regular morphology. A potential companion visible towards the North is actually a background source. Nonetheless, the \Ha flux map unveils a more complicated morphology. Many detached clouds are detected in the disk surroundings. These clouds are preferentially found in the South-West part of the galaxy, even though some clouds are detected also towards East. A horn extending from the southern part of the galaxy towards East is also detected. This protuberance develops towards the opposite direction of the spiral arms and it is not seen in the rgb image. Most of the ionized gas is powered by star formation. Shocks might be present only in the very outskirts of the galaxy. The ionized gas kinematics is { extremely asymmetric ($A_g$=0.98)} and has very little rotation ($|\Delta(v)|=25$km/s, { $A_v=0.31$}). The locus of zero velocity extends towards the South-West. The gas velocity dispersion is overall quite low ($\sigma v_{gas}\sim 15$km/s).  The stellar component also shows very little velocity, with the locus of zero velocity bent towards South, indicating the presence of a bar (O. Sanchez et al. in prep.) and/or that something has affected the stars. The stellar velocity dispersion map shows a a double-peaked distribution, which can be explained in terms of  edge-on gas rings due to 1:1 merger \citep{Jesseit2007} or due to edge-on barred disks \citep{Bureau2004, Chung2004}. The median stellar velocity dispersion is 52 km/s.
The metallicity map is quite regular, the gradient rather flat, being  $8.9<12 + \log(O/H)<9.1$ throughout the entire galaxy disk. A light drop in the metallicity is observed only in the very outskirts,  where $12 + \log(O/H)\sim 8.7$. Note however that, because of the low S/N, we have no metallicity measures in the galaxy outskirts.  The merger might have therefore redistributed the gas \citep{Kobayashi2004}. The mass density map shows a rather regular gradient, with most of the mass confined in the core. The LWA is also quite flat, suggesting again a redistribution of the stars. Only a few young regions are evident (LWA$<10^{7.5}$yr) and these might correspond to star-forming regions ignited by the merger. Overall, the galaxy has not increased much in size from the oldest age bin, even though the clouds and protuberances are clearly recently formed. 

P877 is located within a three member group (z= 0.04685, velocity dispersion = 124 km/s, virial radius =  0.2948 Mpc, \citealt{Tempel2014_g, Saulder2016, Calvi2011TheGalaxies}), in the surrounding of a 20 galaxy group, therefore the environment might also play a role in shaping its properties. However, the closest galaxy is MGC95889, located at 128\arcsec. Given its mass (10$^{9.5}$\ms), the \cite{Vollmer2005} formulation excludes that tidal interactions from the companion could affect the galaxy. 
To conclude, there is no clear smoking gun for mergers, even though it is the most probable process. 

\subsubsection{P63947}
P63947 is another minor merger candidate. The rgb image in Fig. \ref{fig:merger_P63947}  shows a strong { lopsidedness ($A_o=70\%$)} and a tidal tail encircling the body in the eastern side of the galaxy. This is a clear feature of interaction/merger \citep{Mihos1993}. A deep photometric analysis reveals the galaxy has a bar (O. Sanchez et al. in prep.).  The original body was much smaller, indicating that some events induced a growth of the disk. Few small objects are seen in the surroundings of P63947, but a careful inspection of their spectra revealed they are background objects. The asymmetry is even more visible in the \Ha flux map, which also reveals a very clumpy disk. The ionized gas presents very little and distorted rotation: the maximum relative velocity of the disk is of only 25 km/s and the locus of zero velocity is chaotic { ($A_g=0.61$)}. The velocity field of the stellar component shows the rotation is distorted along the tidal tail { ($A_v=0.45$)}, where the velocity dispersion is also higher. The BPT map unveils that different ionization mechanisms coexist and the gas is not only powered by star formation. P63947 has a quite flat metallicity distribution, indicating that gas of different metallicity might have been mixed during the merger and gradients have been destroyed \citep{Kobayashi2004}. {  Even within the narrow range, some asymmetries are though detected, with the northern side richer than the southern.}
The mass density map shows that the bulk of the stellar mass is located within the original body. Very little mass is formed in the southern part. This region is also the youngest, as highlighted in the LWA map and could be indeed the aftermath of the merger. Given that we have no clear evidence for the merged companion, we label this galaxy as a minor merger candidate. {  The absence of a merger remnants though does not challenge the classification.}

Regarding the environment, P63947 is relatively isolated in the universe, according to both \cite{Tempel2014_g} and \cite{Saulder2016}. The closest  group is at 2169 kpc, its redshift is 0.05612, its velocity dispersion  $\sigma= $246.4.6 km/s. 
The closest galaxy is P64038 (z= 0.06417) at 5.42\arcmin, too far away to advocate tidal interactions. 

\subsection{Ram pressure stripping} \label{sec:RPS_a}

\subsubsection{P20159}
\begin{figure*}
\centering
\includegraphics[scale=0.6]{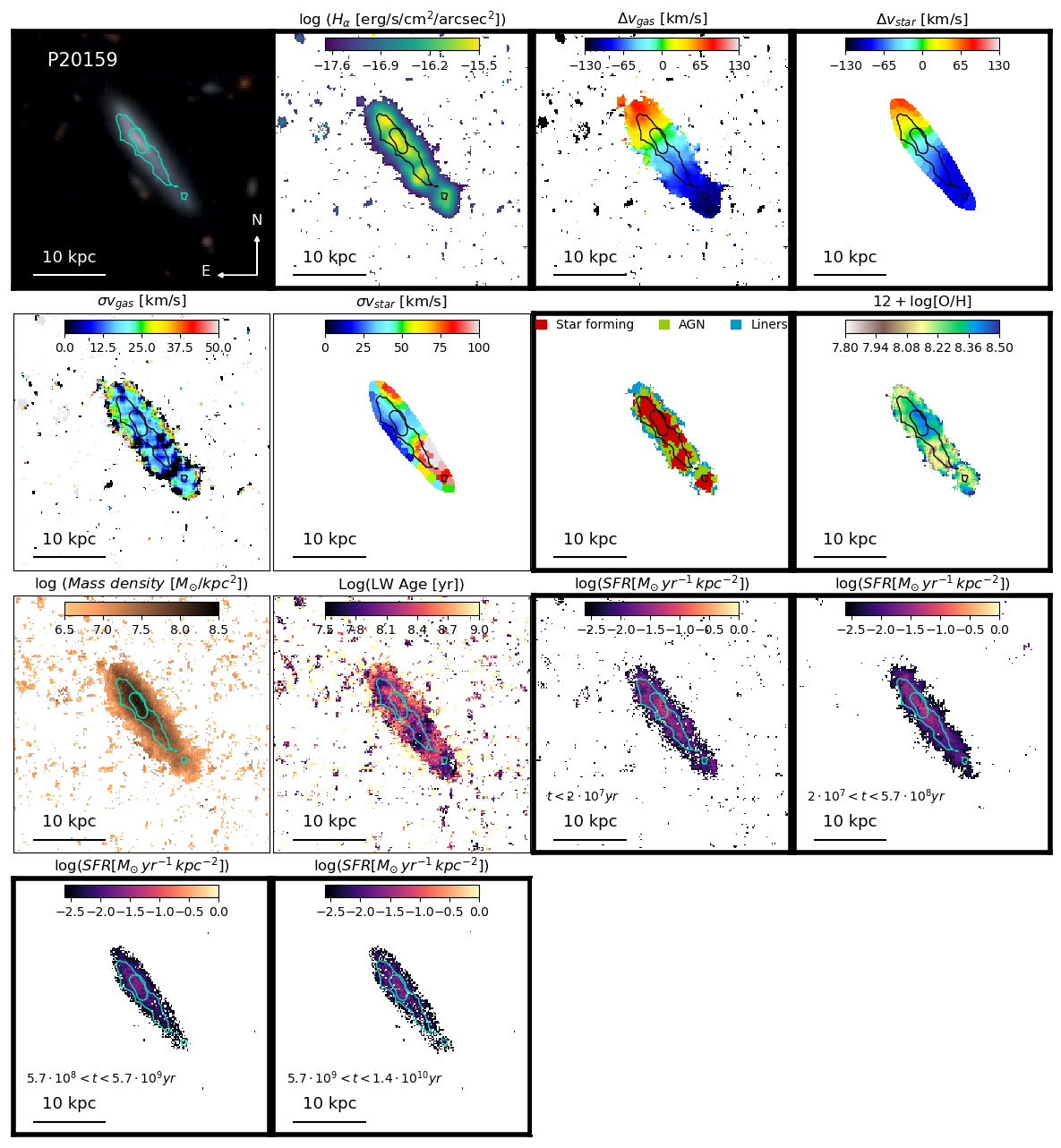}
\caption{P20159: ram pressure stripping candidate. Panels, lines and colors are as in Fig.\ref{fig:P17048_bis}.\label{fig:rps_P20159} }
\end{figure*}

\begin{figure*}
\centering
\includegraphics[scale=0.3]{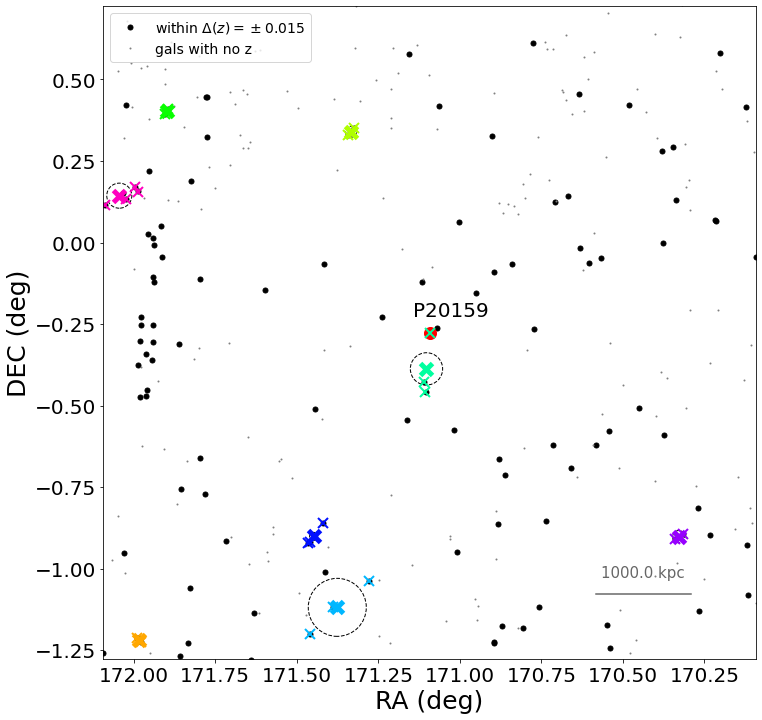}
\caption{Spatial distribution of galaxies around P20159.  Colors  and symbols have the same meaning as in Fig. \ref{fig:rps_env}. \label{fig:rps_env_a} }
\end{figure*}

\begin{figure*}
\centering
\includegraphics[scale=0.6]{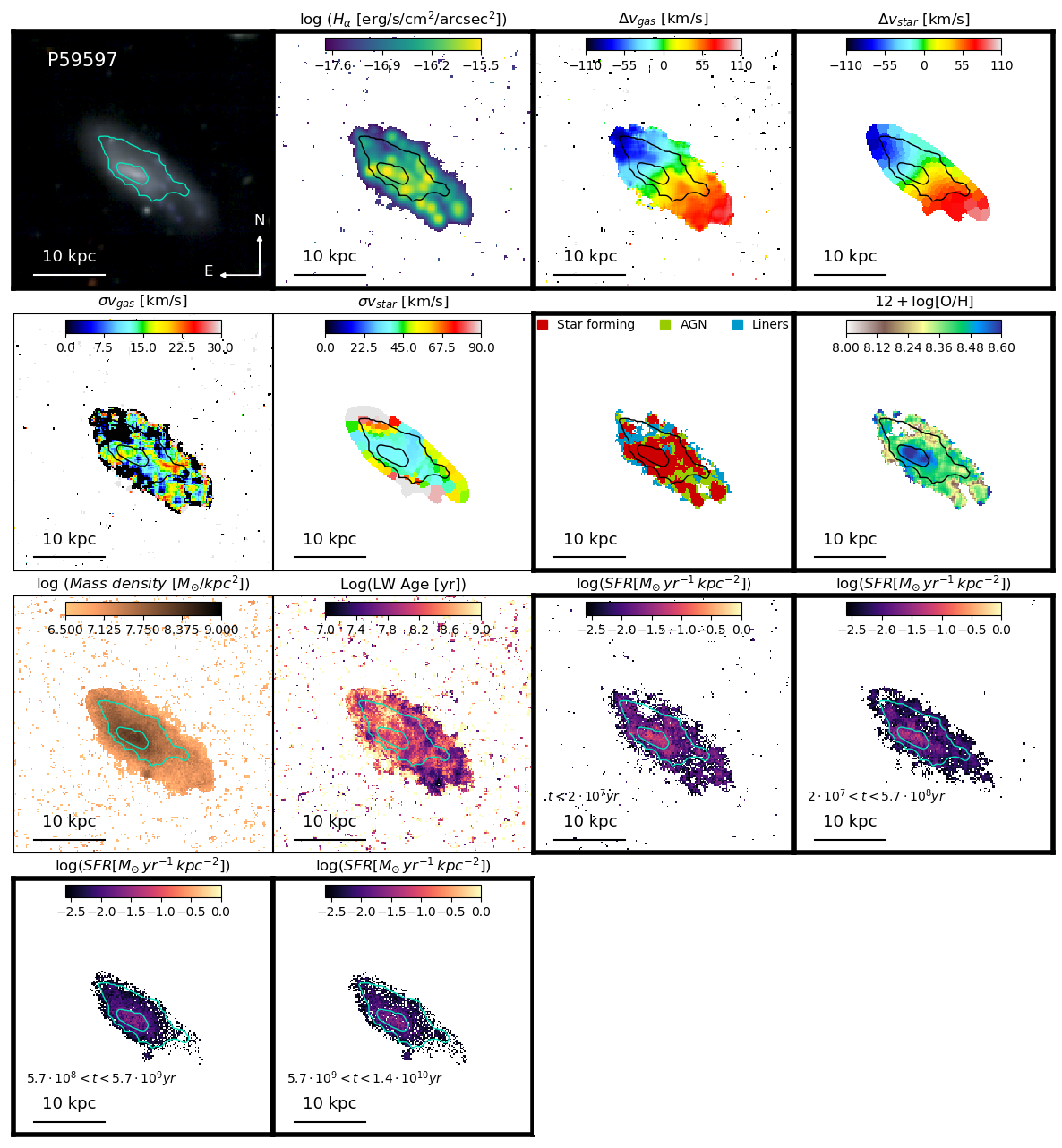}
\caption{P59597: ram pressure stripping candidate. Panels, lines and colors are as in Fig.\ref{fig:P17048_bis}.\label{fig:rps_P59597} }
\end{figure*}

P20159 is a $\log(M_\ast/M_\sun)$=9.6 galaxy found in a small 3 member group \citep{Tempel2014_g, Saulder2016}.  Its environment is shown in Fig. \ref{fig:rps_env_a}. The group center is at z = 0.04993, its measured velocity dispersion is $\sigma =$85.8 km/s, the virial radius is $R_{vir}$=0.1705 Mpc. Taking all these values with caution, we can estimate that P20159 is at 2.2$R_{vir}$, and has a velocity difference from the group center of 318 km/s, corresponding to 3.7$\sigma$. {  If indeed the galaxy is part of the group, it is though found quite far from its center, both in terms of spatial location and relative velocity.}
The closest galaxy to P20159 is P20139, found at 97.63\arcsec. P20139 is at z=0.0488 and has a mass of $\log(M_\ast/M_\sun)$= 8.90. It is not included in the \cite{Poggianti2016JELLYFISHREDSHIFT} catalog of stripping candidates. Given the distance between two galaxies, the approach developed by \cite{Vollmer2005} excludes that tidal interactions could occur.

The rgb image of P20159 in Fig.\ref{fig:rps_P20159}  highlights the high inclination of the galaxy ($i=80^o$) and shows the presence of an elongated tail extending towards South-West.
Focusing on the contours of the original body, the asymmetry of the tail is clearly evident. A number of small objects are visible around the galaxy, but a careful inspection of their spectra revealed that they are all foreground/background systems. The tail is also clearly visible in \Ha and the BPT map suggests that in that region star formation is not the dominant ionizing mechanism. Gas and stars rotate coherently, { with the gas showing a higher asymmetry than the stars ($A_g =0.41 $, $A_v=0.18$)}. Stars are present in the tail, but as suggested by the analysis of the SFH they are mostly young, born out of the stripped gas. The SFH maps also show that the galaxy had an enhancement of the star formation in the core for $t<6\times 10^8$ yr. This is expected when the gas is compressed by ram pressure stripping \citep{Roediger2014}.  The galaxy has a peculiar metallicity distribution:  the metallicity profile of the leading part of the galaxy is almost flat, while the trailing part shows a quite steep metallicity gradient. This is consistent with ram pressure stripping (A. Franchetto et al. in prep).  
However, given the properties of the groups, this classification can not be considered certain. The galaxy could also be affected by cosmic web stripping instead or undergoing accretion.

\subsubsection{P59597}
P59597 is a $\log(M_\ast/M_\sun)$=9.7  galaxy found in proximity of a large group and close to a binary system. According to \cite{Tempel2014_g}, the group has the following properties: z=0.053, 26 members, halo mass of 10$^{13.65}$\ms. However, Fig. \ref{fig:rps_env}  shows that the automatic approach might have broken the same structure into many smaller systems and P59597 could actually be part of this larger group. 
Note that the closest galaxy at similar redshift is P59630, 
which has a mass of 10$^{9.7}$\ms \citep{Calvi2011TheGalaxies} and is located at 5.1\arcmin, from P59597. According to \cite{Vollmer2005}, the galaxies are not interacting. 

P59597 is seen with a quite high inclination ($i=70^o$). The rgb image in Fig.\ref{fig:rps_P59597} shows a trail of bright knots extending towards the North-West. As highlighted by the contours, the main body of the galaxy is located in the southern part of the galaxy, the upper part could be a spiral arm opening up. The \Ha flux map reveals two parallel sequences of bright knots. The Northern sequence presents a shifted velocity field of the \Ha flux compared to the Southern one, suggesting again this is a spiral arm seen almost edge on. The gas kinematic is stretched towards South and asymmetric. The stellar kinematics presents a similar extent, but rotation is more regular { ($A_g =0.38 $, $A_v=0.13$)}. The observed bending towards West might be due to a spiral arm seen  almost edge on. The presence of stars also in the southern part of the galaxy indicates that these galaxies have formed out of the stripped gas. Indeed, the SFH shows that the tail was not as clear in the past as it is today. The SFH maps indicate the galaxy had a central  burst in the recent epochs. This could be due to the compression of the gas as a consequence of the ram pressure stripping \citep{Roediger2014}. P59597 has a metallicity that decreases from the center towards the outskirts, and the metallicity in the tail is the lowest (12+$\log[O/H]=8.3$). In the center, two peaks in metallicity are seen, but this again might be an effect of the inclination, with a metal rich region present in one arm. While the gas in the central regions of the galaxy is ionized by star formation, the BPT maps suggest that in the tail additional mechanisms are taking place. 

To conclude, this galaxy can be considered a certain case of ram pressure stripping induced by the hosting group. 

\subsection{Cosmic web stripping} \label{sec:cws_a}
\begin{figure*}
\centering
\includegraphics[scale=0.6]{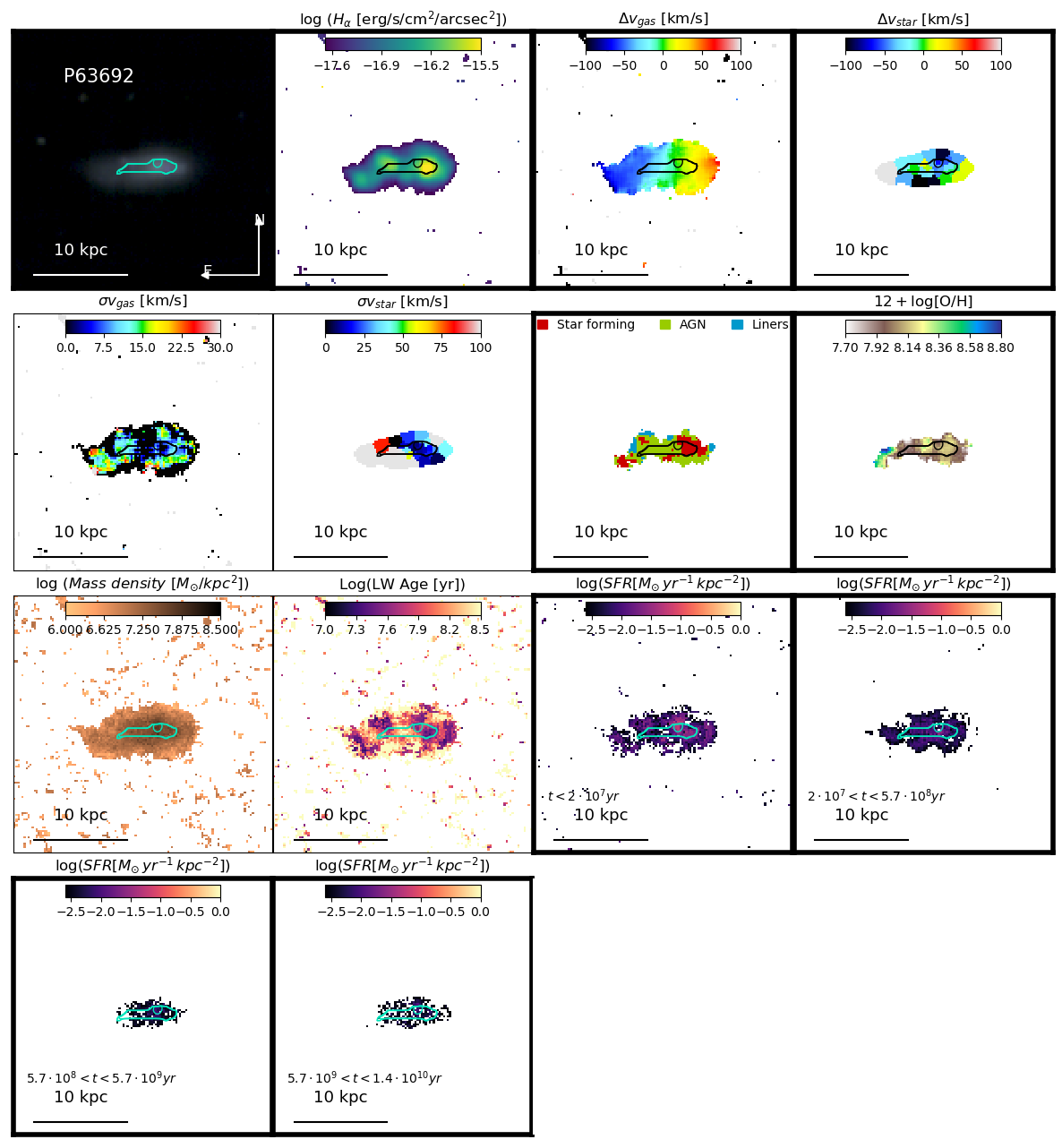}
\caption{P63692: Cosmic web stripped candidate. Panels, lines and colors are as in Fig.\ref{fig:P17048_bis}.\label{fig:cws_a} }
\end{figure*}

\begin{figure*}
\centering
\includegraphics[scale=0.3]{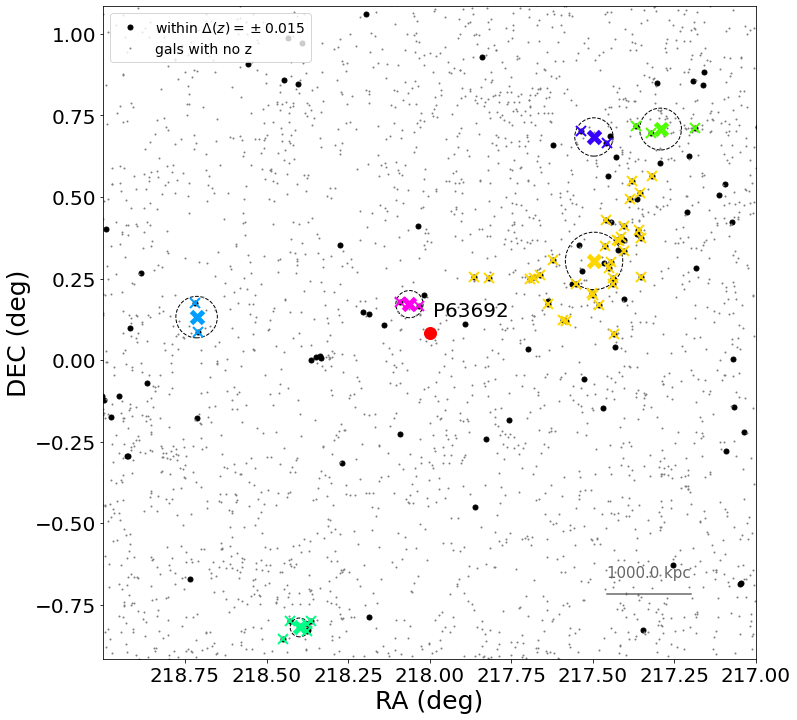}
\caption{Spatial distribution of galaxies around P63692.  Colors  and symbols have the same meaning as in Fig. \ref{fig:P17048_bis}. \label{fig:cws_env_a} }
\end{figure*}

\subsubsection{P63692}
P63692 is the least massive galaxy in GASP, with a mass of $10^{8.9}$\ms. The rgb in Fig.\ref{fig:cws_a} does not highlight a clear central structure but highlights a lopsidedness towards the East {  that resembles a gas tail and whose presence is supported by the analysis of the \Ha extent (Fig.\ref{fig:tail})}. The same lopsidedness {  and tail are} confirmed in the \Ha flux map, where a bright clump is detected. This clump does not perfectly correspond to the original body, though, but it is shifted West.  Few additional bright clumps are also seen. The gas velocity field is  asymmetric { ($A_g=0.35 $)}, spanning a range -80$<v_{gas} [km/s]<$30. The stellar kinematics seems chaotic, but is most likely due to the low S/N. The galaxy has very low metallicity ($12+\log[O/H]\sim 8$), but in the very end of the tail, where it reaches $12+\log[O/H]\sim 8.6$. The BPT map highlights the presence of shocks and other mechanisms ionizing the gas. The SFH maps shows that the galaxy acquired the current extent only for $t<5.7\times 10^8$ yr. At earlier epochs the galaxy was more symmetric. 

P63692 is not connected to any bound structure, so ram pressure stripping might not be advocated. 
The direction of the tail though is consistent with the galaxy approaching a massive group found nearby. This group is at 2155 kpc, its redshift is 0.05612, its velocity dispersion  $\sigma= $246.4.6 km/s. The velocity difference between the galaxy and the group is very small ($\sim 14 km/s$), but the galaxy is found $\sim 6$  R$_{vir}$ from the group center. 
P63692 is only 420 kpc away (2.6R$_{vir}$ given the size of the group) from another small group, whose size ($\sigma =120 km/s$) means it is most likely too small to exert any ram pressure stripping. 
The galaxy is therefore most likely experiencing cosmic web stripping, {  even though we can state it clearly, mainly due to the high uncertainties in the stellar velocity field.}

\subsection{Starvation} \label{sec:passive_a}
\begin{figure*}
\centering
\includegraphics[scale=0.5]{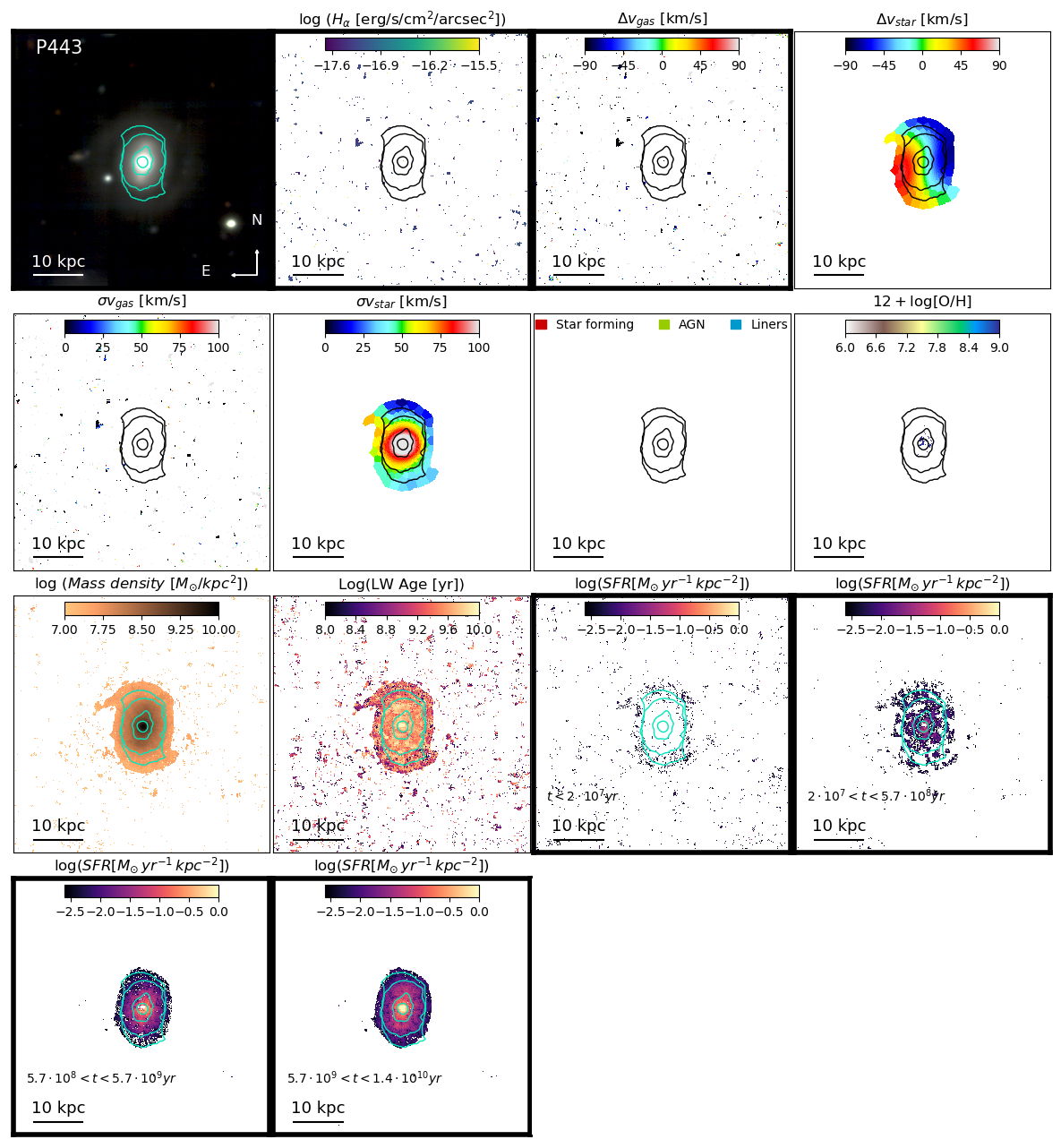}
\includegraphics[scale=0.5]{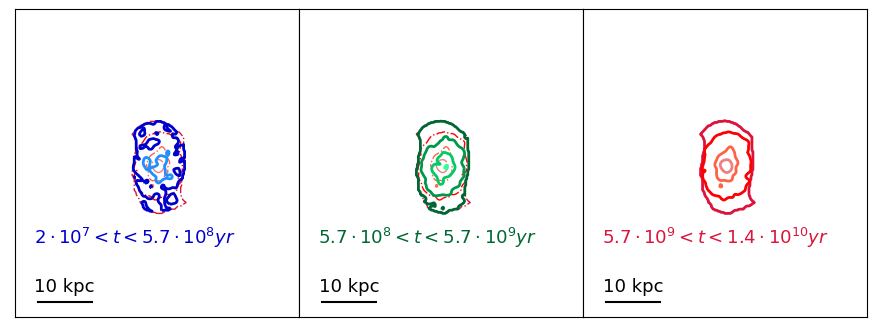}
\includegraphics[scale=0.33]{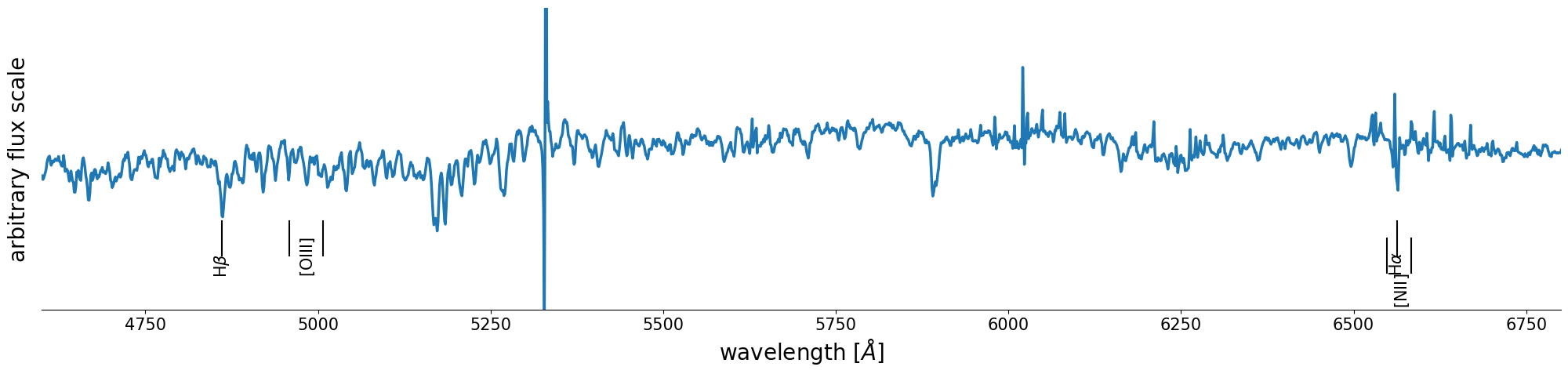}
\caption{P433: starvation candidate. Panels, lines and colors are as in Fig. \ref{fig:passive} and \ref{fig:P17048_bis}.\label{fig:passive_a} }
\end{figure*}

\subsubsection{P443}
P433 is a face on spiral galaxy, characterized by the presence of a bar and a ring, whose formation is usually associated with the interplay between bar-driven inflow and bar resonances \citep{Schwarz1984, Byrd1994, Piner1995}. It is also characterized by boxy isophotes, which are another indication of the presence of a bar \citep{Patsis2002,  Aronica2003, ONeill2003}. 

The stellar kinematics is overall regular {($A_v=0.23$)}, but disturbed by the presence of the bar, with respect to which it is tilted. The \Ha flux map (not shown) shows no flux throughout the entire galaxy disk. This is nicely summarized in the integrated spectrum that reveals no emission lines. This galaxy is therefore a passive disk with no signs of interactions. 

The comparison of the SFHs indicates that the galaxy had a similar extent as long as it was forming stars. The galaxy turned passive only at $t<2\times 10^7$yr. The suppression of the star formation occurred at the same rate at all distances from the center: there are no signs of outside-in quenching. 

The galaxy is found just within the virial radius (0.9 R$_{vir}$) of a four member group ($\sigma =$167.2 km/s, virial radius = 0.0663 Mpc, $\log M_{halo}/M_\odot$= 12.6). The quenching could therefore be due to the change of the environment. Given the similar suppression of the SFR at all distances from the galaxy center, we point to starvation as the most probable mechanism for the quenching. 

On top of that, also the presence of the bar could have quenched the galaxy. Indeed, galactic bars very likely play an important role in both secular evolution of disk galaxies \citep{Combes1981, Combes1990, Debattista2004, Athanassoula2005, Athanassoula2013, DiMatteo2015, Gadotti2015}, and in a dynamical re-distribution of gas \citep{Combes1985,  Athanassoula1992, Athanassoula2000, RomeroGomez2007, Berentzen2007}. 
Their formation and dynamics are potentially important mechanisms for regulating the evolution of the SFR in disk galaxies \citep[e.g.,][]{Laurikainen2004, Jogee2005, Masters2010, Ellison2011} and its quenching \citep{Cheung2013, James2016, Gavazzi2015, Carles2016}. Bars can sweep most of the galaxy's gas into the galactic center, where it is then converted into stars. The strength of the resulting bar-induced starburst depends on the mass of the galaxy, with massive barred galaxies converting all the gas funnelled to their centers into stars \citep{Carles2016}.

However, as the dynamics of gas in a bar potential can also depend on the local environment of the interstellar medium (ISM) \citep[see e.g.,][]{Athanassoula1992, Piner1995, Englmaier2000,  Wada2001, Maciejewski2002, Regan2004,  Fragkoudi2016}, we can not exclude that the environment is also determining the bar's fate, as in \cite{Gadotti2015}.

\end{document}